\documentclass[pra, twocolumn, superscriptaddress, groupaddress, floatfix, 10pt, aps]{revtex4-2}

\usepackage{array}[=2016-10-06]
\usepackage{slashed}
\usepackage{graphicx}
\usepackage{subfigure}
\usepackage[colorlinks=true,linkcolor=blue,citecolor=blue,urlcolor=blue]{hyperref}
\usepackage{amsmath}
\usepackage{amsfonts}
\usepackage[toc]{appendix}
\usepackage{txfonts}
\usepackage[british]{babel}
\usepackage{physics}
\usepackage{multirow}
\usepackage{tabularx}

\everymath{\displaystyle}
\newcommand{\pr}[1]{\ensuremath{\left[#1\right]}}
\newcommand{\pc}[1]{\ensuremath{\left(#1\right)}} 
\newcommand{\px}[1]{\ensuremath{\left\lbrace#1\right\rbrace}} 
\newcommand{\md}[1]{\ensuremath{\left\vert#1\right\vert}} 
\newcommand{\av}[1]{\ensuremath{\left\langle#1\right\rangle}} 
\newcommand{\mi}{\mathrm{i}}

\usepackage{tikz}
\usepackage{comment}
\usepackage{hhline}
\usetikzlibrary{arrows, shapes, decorations.markings, calc, intersections, backgrounds, shapes.misc, external, positioning, fit}
\usepackage{pgfplots}
\pgfplotsset{compat=newest}
\usetikzlibrary{pgfplots.colormaps}

\definecolor{lapislazuli}{rgb}{0.15, 0.38, 0.61}
\definecolor{YKblue}{rgb}{0.0, 0.18, 0.65}
\definecolor{carmine}{rgb}{0.81, 0.09, 0.03}
\definecolor{lavender}{rgb}{0.84, 0.49, 0.87}

\begin{document}

\title{
Many-Body Effects in Dark-State Laser Cooling 
}

\author{Muhammad Miskeen Khan}
    \thanks{These authors contributed equally.}
    \email{muhammadmiskeen.khan@slu.edu}
	\affiliation{JILA, National Institute of Standards and Technology and University of Colorado, 440 UCB, Boulder, Colorado 80309, USA}
	\affiliation{Center for Theory of Quantum Matter, University of Colorado, Boulder, Colorado 80309, USA}
    \affiliation{Department of Electrical and Computer Engineering, Saint Louis University, St. Louis, Missouri, 63103, USA}

\author{David Wellnitz}
    \thanks{These authors contributed equally.}
    \email{d.wellnitz@fz-juelich.de}
	\affiliation{JILA, National Institute of Standards and Technology and University of Colorado, 440 UCB, Boulder, Colorado 80309, USA}
	\affiliation{Center for Theory of Quantum Matter, University of Colorado, Boulder, Colorado 80309, USA}
    \affiliation{Institute for Theoretical Nanoelectronics (PGI-2), Forschungszentrum Jülich, 52428 Jülich, Germany}
    \affiliation{Institute for Quantum Information, RWTH Aachen University, 52056 Aachen, Germany}

 \author{Bhuvanesh Sundar$^\circ$}
	\affiliation{JILA, National Institute of Standards and Technology and University of Colorado, 440 UCB, Boulder, Colorado 80309, USA}

 \author{Ansh Das }
	\affiliation{
Department of Physics, Indian Institute of Technology Madras, 600036 Chennai, India}
\affiliation{Center for Quantum Information, Communication and Computing, Indian Institute of Technology Madras, Chennai 600036, India}

    \author{Haoqing Zhang}
	\affiliation{JILA, National Institute of Standards and Technology and University of Colorado, 440 UCB, Boulder, Colorado 80309, USA}
	\affiliation{Center for Theory of Quantum Matter, University of Colorado, Boulder, Colorado 80309, USA}
 \author{Allison Carter }
	\affiliation{National Institute of Standards and Technology, Boulder, Colorado 80305}

 \author{John J. Bollinger}
	\affiliation{National Institute of Standards and Technology, Boulder, Colorado 80305}

 \author{Athreya Shankar }
	\affiliation{
Department of Physics, Indian Institute of Technology Madras, 600036 Chennai, India}
\affiliation{Center for Quantum Information, Communication and Computing, Indian Institute of Technology Madras, Chennai 600036, India}

	\author{Ana Maria Rey}
\affiliation{JILA, National Institute of Standards and Technology and University of Colorado, 440 UCB, Boulder, Colorado 80309, USA}
	\affiliation{Center for Theory of Quantum Matter, University of Colorado, Boulder, Colorado 80309, USA}

\begin{abstract}
We develop a unified many-body theory of two-photon dark-state laser cooling, the workhorse for preparing trapped ions close to their motional quantum ground state.  For ions with a $\Lambda$ level structure, driven by Raman lasers, we identify an ion-number-dependent crossover between weak and strong coupling where both the cooling rate and final temperature are simultaneously optimized. 
We obtain  simple analytic results in both extremes: In the weak coupling limit, we show a Lorentzian spin-absorption spectrum   determines the cooling rate and final occupation of the motional state, which are both independent of the number of ions. We also highlight the benefit of including an additional spin-dependent force in this case.  In the strong coupling regime, our theory reveals the role of collective dynamics arising from phonon exchange between dark and bright states, allowing us to explain the enhancement of the cooling rate with increasing ion number.
Our analytic results agree closely with exact numerical simulations and provide experimentally accessible guidelines for optimizing cooling in large ion crystals, a key step toward scalable, high-fidelity trapped-ion quantum technologies.
\end{abstract}

\maketitle

\section{ Introduction} 

Coherent quantum control of trapped atomic systems \cite{RevModPhys.93.025001,Gross2021, Zhang_2022,Vilas2024} enables a wide range of quantum technologies, including information processing \cite{Bluvstein2023, Castelvecchi2023, RevModPhys.82.1041}, computation \cite{Egan2021}, simulation \cite{Zhang2017,PhysRevLett.128.120404,Britton2012,PhysRevLett.121.040503,PRXQuantum.3.040324}, sensing \cite{RevModPhys.89.035002,RevModPhys.87.637,Gilmore2021}, and fundamental tests of physics \cite{Roussy2023}. In trapped-ion systems, interactions and entanglement between ions are typically mediated by their collective motion \cite{RevModPhys.93.025001,foss2025progress}. Thermal excitations of this motion introduce decoherence, limiting the fidelity and scalability of quantum technologies \cite{bharti2024strong,picard2025entanglement}. Near-ground-state cooling of large ion crystals is therefore essential for advancing high-performance quantum simulation, computation, and metrology, but often takes a substantial fraction of experimental runtime~\cite{moses2023race}.

Electromagnetically induced transparency (EIT) cooling~\cite{PhysRevLett.85.4458,PhysRevLett.85.5547,PhysRevA.67.033402} is one of the workhorses to cool ions below the Doppler limit close to the ground state.
Using EIT cooling, experiments can reach average phonon occupations $\langle \hat n \rangle \sim 0.1$ within a few hundred $\mu$s~\cite{scharnhorst2018experimental,PhysRevLett.125.053001,PhysRevLett.122.053603,qiao2021double}.
Beyond speed and efficiency, a broad cooling bandwidth enables the simultaneous cooling of many modes.
This feature is especially appealing for large ion crystals with many modes~\cite{PhysRevA.93.053401,PhysRevLett.125.053001,PhysRevLett.122.053603,qiao2021double,zhang2022parallel,huang2024electromagnetically}, and for sympathetic cooling using multi-species architectures~\cite{sun2023sympathetic,dawel2025a,wu2025electromagnetically}.
\begin{figure}[t!]
\includegraphics[width=1\columnwidth]{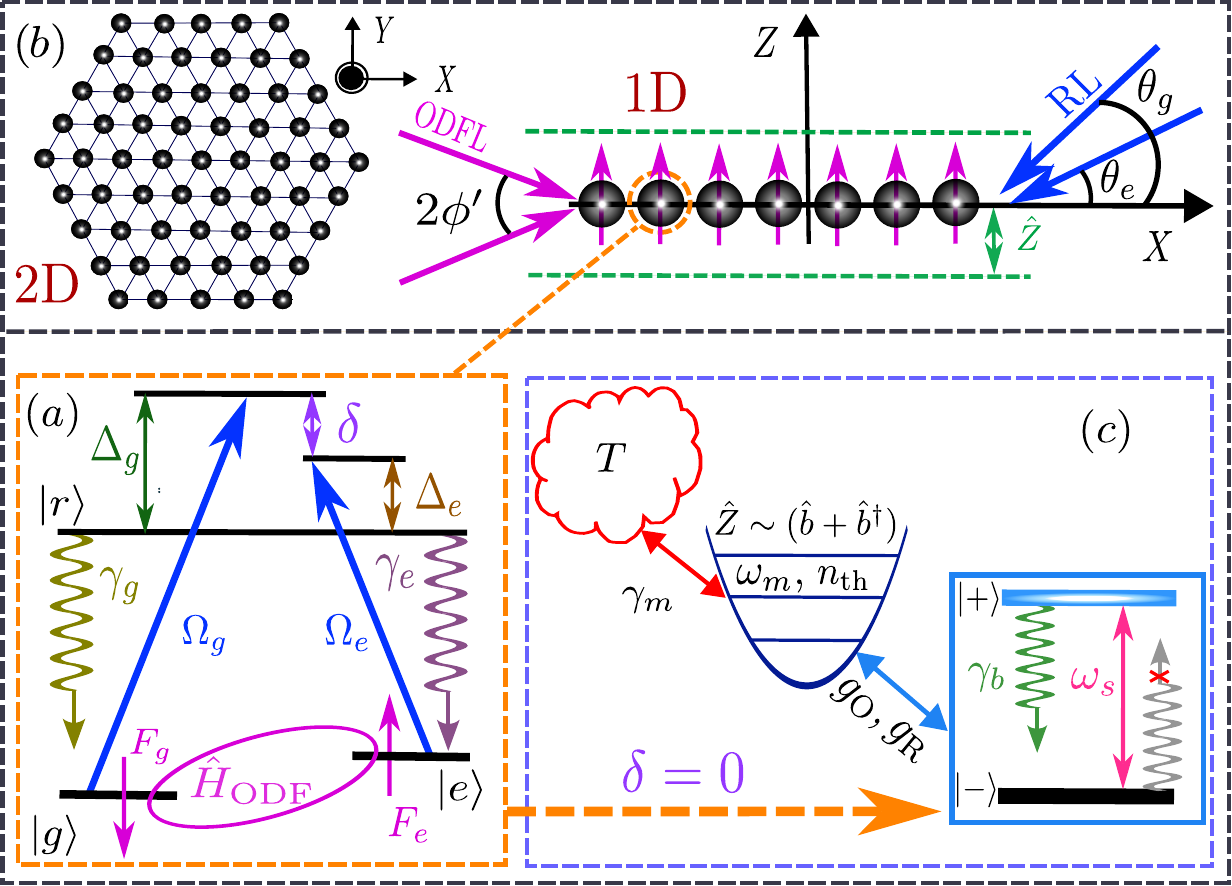}
\caption{Schematics of the cooling scheme. (a): A generic three level $\{\ket{g},\ket{e},\ket{r}\}$ atomic system is employed for cooling and it is driven by two Raman lasers (RL). (b) Such a level scheme can be realized in either 1D or 2D trapped ion crystals, where the RL couple (with rate $g_{\rm R}$) the internal states to the external collective motion of the ion crystal. Optical dipole-force lasers (ODFL) can add additional coupling (with rate $g_{\rm O}$) of the internal states $\{\ket{g},\ket{e}\}$ to the collective motion. (c) Effective two-level (dark $\ket{-}$ and bright $\ket{+}$ spin states) after adiabatically eliminating the state $\ket{r}$. The effective two-level spin system is coupled to the mechanical mode of the ions with rate $g_{\rm R}$ and $g_{\rm O}$. The mode is further coupled to the intrinsic thermal bath at temperature $T$.
}
\label{Fig:Fig1}
\end{figure}
EIT cooling, however,  is typically analyzed for tightly confined ions driven by two Raman beams in the two-photon Lamb–Dicke regime with highly imbalanced Rabi frequencies~\cite{PhysRevLett.85.4458,PhysRevLett.85.5547,PhysRevA.67.033402,scharnhorst2018experimental,wu2025electromagnetically,PhysRevA.93.053401,PhysRevLett.125.053001,qiao2021double,zhang2022parallel,huang2024electromagnetically,dawel2025a}. These assumptions can be experimentally restrictive, e.g., requiring very small Raman beam angles. Moreover, outside this regime, a collective speed-up of cooling can occur for balanced Rabi frequencies ~\cite{PhysRevLett.122.053603}. Numerical approaches can describe such behavior~\cite{PhysRevA.99.023409,bartolotta2024laser,fouka2025multilevel,PhysRevLett.73.2829} but offer limited intuition. Therefore, analytical understanding beyond the traditional EIT cooling is needed to better support experimental optimization.

In this article, we develop a unified framework for dark-state laser cooling theory that extends beyond the traditional EIT regime and explicitly investigates the role of many-body effects in cooling. This is done by adiabatically eliminating the optically excited state in a $\Lambda$ level structure, which allows us to derive an effective two-level picture of cooling valid for arbitrary spin-motion coupling strength. In our reduced two-level setting, we  find analytical expressions for the cooling rates and final temperatures for arbitrary ion numbers in the weak and strong coupling regimes, given in Table~\ref{tab:analytic_results}. All our analytic  results agree well with exact simulations.

The  analysis allows us to conclude  that while cooling is fastest in the strong coupling regime, in the weak coupling regime one can  reach a lower temperature. As such, temperature and cooling rate are simultaneously optimized at the value of the Lamb–Dicke parameter corresponding to the crossover between the two regimes.
Regarding many-body effects, our theory predicts that in the weak coupling regime the cooling performance is independent of the number of ions. In contrast, in the strong coupling regime the cooling rate does improve with increasing ion number. The optimal crossover point  therefore changes with the number of ions.

This article is structured as follows.
We start in Sec.~\ref{sec:review} by reviewing some relevant predictions of EIT cooling.
In Sec.~\ref{sec:single-ion}, we introduce the model and discuss cooling dynamics for a single ion.
In Sec.~\ref{sec:multi-ion}, we generalize to the multi-ion case and discuss the collective effects in the strong coupling regime.
In Sec.~\ref{sec:IV} we then numerically benchmark our analytic results against exact numerics.
In Sec.~\ref{sec:optimal-parameters} we  identify the optimal cooling parameters based on our analytic derivations. We finally conclude in Sec.~\ref{sec:conclusion}.

\section{EIT Cooling}
\label{sec:review}

In the standard EIT cooling picture~\cite{PhysRevLett.85.4458,PhysRevA.67.033402}, two lasers drive a $\Lambda$ three-level  system, between  two long lived $\ket{g},\ket{e}$ states and an optical  state $\ket{r}$, in the deep Lamb-Dicke regime. The lasers are set  on two-photon resonance, $\delta = 0$, with highly imbalanced Raman Rabi frequencies: a strong coupling $\ket{ g} \leftrightarrow \ket{r}$ Rabi frequency $\Omega_g$,  and a weak probe field, $\ket{e}\leftrightarrow\ket{r}$, Rabi frequency $\Omega_e$, so that  $\Omega_g \gg \Omega_e$ (see Fig.~\ref{Fig:Fig1}(a)).
Without motion, the system is trapped in a dark state ($\ket{d} \approx \ket{e}$) which does not scatter photons~\cite{RevModPhys.77.633}.
This is commonly explained through the Fano-like photon scattering spectrum of the weak probe laser, which is zero exactly on two-photon resonance.
Then, only motional sidebands away from two-photon resonance can scatter photons.
Because the Fano profile is highly asymmetric, it can be tuned so that during the change of the internal state,  phonon absorption is much more likely than phonon emission, thus cooling the motion.
If the internal state after this step  is  transferred to $\ket g$, then it  is   quickly reset to $\ket{d}$ through absorption of a photon from the strong coupling laser, followed by spontaneous emission without changing the motion.

Recent progress has extended EIT cooling beyond this regime.
For example, for  balanced Raman Rabi frequencies where the Fano-like scattering rate becomes a coherent population trapping (CPT) profile \cite{Lounis1992,Gray:78}, there is no clear distinction between coupling and probe lasers, yet cooling can be very efficient~\cite{PhysRevLett.122.053603,PhysRevA.99.023409}.
Recently, theory was extended to perturbatively capture the effects beyond the deep Lamb-Dicke regime for a single ion, finding a speed-up and increasing steady-state temperature as the Lamb-Dicke parameter is increased~\cite{Zhang2021,PhysRevA.104.013117}.

Below we  develop a general framework for understanding EIT-like (dark state) laser cooling over a wider experimental parameter space.

\section{ Single-ion Cooling} 
\label{sec:single-ion}
 We consider first  a single  ion  with a set of three internal levels arranged in a $\Lambda$-type configuration, with two low energy hyperfine spin states $\ket{g}$ and $\ket{e}$ and an optically excited state $\ket{r}$ as shown in Fig.~\ref{Fig:Fig1} (a). Two off-resonant Raman lasers ${\rm (RL)}$, with wave vector $\vec{k}_{l}$ and laser frequency $\omega_{l}^{L}$, are used to drive the transitions $\ket {l}\rightarrow \ket{r}$, with $l={g,e}$. Their corresponding detunings from the transition are $\Delta_{l}$ and their Rabi frequencies $\Omega_{l}$. The difference $\delta\equiv \Delta_{e}-\Delta_{g}$ sets  the two-photon-detuning. In the rotating frame of the lasers frequencies (see Appendix \ref{Ap:AppendixA}), the laser-ion Hamiltonian reads (with  $\hat{\sigma}_{mn}\equiv\ket{m}\bra{n}$) 
\begin{align}
\hat{H}_{\rm RLI}/\hbar &=\sum_{l=g,e}\pr{\Delta _{l}\hat{\sigma}_{ll}+
\frac{\Omega_{l}}{2}\pc{\hat{\sigma}_{rl} e^{i \vec{k}_{l}.\vec{\hat{R}}}+ {\rm H.C.}}}.
\label{eq:hamoltoninaSystem2}
\end{align}
The goal here  is to cool the ion's   motional degrees of freedom. For simplicity we will  restrict our analysis to a single mode describing  motion along $Z$ direction characterized by the bosonic operator $\hat{b}$, with  mode frequency $\omega_{m}$. We will later adapt it to describe cooling of the Center of Mass mode (CM)  of an   ion array, and discuss multimode effects at the end of the individual parts in Sec.~\ref{sec:multi-ion}.  One can therefore  expand the position  operator  of the ion as $\hat{Z}=Z_{zpf}(\hat{b}+\hat{b}^{\dagger})$, with  a phonon annihilation (creation) operator $\hat{b}~(\hat{b}^{\dagger})$. The quantity $Z_{zpf}\equiv \sqrt{\hbar/(2 m _{I}\omega_{m})}$ is the zero point fluctuation (ZPF) and $m_{I}$ the ion mass (see Tables~\ref{tab:weak_symbols} and \ref{tab:strong_symbols} for a summary of all symbols). We additionally introduce a spin-dependent force acting on the ion in consideration. For ions this can be  associated with the  so called optical dipole force described by the Hamiltonian $\hat{H}_{\rm ODF}=g_{\rm O}(\hat{\sigma}_{ee}-\hat{\sigma}_{gg})(\hat{b}+\hat{b}^\dagger)$. This spin-phonon interaction  with coupling rate $g_{\rm O}$, can be generated by  an additional pair of lasers tuned to generate  a force that is identical and opposite for the two low energy states, as schematically shown in Fig.~\ref{Fig:Fig1} (a). Such ODF-type interactions have been experimentally engineered in trapped-ion systems \cite{PhysRevA.91.012319, PhysRevLett.121.040503,Britton2012}. Moreover, they have been theoretically proposed as a means to enhance the cooling rate and reduce the final temperature toward a vanishing mean phonon occupation, $\av{\hat n}(t\to\infty)\to0$ when combined with Raman lasers driving a $\Lambda$-type three-level system \cite{retzker2007fast,PhysRevLett.104.043003, Albrecht_2011}.

The total Hamiltonian for the ion is then 
\begin{align}
\hat{H}=\hat{H}_{\rm RLI}+ \hat{H}_{\rm ODF}+\omega_{m}\hat{b}^{\dagger}\hat{b}.
\label{eq:hamoltoninaSystemLabeled}
\end{align} We further consider spontaneous decay of the excited state $\ket{r}\rightarrow\ket{l} $ at a rate   $\gamma_{l}$, described  by the two jump operators
\begin{align}
    \hat{L}_{l} &= \sqrt{\gamma_{l}}\hat{\sigma}_{lr}
\label{eq:SystemBasicJumpoperators}
\end{align}
For now, we neglect any recoil effect of the spontaneously emitted photon, and consider their effects on cooling in the section \ref{sec:IV}.

In the  far detuned limit, $\Delta_{ g,e}\gg \px{\Omega_{g},\Omega_{e},\gamma_{g},\gamma_{e}}$  one can  adiabatically eliminate the optical state $\ket{r}$ and obtain an effective master equation (see Appendix \ref{Ap:AppendixA}) for the two lower spin states, given by 
\begin{align}
\partial_{t}\hat{{\rho}}=-i\pr{\hat{{H}}_{\rm eff},\hat{{\rho}}}+\textstyle{\sum_{\hat{L}_{\beta}}}\mathcal{D}_{\hat{L}_{\beta}}[\hat{{\rho}}].
\label{eq:MasterEqDressedstate1}
\end{align}
Here the dissipative part is $\textstyle{\mathcal{D}_{\hat{O}}[\hat{{\rho}}]=\hat{O}\hat{{\rho}} \hat{O}^{\dagger}-\frac{1}{2}[\hat{O}^{\dagger}\hat{O}\hat{{\rho}}+\hat{{\rho}}\hat{O}^{\dagger}\hat{O}]}$, where $\hat{O}$ are the denoted jump operators. The effective  Hamiltonian is given by: $\hat{H}_{\rm eff}=\hat{H}_{\rm ODF}+\omega_{m}\hat{b}^{\dagger}\hat{b}+ \hat{H}_{\rm R}$, where
\begin{align}
\hat{H}_{\rm R}/\hbar=(-\delta+\omega_{{\rm LS}})\frac{\hat{\sigma}_{ee}-\hat{\sigma}_{gg}}{2}
+\frac{ \Omega_{\rm R} e^{i{\Delta k_z}\hat{Z}}\hat{\sigma}_{ge}+{\rm H.C.} }{2}.
\label{eq:SpinhamoEffRaman}
\end{align}
Here $\omega_{{\rm LS}}= \Delta_{e}\Omega_{e}^{2}/(4\Delta_{e}^{2}+\gamma^{2})- \Delta_{g}\Omega_{g}^{2}/(4\Delta_{g}^{2}+\gamma^{2})$  is the induced differential light-shift, and $\Omega_{\rm R}=\Omega_{g}\Omega_{e}(\Delta_{g}+\Delta_{e})/[(4 (\Delta_{g}-i\gamma/2)(\Delta_{e}+i\gamma/2)]$ an effective Raman coupling, and ${\Delta k_z}=k_{ez}-k_{gz}$ is the wave vector difference along the cooling direction. The effective light induced incoherent spin scattering processes are determined by the jump operators
\begin{align}
\hat{L}_{g}^{{\rm eff}}=\alpha_{gg}\hat{\sigma}_{gg}+\alpha_{ge}\hat{\sigma}_{ge},~ \hat{L}_{e}^{{\rm eff}}=\alpha_{eg}\hat{\sigma}_{eg}+\alpha_{ee}\hat{\sigma}_{ee},
\label{eq:SpinJumpEff}
\end{align}
with $\textstyle{\alpha_{kl}=(\sqrt{\gamma_{k}}\Omega_{l})/(2\Delta_{l}-i\gamma})$, $k,l=\{g,e\}$, and  $\gamma=\gamma_{g}+\gamma_{e}$. 

To capture various additional heating processes that can increase the phonon mode temperature without affecting the ion's internal state, we introduce generic jump operators \cite{PhysRevLett.82.1971, PhysRevA.62.022311}
\begin{align}
\hat{L}^{m}_{\hat{b}}=\sqrt{\gamma_{m}\pc{n_{\rm th}+1}}~\hat{b},\quad \hat{L}^{m}_{\hat{b}^{\dagger}}=\sqrt{\gamma_{m}n_{\rm th}}~\hat{b}^{\dagger}.
\label{eq:JumThermalPhonon1}
\end{align}
In the absence of cooling, these jump operators will increase the phonon mode occupation to a thermal state with $\langle \hat n \rangle = n_{\rm th} \gg 1$ at rate $\gamma_m$. For convenience, we choose $n_{\rm th}$ to coincide with our initial occupation.

By setting $\delta=0$, i.e. $\Delta_{g}=\Delta_{e}\equiv\Delta_{\rm R}$ (which will be our operation point), and in the  Lamb-Dicke regime, $\eta_{z}^{\prime}\equiv \eta_{z}\sqrt{n_{\rm th}+1}={\Delta k_z}Z _{zpf}\sqrt{n_{\rm th}+1}\ll 1$, the total effective Hamiltonian acting on the lower two spin levels (can be written in Pauli spin operators $\hat{\sigma}_{x,y,z}$) to linear order in $\eta_{z}$ simplifies to: $\hat{H}_{\rm eff}=\hat{H}_{s}+\hat{H}_{\rm int}$, with 
\begin{align}
    &\hat{H}_{s} =
    \frac{\omega_{{\rm LS}}}{2} \hat{\sigma}_{z} + \frac{\Omega_{\rm R}}{2}\hat{\sigma}_{x},
    \label{eq:Hamiltonian4}\\
    \hat{H}_{\rm int} &=(g_{\rm R}\hat{\sigma}_{y}+g_{\rm O}\hat{\sigma}_{z})(\hat{b}+\hat{b}^{\dagger})\equiv\hat{F}(\hat{b}+\hat{b}^{\dagger})
    \label{eq:EffIntarctionhamTerm}
\end{align}
Here, we have defined the Heisenberg spin force operator $\hat{F}=(g_{\rm R}\hat{\sigma}_{y}+g_{\rm O}\hat{\sigma}_{z})$ and $\textstyle{\Omega_{\rm R}=2\Omega_{g}\Omega_{e}\Delta_{\rm R}/(4\Delta_{\rm R}^2+\gamma^2)}$, where $g_{\rm R}=(\eta_{z}\Omega_{\rm R})/2$ is the effective Raman laser induced spin-motion coupling. 
We introduce the dressed basis (with tilde notation) i.e. $\hat{\tilde{\sigma}}_{x,y,z} \equiv e^{-i\frac{\alpha}{2} \hat{\sigma}_y} 
\hat{\sigma}_{x,y,z}
e^{i\frac{\alpha}{2}\hat{\sigma}_{y}}$, that diagonalize $\hat{H}_{s}$ and rewrite the Hamiltonian in this dressed basis as $\hat{\tilde{H}}_{\rm eff}=\hat{\tilde H}_{s}+\hat{\tilde{H}}_{\rm int}$:
\begin{align} \label{eq:barediagonalTLHam}
\hat{\tilde H}_{s}= (\omega_{s}/2)\hat{\tilde{\sigma}}_{z},
\end{align}
\begin{align}
 \hat{\tilde{H}}_{\rm int}=[g_{\rm O}(\cos{\alpha}\hat{\tilde{\sigma}}_{z}+\sin{\alpha}\hat{\tilde{\sigma}}_{x})+g_{\rm R} \hat{\tilde{\sigma}}_{y}](\hat{b}+\hat{b}^{\dagger}),
 \label{eq:H_int_singleion}
\end{align}
Here, $\omega_{s}=\textstyle{\sqrt{\omega_{\rm LS}^{2}+\Omega_{\rm R}^{2}}=\Delta_{\rm R}(\Omega_{g}^2+\Omega_{e}^2)/(\gamma^2+4 \Delta_{\rm R}^2)}$ is the frequency splitting of the eigenstates of $\hat H_s$ with $\tan{\alpha}=(\Omega_{\rm R}/\omega_{\rm LS})$. These eigenstates,
$\hat {\tilde{H}}_s \ket{\pm} =\pm \omega_s/2 \ket{\pm}$, are the bright $\ket{+}=(\Omega_{g}\ket{g}+\Omega_{e}\ket{e})/\Omega_{s}$ and dark $\ket{-}=(\Omega_{e}\ket{g}-\Omega_{g}\ket{e})/\Omega_{s}$ states, with $\Omega_{s}=\sqrt{\Omega_{g}^2+\Omega_{e}^2}$. Furthermore, in the dressed basis $\hat{\tilde{H}}_{\rm int}$ offers phonon changing side-band transitions, without a carrier term. Importantly, in this dressed basis the spin jump operators become
\begin{align} \label{eq:DissipatinJumpDressedframe12} 
\hat{\tilde{L}}_{\rm 1}=(\tilde{\alpha}_{gg}\ket{+}+\tilde{\alpha}_{ge}\ket{-})\bra{+},\\
\label{eq:DissipatinJumpDressedframe13}
\hat{\tilde{L}}_{\rm 2}=(\tilde{\alpha}_{ee}\ket{+}-\tilde{\alpha}_{eg}\ket{-})\bra{+}.
\end{align}
where $\tilde{\alpha}_{kl=e,g}=(i\sqrt{\gamma_{k}}\Omega_{l})/(\gamma+2 i\Delta_{\rm R})$, with an effective bright state decay $\textstyle{\gamma_{b}=\sum_{k,l}|\tilde{\alpha}_{kl}|^2=\gamma(\Omega_{g}^2+\Omega_{e}^2)/(\gamma^2+4\Delta_{\rm R}^2)}$. In this basis, $\hat{\tilde{L}}_{\rm 1,2}\ket{-}=0$, therefore $\ket{-}$ is a genuine dark state. The dynamics of the spin alone obeys  the following  optical-Bloch (OB) master equation

\begin{align}
\partial_{t}\hat{\tilde{\rho}}=-i[\hat{\tilde{H}}_{s},\hat{\tilde{\rho}}]+\sum\nolimits_{\hat{\tilde{L}}_{s}=\hat{\tilde{L}}_{\rm 1},\hat{\tilde{L}}_{\rm 2}}\mathcal{D}_{\hat{\tilde{L}}_{s}}[\hat{\tilde{\rho}}]\equiv \mathcal{L}_{\rm OB}\hat{\tilde{\rho}},
\label{eq:MasterEqDressedstate1OB}
\end{align}
with $\hat{\tilde{\rho}}$ being the density operator of the spin in the dressed basis.  

\subsection{ Weak coupling regime} 
\begin{table}[]
    \renewcommand{\arraystretch}{1.4}
    \centering
    \begin{tabularx}{\columnwidth}{
        c |
        c |
        >{\raggedright\arraybackslash}X
    }
        Symbol & Definition & Description \\
        \hline
        $\gamma $ & $\gamma_g + \gamma_e$ & Total spontaneous decay rate\\[4pt]
        $\Omega_{\rm R}$ & $\frac{\Omega_g \Omega_e(\Delta_g + \Delta_e)}{4 (\Delta_g - \mi \gamma/2) (\Delta_e - \mi \gamma/2)}$ & Raman Rabi frequency \\[8pt]
        $\omega_{\rm LS}$ & $\frac{\Delta_{e}\Omega_{e}^{2}}{4\Delta_{e}^{2}+\gamma^{2}}- \frac{\Delta_{g}\Omega_{g}^{2}}{4\Delta_{g}^{2}+\gamma^{2}}$& Total light shift in spin states\\
        $\alpha$ & $\arctan[\Omega_{\rm R}/\omega_{\rm LS}]$ & Dressed states mixing angle\\
        $\alpha_{kl=e,g} $ & $(\sqrt{\gamma_{k}}\Omega_{l})/(2\Delta_{l}-i\gamma)$ & Spin scattering rates in bare basis \\
        $\tilde \alpha_{kl=e,g}$ & $(\sqrt{\gamma_{k}}\Omega_{l})/(2 \Delta_{\rm R} - i \gamma )$& Spin scattering rates in dressed basis \\
         $\delta $ & $\Delta_g-\Delta_e$& Two-photon detuning \\
        $Z_{zpf}$ & $\sqrt{\hbar/(2 m _{I}\omega_{m})}$& Zero point fluctuations\\
        $\eta_{z}$ & $(|k_{ez}|-|k_{gz}|)Z_{zpf}$&Two-photon Lamb-Dicke parameter\\
        $g_{\rm R}$ & $(\eta_{z}\Omega_{\rm R})/2$ & Raman spin-motion coupling\\
        $\omega_s$ &$\Delta_{\rm R} (\Omega_{g}^2+\Omega_{e}^2)/(\gamma^2+4\Delta_{\rm R}^2)$& Dark-bright spin frequency splitting \\
        $\gamma_b$ &$ \gamma(\Omega_{g}^2+\Omega_{e}^2)/(\gamma^2+4\Delta_{\rm R}^2)$& Bright state linewidth\\
        $A_\pm$ & Eq.~\eqref{eq:Spin-Absorption}& Cooling and heating rates \\
        $\Gamma_c$ & $A_{-}-A_{+}$ & Net cooling rate \\
        $Q_{s}$ & $\Delta_{\rm R}/\gamma$ & Dark-bright spin quality factor \\
        $n_{\rm BA}$& $(\gamma/4\Delta_{\rm R})^2$& Quantum back action limit\\
        $\gamma_{s,\rm opt}$ & Eq.~\eqref{Eq:rates}& Optimal cooling rate \\
  \end{tabularx}
    \caption{Summary of the symbols used in the manuscript in the weak coupling regime. Longer equations are referenced.}
    \label{tab:weak_symbols}
\end{table}

\subsubsection{Analytic model}In the  weak spin-phonon coupling limit,  $\gamma_{b}\gg \{g_{\rm R},g_{\rm O}\} $ (or $\eta_{z}< [\gamma(\Omega_g^2+\Omega_{e}^2)]/[\Delta_{\rm R} \Omega_g \Omega_e] $ in terms of the free parameters), the spin is decoupled from the motion and decays at a much faster rate than any motional dynamics by following an independent dynamics under Eq. \eqref{eq:MasterEqDressedstate1OB}. Such decay manifests the optical pumping of the spin to a dark steady state $\textstyle{\hat{\tilde{\rho}}_{ss}=\ket{-}\bra{-}}$ (with $\mathcal{L}_{\rm OB}\hat{\tilde{\rho}}_{ss}=0$), a process known as coherent population trapping (CPT) \cite{RevModPhys.77.633}.
Therefore in the regime (i) $\{g_{\rm R},g_{\rm O}\} <\gamma_{b}<\omega_{m}$ and (ii) large cooperativity $\textstyle {\mathcal{C}}\equiv g_{\rm R,O}^2/(\gamma_{b}\gamma_{m}n_{\rm th})\gg1$\cite{PhysRevA.46.2668, PhysRevA.49.2771, PhysRevA.102.033115} the effective two-level system  behaves as a pristine model for cooling (cf. Fig.~\ref{Fig:Fig1} (c)). The first inequality in (i) ensures the scattering away of a phonon from the mode to an electromagnetic vacuum bath through fast spin relaxation to dark state, and the second inequality in (i) requires that the sidebands (if present) are sufficiently resolved to avoid heating mechanisms due to undesired sideband transitions. Condition (ii) demands a large ratio of the spin-induced cooling rate $\sim g_{\rm R,O}^2/\gamma_{b}$ to the re-thermalization rate $\gamma_{m}n_{\rm th}$ of the mode.
\begin{figure}[t!]
\includegraphics[width=\columnwidth]{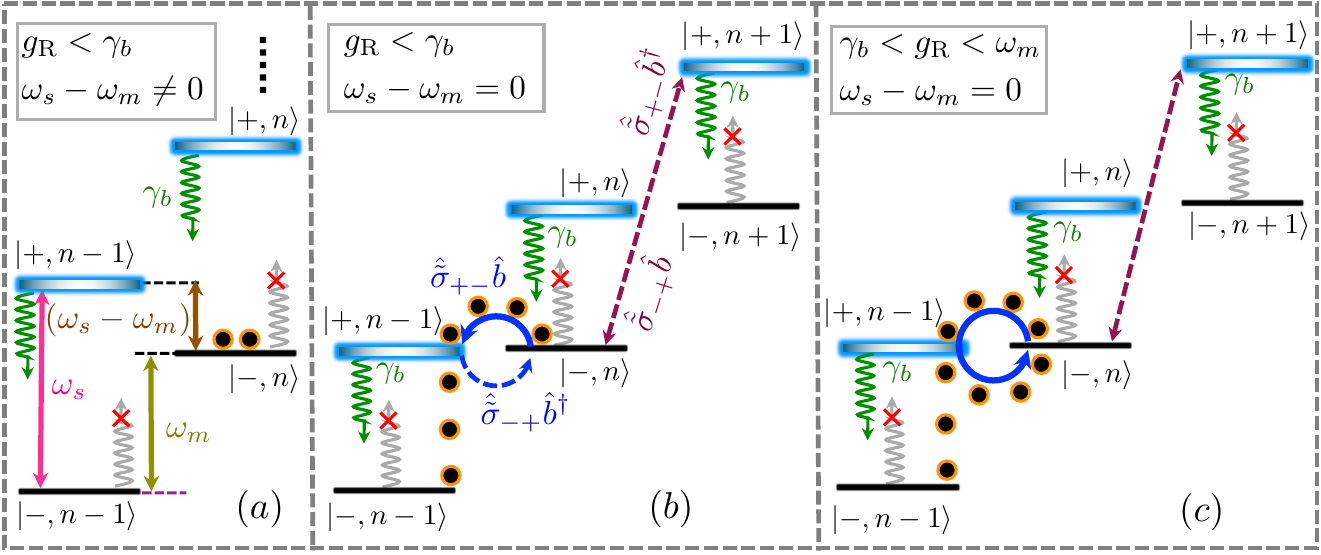}
\caption{Effective single-ion cooling dynamics. (a): The dark steady state is populated far from resonance $\omega_{s}-\omega_{m}\neq 0$. (b) Cooling cycle starts for the resonant case $\omega_{s}-\omega_{m}=0$ in the weak spin-phonon coupling limit, where the spin escapes from dark steady state by absorbing a phonon from the mode, followed by a fast reinitialization to the same dark state. The dominant processes are shown with solid arrowhead lines, while weak processes are shown with dashed arrowhead lines. The red cross mark represents the absence of such a process. (c) Cooling cycle in the strong spin-phonon coupling regime. Both Rabi flopping (shown as blue circles) and subsequent cooling steps take place (see text).
}
\label{Fig:Fig2}
\end{figure}

Out of resonance  ($\omega_{s}-\omega_{m})\ne0$,  the dark steady state is stable   as shown in Fig.~\ref{Fig:Fig2} (a). However, by setting the system at the spin-mode resonance condition, and $|\omega_{s}-\omega_{m}|\ll g_{\rm R,O}$, a cooling cycle (See Fig.~\ref{Fig:Fig2} (b)) starts, where  the red-side band term   $\hat{\tilde{\sigma}}_{+-} \hat{b} + {\rm H.C.}$ dominates over other terms in $\hat{\tilde{H}}_{\rm int}$. This results into the transition $\ket{-,n}\rightarrow \ket{+,n-1}$. Given  $g_{\rm R,O}<\gamma_{b}$, immediately the bright state  decays to the dark state $\ket{+,n-1} \rightarrow \ket{-,n-1}$ (i.e. fast optical pumping to dark state) instead of  performing spin-phonon Rabi oscillations. The coupled spin-phonon dynamics therefore follow the overall cooling cycle: $\ket{-,n}\rightarrow \ket{+,n-1}\rightarrow\ket{-,n-1}$.  We note that the present trapped particle cooling scheme with dark and bright dressed states is a direct analogue of a cooling scheme for free particles as performed using velocity selective coherent population trapping \cite{Aspect:89}, where the possibility of the dark to bright state transition is determined by the velocity of the free particles. The final cooling efficiency in the present scheme is set by $\gamma_m$ as we now proceed to discuss. 
\begin{figure}[t!]
	\includegraphics[width=0.46\columnwidth]{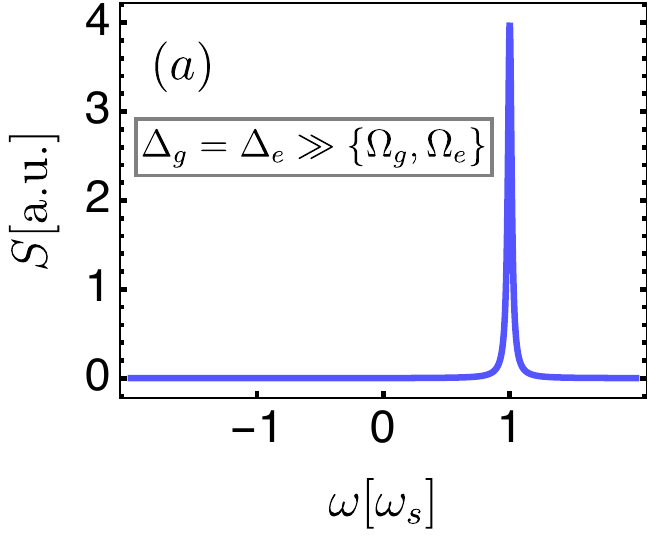}
		\includegraphics[width=0.49\columnwidth]{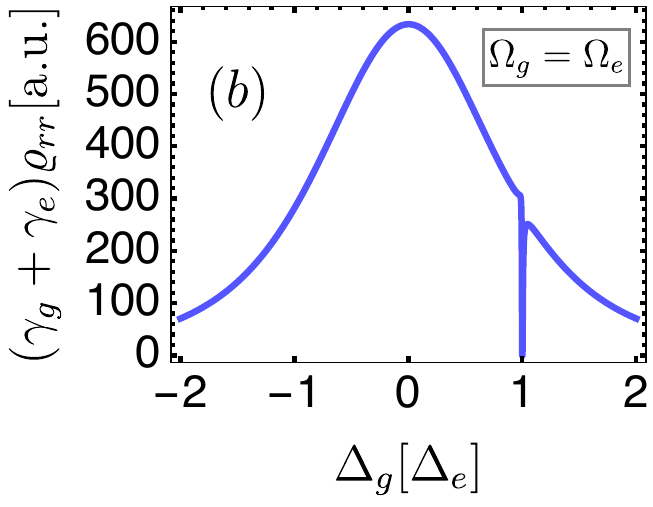}
\caption{Cooling spectra and excited state scattering rate. (a) Spin in a dark steady state is characterized by its Lorentzian absorption spectrum which determines the cooling rate and final occupation number of the motional state (see text). (b) Steady-state (SS) scattering rate $(\gamma_{g}+\gamma_{e})\varrho_{rr}$ of the optically excited state $\ket{r}$: It features an underlying coherent population trapping profile for  $\Omega_{g}=\Omega_{e}$  (see text).}
\label{Fig:Fig3}
\end{figure}
To determine the cooling rate and minimum possible occupation number, the two figures of merit of a cooling scheme, we use the
separation of the time scales $g_{\rm R,O}^{-1}\gg \gamma_{b}^{-1}$, between the two subsystems (spin and phonons), which   allows us to adiabatically eliminate the spin degrees of freedom and derive a master equation for the  phonons only, which is described by a Hamiltonian $\omega_{m}\hat{b}^{\dagger}\hat{b}$ with modified jump operators for the reduced mechanical mode given by
(see Appendix \ref{Ap:AppendixB} ,~\cite{PhysRevA.86.012126,Jaehne_2008}) 
\begin{align}
\label{eq:Spin-Absorption} 
&\hat{L}_{\hat{b}}^{m,{\rm R}}=
\sqrt{ A_-}~\hat{b}, \quad \hat{L}_{\hat{b}^{\dagger}}^{m,{\rm R}}=\sqrt{A_+}~ \hat{b}^{\dagger}; \nonumber\\
&A_-\equiv S\pc{\omega_{m}}+\gamma_{m}\pc{n_{\rm th}+1}, \quad A_+\equiv  S\pc{-\omega_{m}}+\gamma_{m}n_{\rm th}.
\end{align}
The corresponding rate equation for the mean phonon number,  $\langle\hat{n}\rangle=\langle{\hat{b}^{\dagger}\hat{b}}\rangle$, yields    
\begin{equation}
\langle \dot{\hat{n}} \rangle\equiv \Tr \pr{\dot{\hat{\rho}}_{m}\hat{b}^{\dagger}\hat{b}}=-\Gamma_{c}\langle \hat{n}\rangle+A_{+}.
\label{eq:eq_rate}
\end{equation} where $\Gamma_{c}=A_{-}-A_{+}$ is the net cooling rate, and $n_{f}\equiv \langle{\hat{b}^{\dagger}\hat{b}}\rangle|_{t\rightarrow \infty}= A_{+}/\Gamma_{c}$ the steady state final occupation number. In these expressions, $ \textstyle{S\pc{\omega}=2{\rm Re}[\int_{0}^{\infty}d\tau e^{i\omega \tau}\langle\widehat{\delta F}\pc{\tau} \widehat{\delta F}\rangle_{ss}}]$ is the spin absorption spectrum, where the function $\langle\widehat{\delta F}\pc{\tau} \widehat{\delta F} \rangle_{ss}$ is the steady state correlation function of the spin force fluctuation operators,  with $\widehat{\delta F}\equiv\hat{F}-\langle\hat F \rangle_{ss}$, which we evaluate around the mean field steady state value $\langle\hat{F}\rangle_{ss}$. The spectrum (see its derivation in Appendix \ref{spectrumcal}), as shown in Fig.~\ref{Fig:Fig3} (a), is a Lorentzian peaked at frequency $\omega_{s}$ with a width $\gamma_{b}$:
\begin{align}
S\pc{\omega}=\frac{\gamma_{b} \{4 g_{\rm O}^2 \Omega_{g}^2 \Omega_{e}^2+g_{\rm R}^2 (\Omega_{g}^2+\Omega_{e}^2)^2\}}{(\Omega_{e}^2+\Omega_{g}^2)^2[(\omega-\omega_{s})^{2}+(\gamma_{b}/2)^{2}]}.
\label{eq:Lorentzian}
\end{align}
It accounts for both damping and thermalization effects as provided by the spin-phonon coupling. The positive (negative) frequency region of the spectrum characterizes the ability of the spin to perform a transition from its initial dark steady state to a final bright state \cite{RevModPhys.82.1155, PhysRevA.46.2668, PhysRevB.82.165320, PMeystre2007}, by absorbing (emitting) a phonon through the side-band transition $\hat{\tilde{\sigma}}_{+-} \hat{b} ( \hat{\tilde{\sigma}}_{+-} \hat{b}^{\dagger} ) + {\rm H.C.}$  generated  by $\hat{\tilde{H}}_{\rm int}$, when tuned to $\omega_{s}=\omega_{m}(-\omega_{m})$. Close to resonance, $\omega_{s}=\omega_{m}$, the suppression of the negative part is therefore determined by the condition $\gamma_{b}<4\omega_{s}$.
At
 the optimal cooling point, $\Omega_{g}=\Omega_{e}\equiv \Omega$,  for $\gamma_{m}\ll\gamma_{s,{\rm opt}}$ (see Appendix \ref{Ap:AppendixB2} for derivation and full expressions), the net cooling rate is $\Gamma_{c,{\rm opt}}=\gamma_{s,{\rm opt}}+\gamma_{m}$, wherein $\gamma_{s,{\rm opt}}$ and the minimum occupation number are,
\begin{equation}\gamma_{s,{\rm opt}}=\frac{64 \Delta_{\rm R}^2(g_{\rm O}^2 +g_{\rm R}^2)}{(16\Delta_{\rm R}^2+\gamma^2)\gamma_{b}},\quad   n_{f,\rm min}\simeq\frac{1}{(4Q_{s})^2}+ \frac{\gamma_{m}n_{\rm th}}{\gamma_{s,{\rm opt}}}.
\label{Eq:rates} 
\end{equation} 
Here $Q_{s}\equiv\omega_{s}/\gamma_{b}=\Delta_{\rm R}/\gamma$ is the quality factor of the Lorentzian spin absorption spectrum and sets the quantum back-action limit (QBL) \cite{PhysRevLett.116.063601,PhysRevLett.99.093902} $n_{\rm BA}\equiv(1/4Q_{s})^2=(\gamma_{b}/4\omega_{s})^2=(\gamma/4\Delta_{\rm R})^2$  on minimum occupation number for $\gamma_{m}\rightarrow 0$, that can't be violated. The effective trapped-ion cooling theory described here exhibits a correspondence with optomechanical cooling. In optomechanical cooling, an optical cavity dark vacuum state with a Lorentzian absorption spectrum is fundamentally limited by the quantum back-action optomechanical cooling limit $n_{\rm BA}^{\rm OM}=(\kappa/4\omega_m)^2$ \cite{PhysRevLett.99.093902}, where $\kappa$ is the cavity (Lorentzian) linewidth and $\omega_m$ is the mechanical mode frequency.

Our simplified two level model with Lorentzian absorption spectrum has the great advantage that it further allows us to understand the net effect of the ODF on the cooling. This task has been challenging in treatments that retain the excited electronic level $\ket{r}$ in the model \cite{retzker2007fast,PhysRevLett.104.043003, Albrecht_2011, PhysRevA.98.013423} (with non-Lorentzian spectrum), where the QBL has been violated and $n_{f,min}=0$ has been shown. If we set $g_{\rm O}=p g_{\rm R}$ in Eq.~(\ref{Eq:rates}), we obtain a simple expression,
$\gamma_{s,\rm{opt}}= [(1+ p^2 )\Omega^2 \eta_z^2]/[2\gamma(1+(\gamma/2\Delta_{\rm R})^2)(1+(\gamma/4\Delta_{\rm R})^2)]$. From these expressions, it is clear that for $\Delta_{\rm R}\gg \gamma$, i.e. by operating in the large detuning regime, $\gamma_{s,\rm{opt}}$ becomes independent of $\Delta_{\rm R}$. Therefore, by only choosing $p$, one can enhance the cooling rate by a factor of $1+p^2$ which also reduces $n_{f,{\rm min}}$ in the presence of mode rethermalisation effects (i.e. for  $\gamma_{m}\neq 0$). We also note that the addition of the ODF spin-motion coupling fully respects the QBL for $\gamma_{m}=0$. 
\begin{figure}[t!]
	\includegraphics[width=0.48\columnwidth]{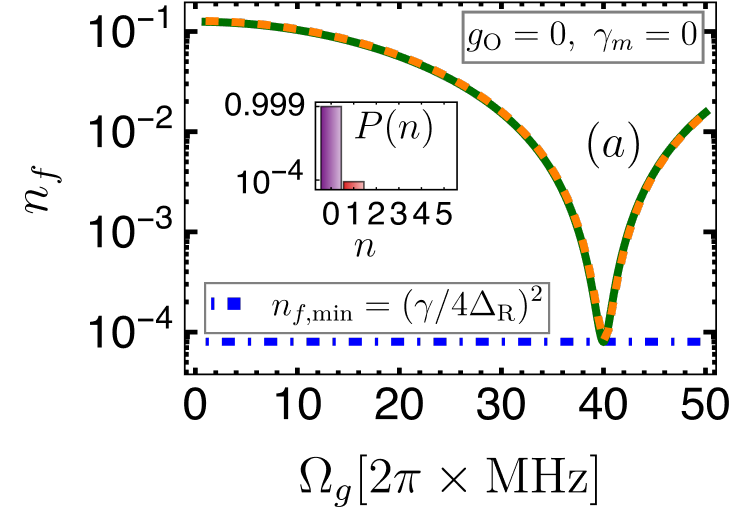}
		\includegraphics[width=0.48\columnwidth]{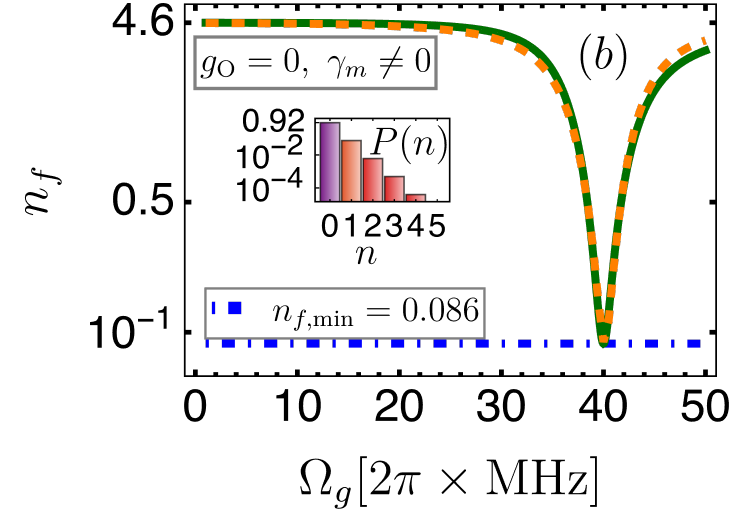}

			\includegraphics[width=0.48\columnwidth]{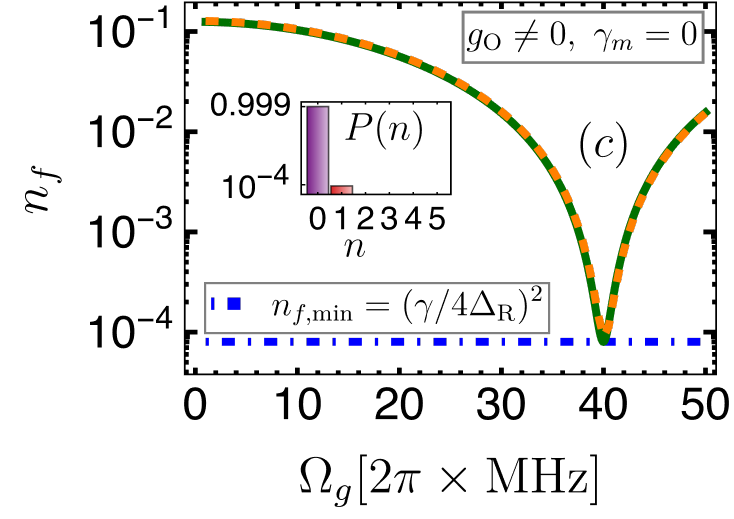}
		\includegraphics[width=0.48\columnwidth]{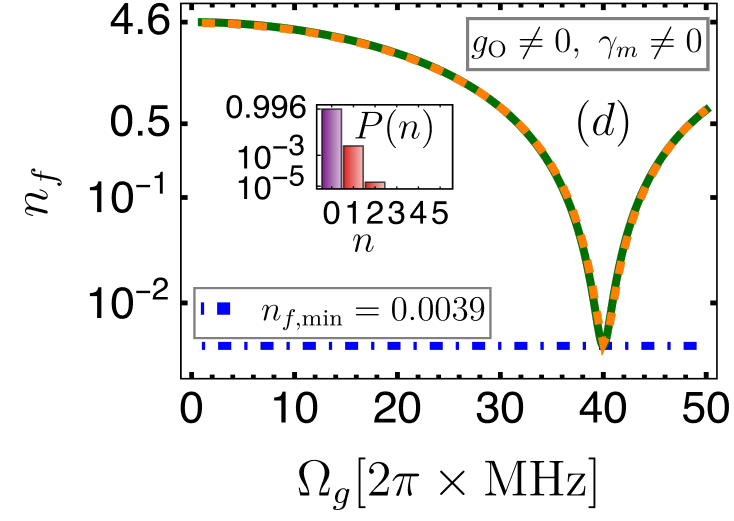}
		
				\includegraphics[width=0.48\columnwidth]{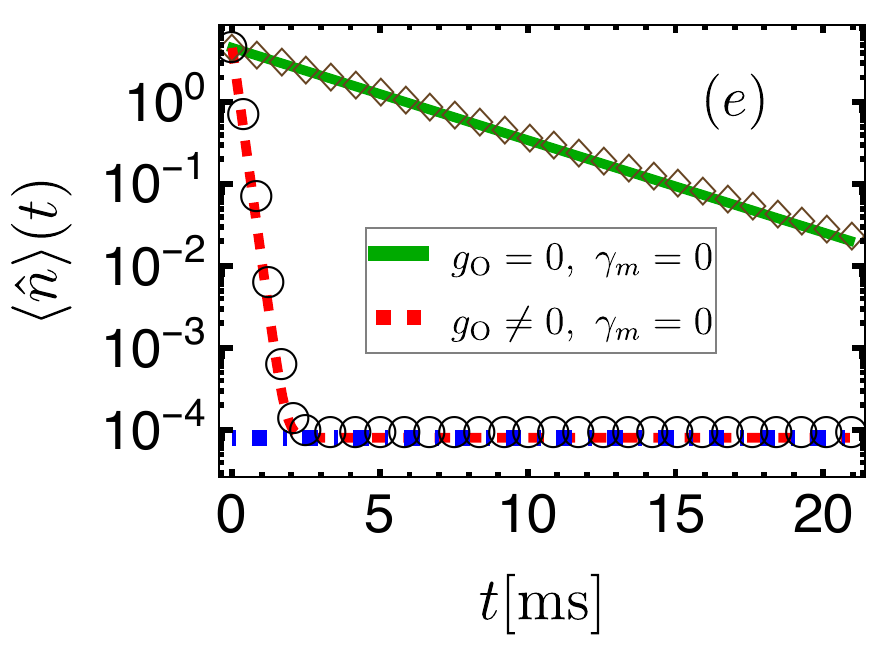}
		\includegraphics[width=0.48\columnwidth]{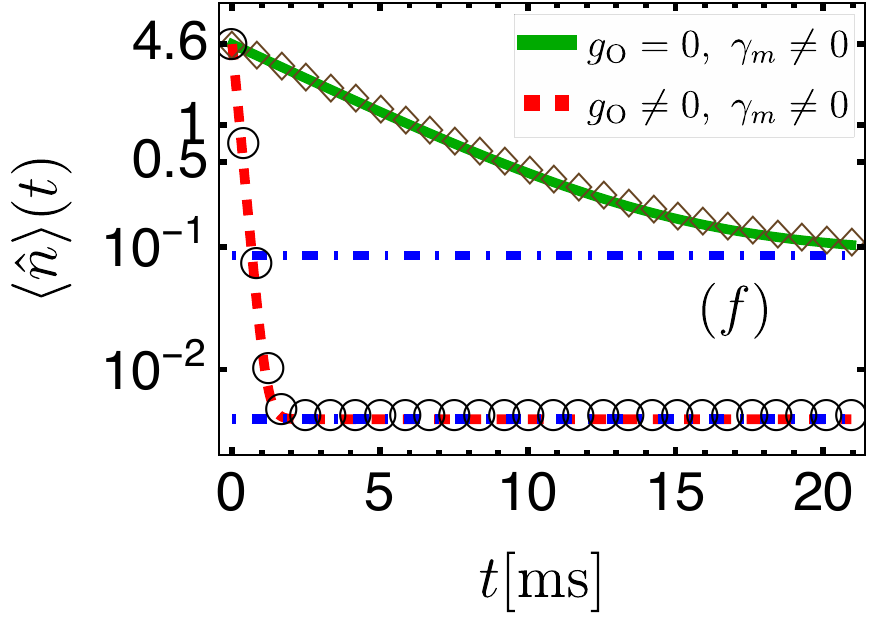}
\caption{Cooling results in weak spin-phonon coupling regime. See text for parameters used. (a-d) Steady state occupation number $n_{f}$ as a function of single-photon Rabi drive strength $\Omega_{g}$ for different cases (see insets). Solid lines (green) represent the analytical solutions as determined by the effective two-level spin absorption spectrum, Eq.~\eqref{eq:Lorentzian}. The dashed lines (orange) are numerically exact simulations of the master equation for the full three-level system using Eq.~\eqref{eq:MasterEq3L}. (e-f) Time resolved dynamics for the mean phonon number at spin-phonon resonance condition. Solid (green) and dashed (red) lines are effective two-level analytical solutions obtained from the rate Eq.~\eqref{eq:eq_rate} for the cases shown (see inset). Open markers, diamond and circle, are the corresponding numerical simulation of the full three-level master equation. In the simulations, we truncated the Fock state space with a cutoff $n_{cut}=30$ for a thermal state with mean occupation $n_{\rm th}=4.6$. In all figures, the horizontal dotted-dashed lines (blue) represent the minimum possible mean phonon number for corresponding cases.}
\label{Fig:fig3abcd}
\end{figure}

While so far we have focused on $\delta=0$, let us make some remarks about operation in different regimes. If $\delta\neq 0$, the cooling is no longer optimal since there is a remaining carrier term similar to the one seen in standard Raman side-band cooling (RSC) \cite{PhysRevLett.75.4011, PhysRevX.2.041014,Lu2024,PhysRevX.14.031002}, which induces an additional off-resonant carrier coupling between the $\{|g\rangle,|e\rangle\}$ states (cf. Eq. \eqref{eq:SpinhamoEffRaman}). The latter makes the eigenstates of the Hamiltonian  different from the dark state  of the jump operator, thus introducing the possible need of repumping beams.
Moreover, we note that in contrast to the usual implementation of EIT cooling \cite{PhysRevLett.85.5547, PhysRevA.93.053401, PhysRevLett.110.153002, PhysRevLett.125.053001}, where one of the Raman lasers, the so-called strong pump laser, has a much higher Rabi frequency, our system optimally operates in the balance case $\Omega_e=\Omega_g$,  i.e. by employing dark and bright states of maximum coherence. 

Regardless of the different operating conditions, if we compare  our approach  with $g_{\rm O}=0~({\rm i.e.~}  p=0 )$, and EIT in the limit $\gamma\ll\Delta_{\rm R}$  at $\delta=0$ \cite{PhysRevA.67.033402}, one obtains that both approaches reach similar cooling rates and minimum occupation numbers. In Fig.~\ref{Fig:Fig3} (b), we also plot the steady state scattering rate $\gamma\mathcal{\varrho}_{rr}=\gamma {\rm Tr}[\hat{\rho}_{e,g,r}\hat {\sigma}_{rr}]$ vs $\Delta_{g}/\Delta_{e}$, with $\hat{\rho}_{e,g,r}$ the density matrix of the three level internal states only. The obtained rate has a sharp decrease when two-photon detuning becomes zero (i.e. $\delta=\Delta_{g}-\Delta_{e}=0$) akin to the CPT profile \cite{Gray:78}. Only when $\Omega_{g}\ll\Omega_{e}$, this profile becomes the EIT Fano-like profile \cite{Lounis1992,PhysRevLett.85.4458,PhysRevLett.85.5547, PhysRevA.93.053401, PhysRevLett.110.153002, PhysRevLett.125.053001}. 

\subsubsection{ Numerical results } In the following we present numerical calculations solving the full three-level dynamics (see Appendix \ref{Ap:AppendixC}), in the weak coupling perturbative regime $\{g_{\rm R},g_{\rm O}\}<\gamma_{b}$ in which the analytic expressions derived above from the effective two-level model are expected to be valid. The numerical  simulations  confirm the validity of the analytic expressions. 
In Fig.~\ref{Fig:fig3abcd}(a-d), we plot the steady state phonon occupation number as a function of $\Omega_{g}$ for different values of $g_{\rm O}$ and $\gamma_{m}$ and setting $\Delta_{\rm R}= 2\pi \times 503.1~{\rm MHz}$, $\Omega_{e}=2\pi\times 40 ~{\rm MHz}$, $\omega_{m}= 2\pi \times 1.59~{\rm MHz}$, $\gamma_g = 2\pi \times 6$ \textmu s$^{-1}$, $\gamma_e = 2\pi \times 12$ \textmu s$^{-1}$ and $\eta_{z}\sim  0.001$. We assume a mean initial quanta $n_{\rm th}=4.6$.

In Fig.~\ref{Fig:fig3abcd} (a,b) we consider $g_{\rm O}=0$ and compare the cases, $\gamma_{m}= 0$ (a) and  $\gamma_{m} =2\pi \times  0.75~ {\rm s^{-1}}$ (b) where we observe the important role played by a small but non-zero $\gamma_{m}$. Adding a finite $\gamma_{m}$ not only reproduces the expected absence of cooling in the far off resonance regime, but, also sets the limit on the lowest achievable temperature in the presence of mode rethermalization, $n_{f, {\rm min}}=0.086 > 1/(4Q_s)^2 \lesssim 10^{-4}$. Moreover, to illustrate the relevant role played by $g_{\rm O}\neq 0$, in Fig.~\ref{Fig:fig3abcd} (c,d), we set $g_{\rm O}= 2\pi \times 3.6 ~{\rm kHz}$ and plot cases when $\gamma_{m}= 0$ (c), and $ \gamma_{m}=2\pi \times 0.75 ~ {\rm s^{-1}}$ (d). While similar inconsistencies to the one with $g_{\rm O}= \gamma_{m}=0$ are observed in the off-resonant regime, for the more appropriate condition depicted in (d), we do recover the absence of cooling  far from resonance. In addition, we observe the benefit of the ODF, whose  role is to   effectively increase $g_{\rm R}$.
Therefore, it 
leads to  a lower phonon steady state population which reaches $n_{f,\rm {min}}=0.0039$, an order of the magnitude less than the case  $g_{\rm O}= 0$. The reduced final phonon number due to the ODF can also be seen in the phonon probability distribution shown in the insets. From Fig.~\ref{Fig:fig3abcd} (a,c), we note that, when ignoring the mode rethermalization effects (i.e. $\gamma_{m}=0$), the minimum possible occupation number follows the quantum back action limit $n_{f,\rm {min}}=1/(4 Q_{s})^2$  both for $g_{\rm O}=0$ and $g_{\rm O}\neq 0$ as predicted by our analytical calculations. 

Thus, contrasting prior work~\cite{retzker2007fast,PhysRevLett.104.043003, Albrecht_2011}, within the regime of validity of our theory, the ODF cannot push the phonon number below the quantum back action limit.

In Fig.~\ref{Fig:fig3abcd} (e-f), we also display dynamics of the mean phonon occupation at the resonant point. The ODF not only  helps to  reach colder temperatures (without violating QBL), but also to speed up the cooling process. 
\begin{figure}[t!]
	\includegraphics[width=0.495
    \columnwidth]{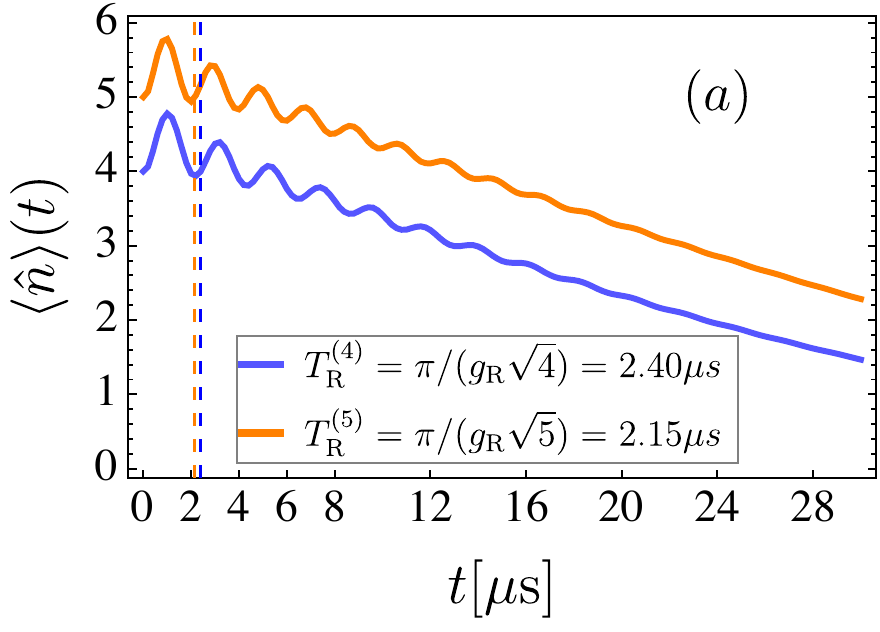}
    \includegraphics[width=0.495\columnwidth]{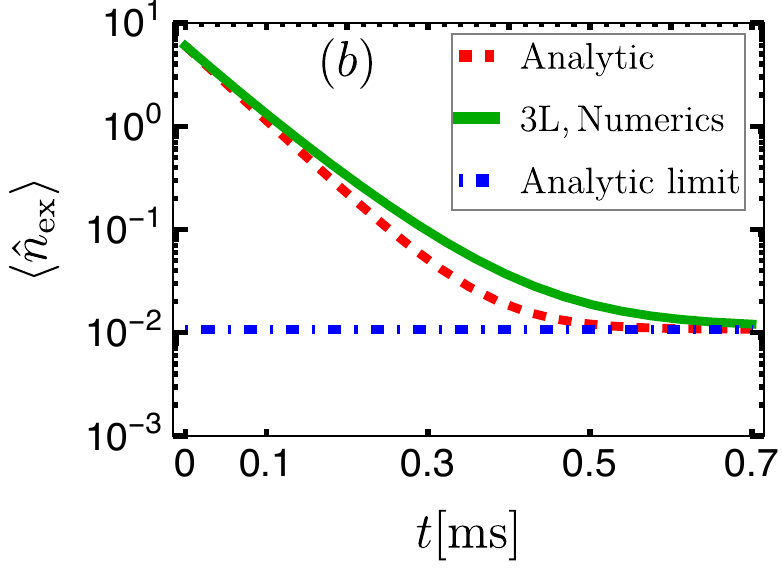}
\caption{(a) Real-time dynamics of the phonon occupation following Eqs.~\eqref{eq:barediagonalTLHam}-\eqref{eq:DissipatinJumpDressedframe13} (with $g_{\rm O}=0$) starting from $\ket +$ and a Fock state in the phonon mode. The orange and blue curves represent initial Fock states $\ket{n=4}$ and $\ket{n=5}$, respectively. The vertical dashed lines indicate the analytical values for Rabi oscillation periods $T_{\rm R}^{(n)}$ computed in the main text.
(b) Comparison of the analytic result Eq.~\eqref{eq:dpex_dt} (red dashed line) to the numerical solution of Eq.~\eqref{eq:hamoltoninaSystemLabeled} and \eqref{eq:SystemBasicJumpoperators} for $N=1$ (green solid line).
The blue dash-dotted line is the analytic solution of the final ground-state population Eq.~\eqref{eq:analytic_final}.
Here we take parameters close to Ref.~\cite{PhysRevLett.122.053603} ($\eta_{z}=0.13$, $\Delta_{\rm R}=2\pi \times 385~{\rm MHz}$, $\Omega_{e}= \Omega_{g}= 2\pi \times  35~ {\rm MHz}$).
}
\label{Fig:RabiabtotalEx}
\end{figure}

\subsection{Strong Coupling}
\label{ssec:single-strong}

\begin{table}[]
    \renewcommand{\arraystretch}{1.4}
    \centering
    \begin{tabularx}{\columnwidth}{
        c |
        c |
        >{\raggedright\arraybackslash}X
    }
        Symbol & Definition & Description \\
        \hline
        $\varphi$ & $\arctan(g_{\rm O}\sin(\alpha) / g_{\rm R})$ & ODF mixing angle \\
        $T_{\rm R}^{(n)}$ & $\pi/(g_{\rm R}\sqrt{n})$ & Dark-bright Rabi oscillation period \\
        $n_{\rm ex}$ & $\left\langle \sum_\nu \hat b_\nu^\dagger \hat b_\nu + \sum_j \ket{+}\bra{+}_j \right\rangle$ & Total excitation number \\
        $\ket{\mathcal X_{n_{\rm ex}}}$ & Eqs.~\eqref{eq:definition_symmetric_eigenstate} and \eqref{eq:definition_antisymmetric_eigenstate} & Single-ion eigenstates \\
        $\kappa_{-,l}^{(\mathcal X \mathcal Y)}$ & $\abs{\bra{\mathcal X_{n_{\rm ex}-1}} \hat L_l \ket{\mathcal Y_{n_{\rm ex}}}}^2$ & Dressed decay rate \\
        $\kappa_{z,l}^{(\mathcal X \mathcal Y)}$ & $\abs{\bra{\mathcal X_{n_{\rm ex}}} \hat L_l \ket{\mathcal Y_{n_{\rm ex}}}}^2$ & Dressed dephasing rate \\
        $\ket{n_{\rm ex}, \zeta}$ & --- & Many-ion eigenstates \\
        $\mathcal N_{N,n_{\rm ex}}$ & Eq.~\eqref{eq:definition_degeneracy_of_states} & Number of many-ion eigenstates \\
        $\hat L_{l;\rm eff}^{(j)}$ & Eq.~\eqref{eq:Leff_collective_eta0} & Many-ion effective jump operator \\
        $\kappa_{-,l}^{(n_{\rm ex, \zeta', \zeta})}$ &  $\abs{\bra{n_{\rm ex}-1, \zeta'} \hat L_{l;\rm eff}^{(j)} \ket{n_{\rm ex}, \zeta}}^2$ & Many-ion dressed decay rate \\
        $\kappa_{z,l}^{(n_{\rm ex, \zeta', \zeta})}$ & $\abs{\bra{n_{\rm ex}, \zeta'} \hat L_{l;\rm eff}^{(j)} \ket{n_{\rm ex}, \zeta}}^2$ & Many-ion dressed dephasing rate \\
        $\gamma_{N=1,\rm sc}$ & $\frac{1}{2} \frac{\Omega_g^2 \gamma_e + \Omega_e^2 \gamma_g}{4\Delta_{\rm R}^2 + \gamma^2}$ & Single-ion cooling rate \\
        $\gamma_{N,\rm sc}^{(n_{\rm ex})}$ & $2\overline{N_+}(n_{\rm ex}, N) \times \gamma_{N=1,\mathrm{sc}}$ & Many-ion cooling rate \\
        $\gamma_{N,M,\rm mm}^{(n_{\rm ex})}$ & $2\overline{N_{+,\rm mm}}(n_{\rm ex}, N,M) \gamma_{N=1,\mathrm{sc}}$ & Multimode cooling rate \\
        $\overline{N_+}(n_{\rm ex}, N)$ & Eq.~\eqref{eq:n_bright} & Average bright state number \\
        $\overline{N_{+,\rm mm}}(n_{\rm ex}, N, M)$ & Eq.~\eqref{eq:nplus_mm} & Multimode average bright state number \\
        $\gamma_{n_{\rm ex}, \rm p}$ & Eq.~\eqref{eq:definition_gamma_1_p} & Reheating rate \\
        $\gamma_{n_{\rm ex}, \rm p}^{(M)}$ & Eq.~\eqref{eq:definition_gamma_1_p_manymode} & Multimode reheating rate \\
        $\mathcal K_{\nu, j}$ & Eq.~\eqref{eq:hamiltonian_raman_multimode} & Mode profile \\
        $\gamma_{\rm scat}$ & Eq.~\eqref{eq:definition_gamma_scat} & Photon scattering rate \\
    \end{tabularx}
    \caption{Summary of the symbols used in the manuscript related to strong coupling. Longer equations are referenced.}
    \label{tab:strong_symbols}
\end{table}

For $g_{\rm R} > \gamma_b$ (i.e.~$\eta_{z}> [\gamma(\Omega_g^2+\Omega_{e}^2)]/[\Delta_{\rm R} \Omega_g \Omega_e] $), the ion's internal state cannot be treated as an effective Markovian bath, but instead is an integral part of the dynamics.
We now introduce a framework to study cooling in this strong-coupling limit, which will translate to the many-body case.
The results we derive for $N=1$ are consistent with prior work in Refs.~\cite{Zhang2021,PhysRevA.104.013117}, further supporting our methodology.
We start from the effective two-level Hamiltonian
\begin{align}
    \hat{\tilde H}_{\rm eff} &=
    \omega_m \hat b^\dagger \hat b + \frac{\omega_s}{2} \hat{\tilde \sigma}_z + \hat{\tilde H}_{\rm int} \, ,
\end{align}
where $\hat{\tilde H}_{\rm int}$ is given by Eq.~\eqref{eq:H_int_singleion}.
We assume $(g_{\rm R}^2 + g_{\rm O}^2)/\omega_s^2 \ll 1$ and $\omega_m = \omega_s$, such that we can, at leading order, neglect off-resonant couplings in $\hat{\tilde H}_{\rm int}$ and focus on
\begin{align}
    \hat{\tilde H}_{\rm JC} &=
    \omega_m \hat n_{\rm ex} + \sqrt{g_{\rm R}^2 + \sin^2(\alpha) g_{\rm O}^2} \qty(e^{\mi \varphi} \hat{\tilde \sigma}_+ \hat b + h.c.)
    \label{eq:jaynescummings}
    \\
    \hat n_{\rm ex} &=
    \hat b^\dagger \hat b + \ket + \bra +
\end{align}
for some phase $\varphi = \arctan(g_{\rm O}\sin(\alpha)/g_{\rm R})$, which is the well-known Jaynes-Cummings Hamiltonian from quantum optics up to a frame transformation.
Since $g_{\rm R}$ and $g_{\rm O}\sin(\alpha)$ enter in the same way, we absorb $g_{\rm O}$ in a redefinition of $g_{\rm R}' = \sqrt{g_{\rm R}^2 + \sin^2(\alpha) g_{\rm O}^2}$ including a trivial rotation by $\varphi$ about $\hat{\tilde \sigma}_z$.

$\hat H_{\rm JC}$ induces Rabi oscillations $\ket{+,n}\leftrightarrow\ket{-,n+1}$ at rate $2g_{\rm R} \sqrt{n}$ and period $T_{\rm R}^{(n)}=\pi/(g_{\rm R} \sqrt{n})$ [see Figs.~\ref{Fig:Fig2} (c) and \ref{Fig:RabiabtotalEx} (a)].
At a slower time scale $\gamma_b^{-1}$, the bright state $\ket{+,n}$ decays to $\ket{-,n}$ through Eqs.~\eqref{eq:DissipatinJumpDressedframe12} and \eqref{eq:DissipatinJumpDressedframe13}, which resets the system.
Following this reset, the cycle restarts with Rabi oscillations $\ket{+,n-1}\leftrightarrow\ket{-,n}$.
This cycle continuously cools the ion and reduces the excitation number $\hat n_{\rm ex}$, ultimately trapping the population in $\sim \ket{-,0}$.
In Fig.~\ref{Fig:Fig2}(c), this describes a top$\rightarrow$down decay (instead of the right$\rightarrow$left decay in panel (b)).

\subsubsection{Cooling rate}
To estimate the cooling rate, we consider the eigenstates of Eq.~\eqref{eq:jaynescummings}, 
\begin{align}
    \ket{\mathcal S_{n_{\rm ex}}} &= (\ket{+,n_{\rm ex}-1} + \ket{-,n_{\rm ex}})/\sqrt{2}
    \label{eq:definition_symmetric_eigenstate}
    \, , \\
    \ket{\mathcal A_{n_{\rm ex}}} &= (\ket{+,n_{\rm ex}-1} - \ket{-,n_{\rm ex}})/\sqrt{2} \, .
    \label{eq:definition_antisymmetric_eigenstate}
\end{align}
If the energy splitting of these states is sufficiently large $2 \sqrt{n_{\rm ex}} \hbar g_{\rm R} \gg \hbar \gamma_b$, the oscillations in Fig.~\ref{Fig:RabiabtotalEx}(a) are much faster than the decay, and we can rotate out coherences between $\ket{\mathcal S_{n_{\rm ex}}}$ and $\ket{\mathcal A_{n_{\rm ex}}}$.
Then, the dynamics are well approximated by the diagonal density matrix  $\hat \rho^{(0)}(t) = $
\begin{align}
P_{\rm ex}^{(0)}(0,t) \ket{-,0}\bra{-,0} +  \sum_{n_{\rm ex}=1}^\infty \sum_{\mathcal X = \mathcal S, \mathcal A} \frac{P_{\rm ex}^{(0)}(n_{\rm ex},t)}{2} \ket{\mathcal X_{n_{\rm ex}}} \bra{\mathcal X_{n_{\rm ex}}} \, . \nonumber
\end{align}
Here, the superscript (0) indicates that we neglect higher-order terms in $\gamma_{b}/g_{\rm R}$.

The dynamics of $P_{\rm ex}^{(0)}(n_{\rm ex},t)$ are described by projecting the Lindblad operators in Eqs.~\eqref{eq:DissipatinJumpDressedframe12} and \eqref{eq:DissipatinJumpDressedframe13} into the eigenstates $\ket{\mathcal A_{n_{\rm ex}}}$ and $\ket{\mathcal S_{n_{\rm ex}}}$.
The  relevant Lindblad operators connect $n_{\rm ex} $ to $n_{\rm ex} -1$ as
\begin{align}
    \hat L_{-,l}^\mathcal{(XY)} 
    &\equiv
    \ket{\mathcal X_{{n_\mathrm{ex}}-1}}\bra{\mathcal X_{{n_\mathrm{ex}}-1}}
    \hat L_l
    \ket{\mathcal Y_{n_\mathrm{ex}}}\bra{\mathcal Y_{n_\mathrm{ex}}}
    \nonumber \\
    &=
    \sqrt{\kappa_{-,l}^{(\mathcal{XY})}} \ket{\mathcal X_{{n_\mathrm{ex}}-1}}\bra{\mathcal Y_{n_\mathrm{ex}}}
    \, ,
\end{align}
for $\mathcal X, \mathcal Y \in \{\mathcal A, \mathcal S\}$ and $l \in \{g,e\}$, and $\kappa_{-,l}^{(\mathcal{XY})}$ corresponding decay rates.
Additional jump operators $\hat L_{z,l}^\mathcal{(XY)} \propto \ket{\mathcal X_{n_\mathrm{ex}}}\bra{\mathcal Y_{n_\mathrm{ex}}}$ contribute to the dephasing observed in Fig.~\ref{Fig:RabiabtotalEx}(a), but leave $n_{\rm ex}$ constant and thus neither cool nor heat.

The jump operators describe unidirectional decay ${n_\mathrm{ex}} \rightarrow {n_\mathrm{ex}}-1$ leading to
\begin{align}
    & \frac{d}{dt} P_\mathrm{ex}^{(0)}(n_\mathrm{ex},t) =
    - \gamma_{N=1,\mathrm{sc}} P_\mathrm{ex}^{(0)}(n_\mathrm{ex},t)
    + \gamma_{N=1,\mathrm{sc}} P_\mathrm{ex}^{(0)}(n_\mathrm{ex}+1,t)
    \, .
    \label{eq:analytic_strong_single}
    \\
    & \gamma_{N=1,\mathrm{sc}} 
    =
    \sum_{\mathcal{X=A,S}}\sum_{l=g,e} \kappa_{-,l}^{(\mathcal{XY})}
    = \frac{\abs{\alpha_{eg}}^2 + \abs{\alpha_{ge}}^2}{2}
    =
    \frac{1}{2}
    \frac{\Omega_g^2 \gamma_e + \Omega_e^2 \gamma_g}{4\Delta_{\rm R}^2 + \gamma^2}
    \, .
    \label{eq:gamma_1sc}
\end{align}
This decay rate is independent of $n_{\rm ex}$, and thus does not describe exponential decay to the ground state. Instead, $\langle n_{\rm ex} \rangle = \sum_{n_{\rm ex}} n_{\rm ex} P^{(0)}(n_{\rm ex},t)$ follows the dynamics
\begin{align}
    \frac{d\langle n_{\rm ex}\rangle}{dt} = \sum_{n_{\rm ex}} n_{\rm ex} \frac{d P_{\rm ex}^{(0)}(n_{\rm ex}, t)}{dt} &= -\gamma_{N=1,\rm sc} \qty[1 - P_{\rm ex}^{(0)}(n_{\rm ex}=0, t)] \, .
\end{align}
For sufficiently large $\langle n_{\rm ex}\rangle$ such that $P_{\rm ex}^{(0)}(n_{\rm ex}=0, t) \approx 0$, this fits an instantaneous exponential decay at rate $\gamma_{N=1,\rm sc}/\langle n_{\rm ex}\rangle$.

\subsubsection{Final temperature}
So far we have studied the  cooling   driven by the jump operators  at a rate of order  $\gamma_b \sim \omega_s \gamma/\Delta_{\rm R}$. These processes capture the leading-order dynamics for $n_\mathrm{ex} > 0$.
At this order, the steady state relaxes all the way to  $P_\mathrm{ex}^{(0)}(0,t\to \infty) = 1$ and $P_\mathrm{ex}^{(0)}(n_\mathrm{ex}>0,t\to \infty) = 0$.
We now compute the leading order corrections to the steady state by adding the relevant terms up to order $\mathcal O(\eta_z^2\gamma_b)$ to Eq.~\eqref{eq:analytic_strong_single}.
To compute the full dynamics at this order, we would need to compute the dressed eigenstates $\ket{\mathcal X_{n_{\rm ex}}}^{(1)}$, which diagonalize the perturbed density matrix.
Then, we could compute the dynamics of the populations of the perturbed blocks $P_{\rm ex}^{(1)}(n_{\rm ex},t)$.

For the steady state, however, we only need to compute the corrections for the unique steady state of the leading-order dynamics, $\ket{-,0}$ with $n_\mathrm{ex}=0$, which is substantially simpler to evaluate. First-order corrections of states with $n_\mathrm{ex} > 0$ are weighted by the small steady-state population of these states. Therefore, they will induce higher-order corrections in $P_\mathrm{ex}(n_\mathrm{ex}>0,t\to \infty)\propto \mathcal O(\eta_z^{n>2})$, which can be neglected at second order.

In standard Hamiltonian perturbation theory we find~\cite{sakurai2020modern}
\begin{align}
    \ket{-,0}^{(1)} =& 
    \qty[1 - \sum_{n_{\rm ex}>0,\mathcal X = \mathcal S, \mathcal A} \frac{\abs{\bra{\mathcal X_{n_{\rm ex}}} \hat H_{\rm CR} \ket{0,-}}^2}{\abs{n_{\rm ex}\omega_m}^2}]\ket{-,0} 
    \nonumber \\
    & - \sum_{n_{\rm ex}>0,\mathcal X = \mathcal S, \mathcal A} \frac{\bra{\mathcal X_{n_{\rm ex}}}\hat H_{\rm CR} \ket{-,0}}{n_{\rm ex}\omega_m} \ket{\mathcal X_{n_{\rm ex}}} \, , \\
    \hat H_{\rm CR} =& -\mi g_{\rm R} \hat{\tilde \sigma}_+ \hat b^\dagger + h.c.
\end{align}

We include these corrections in our ansatz for the density matrix
\begin{align}
    \hat \rho^{(1)}(t) =& P_{\rm ex}(0,t) \ket{-,0}^{(1)}\bra{-,0}^{(1)} \\
    & + \sum_{n_{\rm ex}>0,\mathcal X=\mathcal S,\mathcal A} \frac{P_{\rm ex}(n_{\rm ex},t\rightarrow\infty)}{2} \ket{\mathcal X_{n_{\rm ex}}}\bra{\mathcal X_{n_{\rm ex}}} \, ,
    \nonumber \\
    P_{\rm ex}(0,t) \equiv& \bra{-,0}^{(1)} \hat \rho^{(1)} \ket{-,0}^{(1)}
    \\
    P_{\rm ex}(n_{\rm ex}>1,t) \equiv& \bra{\mathcal S_{n_{\rm ex}}} \hat \rho^{(1)} \ket{\mathcal S_{n_{\rm ex}}}
    =  \bra{\mathcal A_{n_{\rm ex}}} \hat \rho^{(1)} \ket{\mathcal A_{n_{\rm ex}}}
    \, .
\end{align}
As we show in Appendix~\ref{app:collective}, while this ansatz only captures leading-order dynamics, it correctly predicts the steady state up to second order.

We now derive a simple linear ordinary differential equation to describe the dynamics of $P_{\rm ex}(n_{\rm ex},t)$.
While states with $n_{\rm ex} > 0$ remain unmodified, the dressed ground state $\ket{-,0}^{(1)}$ can now decay to $\ket{\mathcal X_{n_{\rm ex}=1}}$ and $\ket{\mathcal X_{n_{\rm ex}=2}}$ at rate
\begin{align}
    \gamma_{[n_{\rm ex}=1,2],\rm p} = \sum_{\mathcal X = \mathcal S, \mathcal A,\hat L = \hat{\tilde L}_1, \hat{\tilde L}_2} \abs{\bra{\mathcal X_{n_{\rm ex}}} \hat L \ket{-,0}^{(1)}}^2
    \, ,
\end{align}
which we evaluate in Appendix~\ref{app:collective}.

\begin{figure*}
    \centering\includegraphics[width=0.99\linewidth]{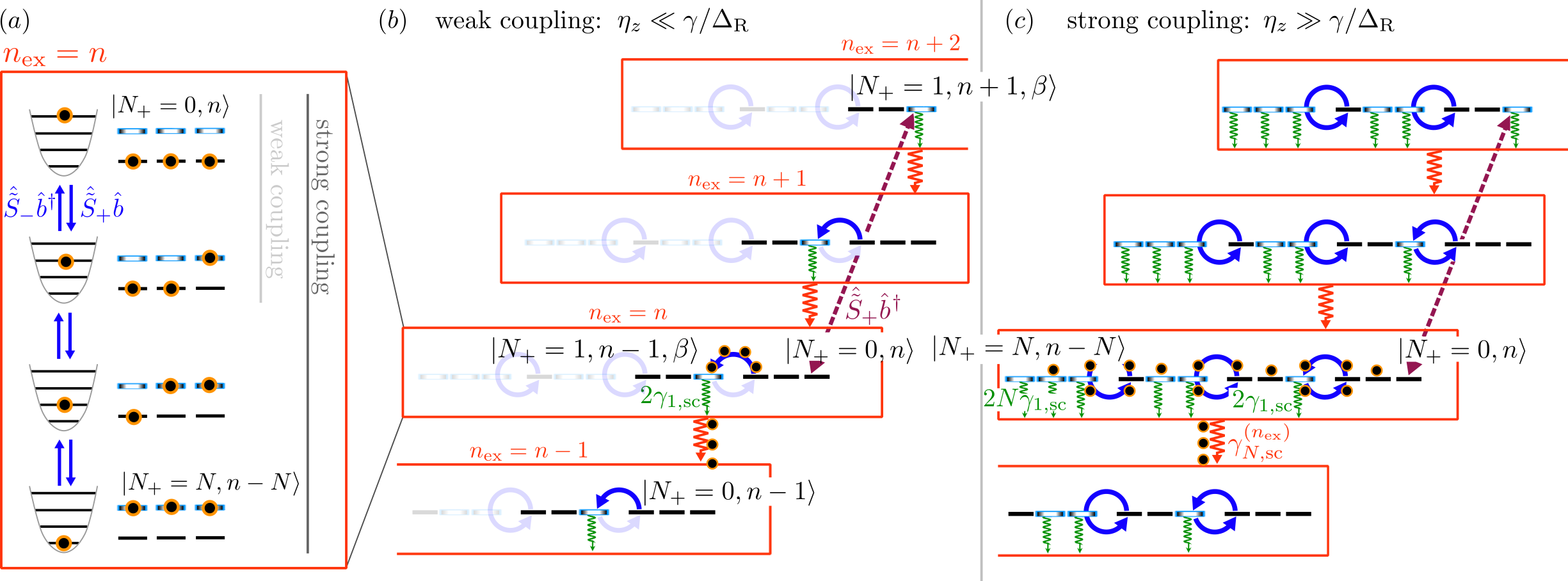}
    \caption{Effective many-ion cooling dynamics analogous to Fig.~\ref{Fig:Fig3}.
    (a) Sketch of the dynamics under the Hamiltonian in Eq.~\eqref{eq:taviscummingsa} for given $n_{\rm ex}$ (red box). The Hamiltonian exchanges spin and motion excitations but cannot change the total number of excitations.
    The grey side bars indicate which states are accessible to weak and strong coupling regimes, respectively.
    (b) Weak coupling.
    Dynamics is restricted to $N_+\approx0$; states that are only populated in high-order perturbation theory are greyed-out.
    Each line indicates a separate ion, which can be in the dark state (black) or the bright state (white).
     The green arrows then indicate independent spontaneous emission of each ion, reducing the number of excitations.
    The purple dashed line indicates off-resonant coupling in Eq.~\eqref{eq:dicke}.
    (c) Strong coupling, analogous to panel (b).
    Here, all states are accessible to the dynamics. 
    The decay rate $2N_+\gamma_{1,\rm sc}$ is proportional to the number of bright ions, as indicated by one arrow per bright state. 
    These combine to an effective decay rate of  $\gamma_{N,\mathrm{sc}}^{(n_\mathrm{ex})}$ from $n_{\rm ex}$ to $n_{\rm ex} - 1$. }
    \label{fig:collective-sketches}
\end{figure*}

The dynamics of the population vector $\vec p_{\rm ex} = \qty[P_{\rm ex}(0,t),P_{\rm ex}(1,t),P_{\rm ex}(2,t),\dots]^T$ then follow the linear first-order ordinary differential equation
\begin{align}
    &\frac{d}{dt}
    \vec p_{\rm ex}(t)
    =
    \mathbf{M} \vec p_{\rm ex}(t) \, ,
    \label{eq:dpex_dt}
    \\
    &\mathbf{M} =
    \begin{pmatrix}
        -\gamma_{1,\rm p} - \gamma_{2,\rm p} & \gamma_{N=1,\rm sc} & 0 & 0 & \cdots \\
        \gamma_{1,\rm p} & - \gamma_{N=1,\rm sc} & \gamma_{N=1,\rm sc} & 0 & \cdots \\
        \gamma_{2,\rm p} & 0 & -\gamma_{N=1,\rm sc} & \gamma_{N=1,\rm sc} & \ddots \\
        0 & 0 & 0 & -\gamma_{N=1,\rm sc} & \ddots \\
        \vdots & & & & \ddots
    \end{pmatrix}
    \nonumber
    \, .
\end{align}
Its steady state has $P_{\rm ex}(n_{\rm ex},t\rightarrow\infty) = 0$ for $n_{\rm ex}>2$, and the normalized populations for $n_{\rm ex}\leq 2$ are
\begin{align}
    \vec p_{\rm ex}(t\rightarrow\infty)
    &=
    \frac{
    \qty[
        \gamma_{N=1,\rm sc},\ 
        \gamma_{1,p} + \gamma_{2,p},\ 
        \gamma_{2,p},\ 0,
        \dots
    ]^T
    }{
    \gamma_{N=1,\rm sc} + \gamma_{1,p} + 2\gamma_{2,p}
    }
    \, .
\end{align}

To compute the steady-state excitation number $\Tr[\hat n_{\rm ex} \hat \rho^{(1)}]$, we take into account that $\bra{-,0}^{(1)} \hat n_{\rm ex} \ket{-,0}^{(1)} = g_{\rm R}^2/2\omega_m^2$, such that we can compute $\langle\hat n_{\rm ex}\rangle = \vec n_{\rm ex} \cdot \vec p_{\rm ex}$ as
\begin{align}
    \vec n_{\rm ex} &=
    \qty(
    \sin^2(\alpha)\eta_z^2/8,
    1,
    2,
    \dots
    )^T
    \, ,
    \\
    \langle n_{\rm ex} \rangle &=
    \frac{\gamma_{N=1,\rm sc}\sin^2(\alpha)\eta_z^2/8 + \gamma_{1,p} + 3\gamma_{2,p}}{\gamma_{N=1,\rm sc} + \gamma_{1,p} + 2\gamma_{2,p}}
    \label{eq:analytic_final}
\end{align}
Fig.~\ref{Fig:RabiabtotalEx} (b) shows good qualitative agreement between the analytic dynamics and the corresponding final ground state population, and the numerical solution of the full dynamics for $N=1$.

\section{  Multi-ion Cooling }
\label{sec:multi-ion}
So far, we have focused on an isolated ion.
However, trapped-ion systems often operate with  many ions. Strong Coulomb interactions between them stabilize the formation of a self-assembled  crystal  when the ions are confined  in Penning or Paul traps.  The  correlated motion of the whole array in a crystal  can be described in terms of collective phonon modes $\nu$ with  angular eigenfrequencies $\omega_\nu$.

As before, we focus on cooling only the transverse motion along $Z$.
We start from Eq.~\eqref{eq:SpinhamoEffRaman} with $\delta=0$. We consider a crystal with $N$ ions,  described by the Hamiltonian
\begin{align}
\hat{H}_{\rm eff}=\sum_{\nu=1}^N\omega_{\nu}\hat{b}^{\dagger}_{\nu}\hat{b}_{\nu}+\sum_{j=1}^N\frac{\omega_{\rm LS}}{2}\hat{\sigma}_{z}^{(j)}+\hat{H}_{\mathrm{R},N}
+ \hat H_{\mathrm{ODF},N}
\, .
\label{eq:SpinhamoEffMMION}
\end{align}
The Pauli matrices $\hat \sigma_{x,y,z}^{(j)}$ act locally on ion $j$, the annihilation operators $\hat b_\nu$ act on the collective phonon modes, and the local displacements $\hat Z_{j}$ are determined by the mode profile amplitudes $\mathcal{K}_{\nu,j}$ as $\textstyle{\hat{Z}_{j}=\sum_{\nu=1}^{N}\mathcal{K}_{\nu,j}\sqrt{\hbar/2 m_{I}\omega_{\nu}}(\hat{b}_\nu+\hat{b}_\nu^\dagger) }$.
The mode $\nu=1$ is the axial center-of-mass mode with $\omega_1 = \omega_m$, and we normalize $\sum_\nu \mathcal K_{\nu,j}^2 = 1$.
The $N$-ion Raman and ODF Hamiltonians $\hat H_{\mathrm{R},N}$ and $\hat H_{\mathrm{ODF},N}$ are generalizations of their single-ion counterparts summed over all $N$ ions
\begin{align}
\hat{H}_{\mathrm{R},N} =&\sum_{j=1}^{N}\frac{\Omega_{\rm R}}{2}\pc{\Pi_{\nu=1}^{N}e^{i{\Delta k_z}\mathcal{K}_{\nu,j}\sqrt{\hbar/2 m_{I}\omega_{\nu}}(\hat{b}_\nu+\hat{b}_\nu^\dagger)}\hat{\sigma}_{ge}^{(j)}+{\rm H.C.}} 
\,,
\label{eq:hamiltonian_raman_multimode}
\\
\hat H_{\mathrm{ODF},N} =&
\sum_{j,\nu=1}^{N} g_{\rm O}
\hat \sigma_z^{(j)} \mathcal{K}_{\nu,j} \sqrt{\frac{\omega_1}{\omega_\nu}} \qty(\hat b_\nu + \hat b_\nu^\dagger)
\, .
\label{eq:SpinhamoEffRI1}
\end{align}
We follow the single-ion recipe and expand $\hat H_{\mathrm{R},N}$ in the Lamb-Dicke parameter $\eta_{\nu,j}\equiv {\Delta k_z}\mathcal{K}_{\nu,j}\sqrt{\hbar/2m_{I}\omega_{\nu}} = \eta_z \mathcal K_{\nu,j} \sqrt{\omega_1/\omega_\nu}$, up to linear order. Higher-order terms are suppressed assuming $\eta_{\nu,j}\ll1$.
We  further move to the dressed basis to obtain
\begin{align}
\hat{H}_{\rm eff} \approx&
\sum_{\nu=1}^{N}\omega_{\nu}\hat{b}^{\dagger}_{\nu}\hat{b}_{\nu}+\sum_{j=1}^N \frac{\omega_{s}}{2} \hat{\tilde \sigma}_{z}^{(j)}
+\sum_{\nu,j=1}^{N} \mathcal K_{\nu,j} \sqrt{\frac{\omega_1}{\omega_\nu}} 
\qty(\hat b_\nu + \hat b_\nu^\dagger)
\nonumber \\
&\quad \times
\qty(g_{\rm R} \hat{\tilde \sigma}_{y}^{(j)} +  g_{\rm O} \cos(\alpha)\hat{\tilde \sigma}_z^{(j)} + g_{\rm O} \sin(\alpha) \hat{\tilde \sigma}_x^{(j)}).
\label{eq:SpinhamoEffMMa}
\end{align}
As for the single-ion case, for balanced Rabi frequencies (such that $\alpha = \pi/2$), $g_{\rm R}$ and $g_{\rm O}$ enter essentially in the same way.
Consequently, up to a trivial rotation around $\hat{\tilde \sigma}_z^{(j)}$, we can absorb $g_{\rm O}$ in a renormalization of $g_{\rm R}' = \sqrt{g_{\rm R}^2 + g_{\rm O}^2}$.
For $\alpha \neq \pi/2$, the term $\sim (\hat b_\nu + \hat b_\nu^\dagger) \hat{\tilde \sigma}_z$ can introduce additional off-resonant heating, making $g_{\rm O}$ suboptimal in the strong coupling regime.
We thus set $g_{\rm O} = 0$ in the following.

The effective jump operators capture local spontaneous emission events and are thus localized to individual ions.
Keeping only the leading-order terms of the Lamb-Dicke expansion, they read
\begin{align}
\hat{L}_{g}^{({\rm eff},j)}=\alpha_{gg}\hat{\sigma}_{gg}^{(j)}+\alpha_{ge}\hat{\sigma}_{ge}^{(j)},~ \hat{L}_{e}^{({\rm eff},{j})}=\alpha_{eg}\hat{\sigma}_{eg}^{(j)}+\alpha_{ee}\hat{\sigma}_{ee}^{(j)},
\label{eq:SpinJumpEffm}
\end{align}
with $j$ being the ion index. These jump operators are identical to the single-ion ones written before---see Eq. (\ref{eq:SpinJumpEff}).

The effective Hamiltonian Eq.~\eqref{eq:SpinhamoEffMMa} includes indirect crosstalk between the modes via coupling to local ion-excitations.
Numerical simulations have confirmed that the many-body dynamics cool multiple modes in parallel~\cite{PhysRevA.99.023409}.
However, describing the coupled dynamics of all modes analytically is beyond the scope of this paper.

Therefore, we focus on cooling only the axial center-of-mass mode $\nu = 1$, which we label $\hat b \equiv \hat b_1$ for simplicity, and discuss multimode cooling at the end of the section.

The center-of-mass mode couples identically to all $N$ ions, such that $\mathcal K_{1,j} = 1/\sqrt{N}$.
The Hamiltonian in the dressed dark and bright states reads
\begin{align}
    \hat H_\mathrm{collective}
    &=
    \frac{2g_{\rm R}}{\sqrt{N}} \hat{\tilde S}_y \qty(\hat b + \hat b^\dagger)
    +
    \omega_m \hat b^\dagger \hat b
    +
    \omega_s \hat{ \tilde{S}}_z
    \, ,
    \label{eq:dicke}
\end{align}
with total spin operators given by  $\hat{\tilde S}_{x,y,z} = \sum_{j=1}^N \hat{\tilde \sigma}^{(j)}_{x,y,z}/2$.
The number of bright ions is then $\hat N_+ = N/2 +\hat{\tilde S}_z$.
On resonance,  $ \omega_m = \omega_s$, and for $\sqrt{Nn_{\rm ex}} g_{\rm R} \ll \omega_m $, we can make a rotating wave approximation to simplify the Hamiltonian to the Tavis-Cummings model, the many-body extension of the Jaynes-Cummings model introduced in Eq.~\eqref{eq:jaynescummings}
\begin{align}
    \hat H_\mathrm{collective,TC}
    &=
    \frac{g_\mathrm{R}}{\sqrt{N}} \qty(\hat{\tilde S}_+ \hat b + \hat{\tilde S}_- \hat b^\dagger)
    +
    \omega_m \qty(\hat b^\dagger \hat b + \hat{\tilde S}_z) 
    \, ,
    \label{eq:taviscummingsa}
\end{align}
which is illustrated in Fig.~\ref{fig:collective-sketches}(a/b).
Note that this rotates out the counterrotating terms $\hat{\tilde S}_+ \hat b^\dagger$ as well as the terms $\hat{\tilde S}_z\hat b^\dagger$ for finite $g_{\rm O} > 0$.
We define collective raising and lowering operators $\hat{\tilde S}_\pm = \hat{\tilde S}_y \mp \mi \hat{\tilde S}_x$ \footnote{This differs by a factor of i from the usual definition to produce the standard form of the Tavis-Cummings model.}.
The coupling strength reduces to $g_\mathrm{R}/\sqrt{N}$.
Physically, this is due  to the increase in the  mass of the array by $N$, which translates into a reduction by $\sqrt{N}$ of the corresponding harmonic oscillator length compared to the one of  a single ion.

\subsection{Weak coupling regime}

In the weak coupling regime $g_\mathrm{R} < \gamma_b$, all bright states can be eliminated, and to an excellent approximation  $\langle \hat{\tilde S}_z \rangle \approx- N/2$.  Therefore, the effective cooling dynamics remains restricted to the phonon degree of freedom as described by the master Eq.~\eqref{eq:Spin-Absorption} and illustrated in Fig.~\ref{fig:collective-sketches}(b). In this regime, the presence of $N$ ions in the array  does not affect the cooling rate [Fig.~\ref{Fig:fig6ab}(a)]. This is because for a $z$-polarized state, the quantum projection noise enhances $\sqrt{\langle \hat{\tilde S}_- \hat{\tilde S}_+ \rangle} \sim \sqrt{N}$. 
This  $\sqrt{N}$ enhancement of the quantum noise exactly compensates the reduced coupling strength. Therefore, the decay rate remains $g_\mathrm{R}^2/\gamma_b$ independent of $N$.

In the weak-coupling regime, crosstalk between the modes is suppressed because the ions quickly relax to their internal steady-state $\hat{\tilde \sigma}_z^{(j)} \approx -1$.
Consequently, each mode can be described individually, and all modes with $\abs{\omega_s - \omega_\nu} \lesssim \gamma_b$ can  cool simultaneously (see Appendix~\ref{Ap:AppendixD}).

\subsection{Strong  coupling regime}
\begin{figure}[t!]

\includegraphics[width=0.485\columnwidth]{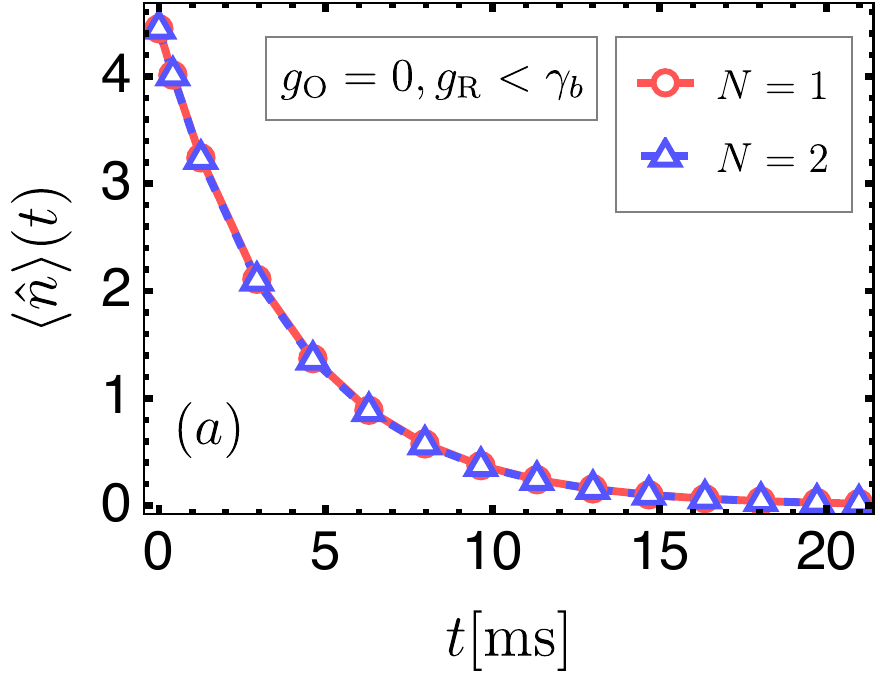}
\includegraphics[width=0.496\columnwidth]{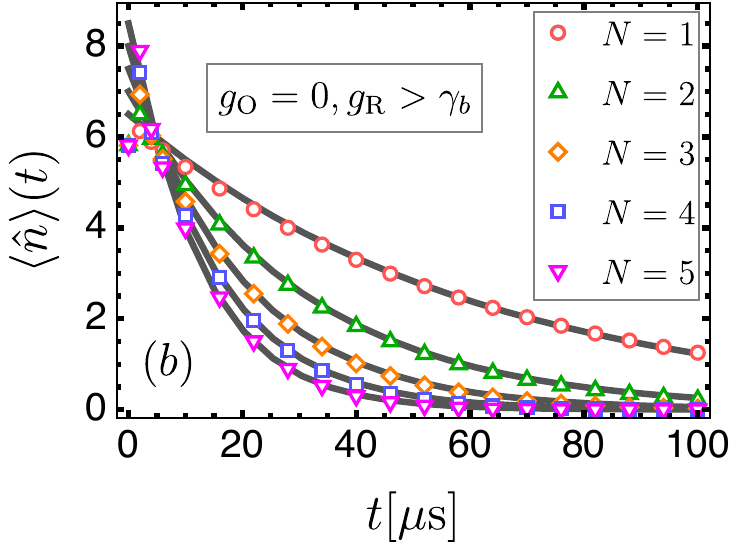}

\includegraphics[width=0.492\columnwidth]{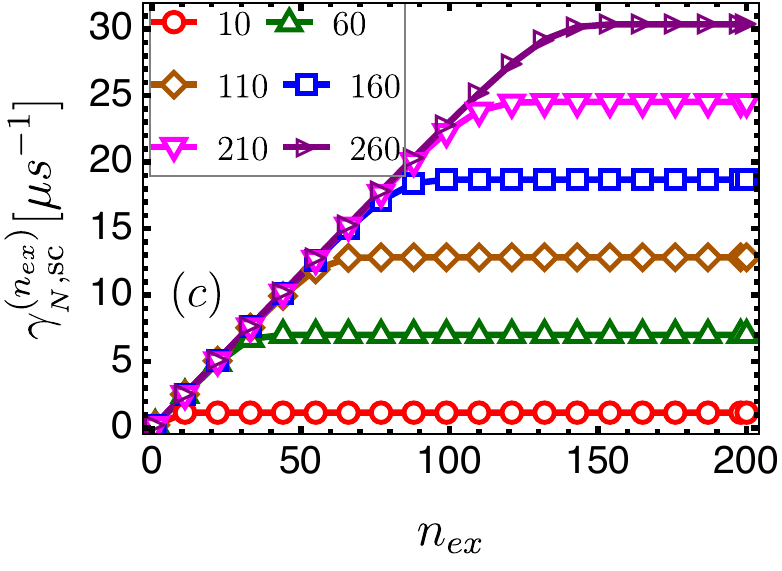}
\includegraphics[width=0.492\columnwidth]{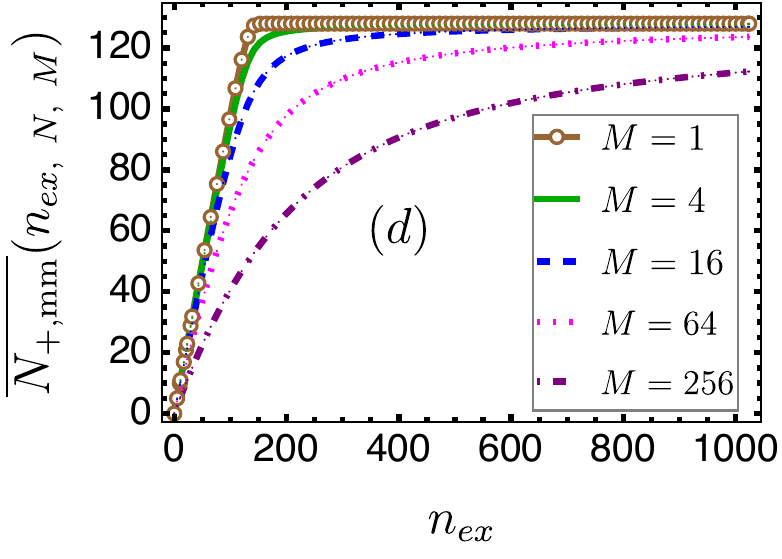}
\caption{Collective cooling dynamics.
(a) Time evolution of the mean phonon number for different $N$ in the weak spin-boson coupling regime ($g_{\rm R}/\gamma_{b}=0.013$) computed numerically with Eqs.~\eqref{eq:barediagonalTLHam}-\eqref{eq:DissipatinJumpDressedframe13}. (b) Same as (a), but in the strong spin-boson regime ($g_{\rm R}/\gamma_{b}\approx1.4$). The curves with markers are numerically computed and represent the different numbers of ions.  The grey curves corresponding to each $N$ are the analytic solutions obtained from Eq.~\eqref{eq:analytic_collective}.
(c) Decay rates $\gamma_{N,\rm sc}^{(n_{\rm ex})}$ given by Eq.~\eqref{eq:gamma_nsc} and \eqref{eq:n_bright} as a function of $n_{\rm ex}$ for different  $N=\{10,60,110,160,210,260\}$ as shown in the inset.
(d) Multimode corrections. Mean number of ions in $\ket{+}$, $\overline{N_{+,\rm mm}}$, following Eq.~\eqref{eq:nplus_mm} as a function of the total number of excitations $n_{\rm ex}$.
The different curves correspond to different numbers of modes $M$, while keeping the total number of ions fixed at $N=256$.
}
\label{Fig:fig6ab}
\end{figure}

In the strong spin-phonon coupling regime $g_{\rm R}>\gamma_{b}$, the situation changes.
As shown in Fig.~\ref{Fig:fig6ab}(b) for $1 \leq N \leq 5$, there is a strong dependence of the evolution of the phonon occupation on $N$.
A similar speed-up of the cooling rate with $N$ was experimentally observed in Ref.~\cite{PhysRevLett.122.053603} and numerically modeled in Ref.~\cite{PhysRevA.99.023409}.
When we  start with an initial state with $\langle \hat{\tilde S}_z \rangle = +N/2$, 
in all cases, we find that the phonon number rises at short times, followed by decay at long times.
Both the initial rise and the subsequent decay rate increase with increasing $N$, clearly indicating some form of  collectively enhanced cooling rate.
In Fig.~\ref{fig:collective-sketches}(c) and below, we provide a  qualitative explanation of the 
enhancement.

The initial peak observed in Fig.~\ref{Fig:fig6ab}(b) originates from Rabi oscillations between spin and phonon degrees of freedom seen in Fig.~\ref{Fig:RabiabtotalEx}(a).
For $N>1$, these Rabi oscillations can now transfer up to $N$ ions from the bright to the dark state, in the process adding $N$ motional quanta.
For example, starting from $\langle \hat{\tilde{S}}_z \rangle = -N/2+M_0$ , with $M_0\in \{0,1,\dots, N\}$, and $n_0$ phonons --- which we label as $\ket{N_+ = M_0,n_0,\beta_0}$ --- the ions can explore the entire manifold  of states, $\ket{N_+=M_0-m,n=n_0+m,\beta_m}$, with $m\in \{-\min(n_0,N-M_0),\dots , M_0\}$.
Here, $\beta_m \in \Bigg\{1,\dots, \binom{N}{M_0-m}\Bigg\}$ accounts for the degeneracy of states with $(M_0-m)$ spin excitations  and with $\binom{N}{M_0-m}$ a binomial coefficient, since we do not care which ions are  excited, just the number of them.
The excitation number directly generalizes from $N=1$ to $\hat n_{\rm ex} = \hat n + \hat N_+$, which is a conserved quantity for $\hat H_{\rm collective, TC}$.
In particular, the energy is simply given by $\omega_m \hat n_{\rm ex}$, with small corrections from $g_R$.
As for $N=1$, the unidirectional right$\to$left motion in the weak coupling regime [Fig.~\ref{fig:collective-sketches}(b)] is replaced by a unidirectional top$\to$bottom motion [Fig.~\ref{fig:collective-sketches}(c)].
This also allows us to understand the collective speed-up more directly.
While for $N=1$ the rate-limiting step was light scattering of the states $\ket{\mathcal S_{n_{\rm ex}}}/\ket{\mathcal A_{n_{\rm ex}}}$, leading to $\gamma_{N=1,\rm sc} \propto \bra{\mathcal X_{n_{\rm ex}}} \ket{+}\bra{+} \ket{\mathcal X_{n_{\rm ex}}} = 1/2$, this will generalize in the many-body case to $\langle \hat N_+ \rangle \propto N$.
This  picture can be more rigorously supported  by 
 an analytical model.


\subsubsection{General dynamics}

For $N>1$, the eigenstates of the Tavis-Cummings are more complicated than the antisymmetric and symmetric superpositions $\ket{\mathcal A_{n_{\rm ex}}}$ and $\ket{\mathcal S_{n_{\rm ex}}}$ introduced in Sec.~\ref{ssec:single-strong}.

In this case, it is convenient to  label the eigenstates as $\ket{n_\mathrm{ex},\zeta}$, with excitation quantum number $\hat n_\mathrm{ex} \ket{n_\mathrm{ex},\zeta} = n_\mathrm{ex} \ket{n_\mathrm{ex},\zeta}$.
Here, $\zeta = 1, \dots, \mathcal N_{N, n_\mathrm{ex}}$ labels the
\begin{align}
    \mathcal N_{N,n_\mathrm{ex}} = \sum_{m=0}^{\min(n_\mathrm{ex},N)} \binom{N}{m}
    \label{eq:definition_degeneracy_of_states}
\end{align}
different eigenstates of Eq.~\eqref{eq:taviscummingsa} for a given $N$ and $n_\mathrm{ex}$.

As before, we project the Lindblad operators into these eigenstates, which is valid for $g_{\rm R} \gg \gamma_b$
\begin{align}
    \hat L_{\mathrm{c}-,j,l}^{(n_\mathrm{ex},\zeta',\zeta)} &=
    \sqrt{\kappa^{(n_\mathrm{ex},\zeta',\zeta)}_{-,j,l}} \ket{n_\mathrm{ex}-1,\zeta'}\bra{n_\mathrm{ex},\zeta}
    \label{eq:collective_decay_definition}
    \, ,
    \\
    \hat L_{\mathrm{c}z,j,l}^{(n_\mathrm{ex},\zeta',\zeta)} &=
    \sqrt{\kappa^{(n_\mathrm{ex},\zeta',\zeta)}_{z,j,l}} \ket{n_\mathrm{ex},\zeta'}\bra{n_\mathrm{ex},\zeta}
    \label{eq:collective_dephasing_definition}
    \, .
\end{align}
The indices $j=1,\dots N$ and $l = g,e$ describe spontaneous emission of the $j$th ion into state $l$.

The first row describes loss of excitation induced by the $\ket{-}\bra{+}$ terms in Eqs.~\eqref{eq:DissipatinJumpDressedframe12} and \eqref{eq:DissipatinJumpDressedframe13},  and the second row describes processes that keep $n_\mathrm{ex}$ fixed induced by the $\ket{+}\bra{+}$ terms in Eqs.~\eqref{eq:DissipatinJumpDressedframe12} and \eqref{eq:DissipatinJumpDressedframe13}.

Following our approach for $N=1$, we made two rotating wave approximations by assuming $\omega_m \gg g_{\rm R}$ (i.e.~$\eta_z \ll 1$) and $g_{\rm R} \gg \gamma_{b}$ and neglect all counterrotating terms.
As a consequence, the density matrix is diagonal in the excitation number basis
\begin{align}
    \hat{\rho}^{(0)}(t) &=
    \sum_{n_{\rm ex},\zeta} \frac{P_{\rm ex}^{(0)}(n_{\rm ex},t)}{\mathcal N_{N,n_{\rm ex}}} \ket{n_{\rm ex},\zeta}\bra{n_{\rm ex},\zeta}
    \, .
\end{align}
We note that this approach is equivalent to degenerate perturbation theory in $\eta_z$ to zeroth order.
While at zeroth order in principle any basis with fixed $n_{\rm ex}$ is valid, at the Hamiltonian level,  $\eta_z$ couples states within a fixed excitation-number sector $n_{\rm ex}$ and between different excitation-number sectors. 
To leading order, we retain only the resonant couplings within a given $n_{\rm ex}$ manifold. 
Additional off-resonant couplings give rise to corrections of order
$\eta_z^2\propto 
g_{\rm R}^2/\omega_s^2
$
which are subleading and will be neglected to expand the jump operators.
As for $N=1$, we also assume that the population is equally distributed among states with the same $n_{\rm ex}$ but different $\zeta$, which is aided by the fact that these states experience similar dynamics.

In the Hamiltonian eigenbasis $\ket{n_{\rm ex},\zeta}$, the slower incoherent dynamics generated by the jump operators, occurring at a rate $\sim \gamma_b$, induces an effective evolution of the excitation-number probability distribution given by 

\begin{align}
    \frac{d}{dt} P_\mathrm{ex}^{(0)}(n_\mathrm{ex},t) =&
    - \gamma_{N,\mathrm{sc}}^{(n_\mathrm{ex})} P_\mathrm{ex}^{(0)}(n_\mathrm{ex},t)
    + \gamma_{N,\mathrm{sc}}^{(n_\mathrm{ex}+1)} P_\mathrm{ex}^{(0)}(n_\mathrm{ex}+1,t)
    \nonumber\\ 
    &+ \mathcal O(\eta_z^2\gamma_b)
    \, .
    \label{eq:analytic_collective}
\end{align}

 In contrast to $N=1$, the decay rates $\gamma_{N,\mathrm{sc}}^{(n_\mathrm{ex})}$ now depend on $n_{\rm ex}$.
We  can then compute them by summing all channels connecting $n_{\rm ex}$ to $n_{\rm ex}-1$ as
\begin{align}
    \gamma_{N,\mathrm{sc}}^{(n_\mathrm{ex})}
    &\equiv
    \frac{1}{\mathcal N_{N,n_{\mathrm{ex}}}}
    \sum_{\zeta,\zeta',\hat L}
    \abs{\bra{n_{\rm ex}-1,\zeta'} \hat L \ket{n_{\rm ex},\zeta}}^2
    \\
    &=
    \sum_{j=1}^N
    \sum_{l=g,e}
    \frac{1}{\mathcal N_{N,n_{\mathrm{ex}}}}
    \sum_{\zeta,\zeta'}
    \kappa_{-,j,l}^{(n_{\rm ex},\zeta',\zeta)}
    \, .
    \label{eq:gamma_N_def}
\end{align}
which, as shown in Appendix \ref{app:collective},
can be  evaluated to be 

\begin{align}
    \gamma_{N,\mathrm{sc}}^{(n_\mathrm{ex})} &= 2\overline{N_+}(n_{\rm ex}, N) \times \gamma_{N=1,\mathrm{sc}} \, ,
    \label{eq:gamma_nsc}
\end{align}
with the state-averaged bright ion number

\begin{align}
    \overline{N_+}(n_{\rm ex}, N)
    &=
    \frac{
    \sum_\zeta \bra{n_{\mathrm ex}, \zeta} \hat N_+ \ket{n_{\mathrm{ex}}, \zeta}
    }{\mathcal N_{N,n_\mathrm{ex}}}
    \nonumber \\
    &=
    \frac{\sum_{m=0}^{\min(n_\mathrm{ex},N)} m 
    \binom{N}{m}
    }{
    \sum_{m=0}^{\min(n_\mathrm{ex},N)} \binom{N}{m}
    } 
    =
    \begin{cases}
      N/2 &\text{for}\ n_\mathrm{ex}\geq N \\    
      \frac{Nn_{\rm ex}}{N+1} + \dots &\text{for}\  n_\mathrm{ex}\ll N
    \end{cases}  \, .
    \label{eq:n_bright}
\end{align}
Here, the dots are small corrections for $1 < n_{\rm ex} \ll N$.
Intuitively, each ion decays independently and therefore  the total decay rate is enhanced by the average number of  bright ions.
The rate $\gamma_{N,\rm sc}^{(n_{\rm ex})}$ is shown in Fig.~\ref{Fig:fig6ab}(c), clearly showing the two regimes.

\subsubsection{Final temperature and cooling rate}

As for $N=1$, we only need to include the first-order corrections to the single leading-order steady state $\ket{n_{\rm ex}=0,\zeta=1}$ to compute the final temperature.
Corrections to other states are weighted by their small steady-state population, and thus enter steady-state observables only at $\mathcal O(\eta_z^{n>2})$.
We find
\begin{align}
&\ket{n_{\rm ex}=0,\zeta=1}^{(1)} = 
\nonumber \\
& \qty[1 - \sum_{n>0,\zeta'} \frac{\abs{\bra{n_{\rm ex}=n,\zeta'} \hat H_1 \ket{n_{\rm ex}=0,\zeta=1}}^2}{\abs{n\omega_m}^2}]\ket{n_{\rm ex}=0,\zeta=1} 
\nonumber \\
&\quad - \sum_{n>0,\zeta'} \frac{\bra{n_{\rm ex}=n,\zeta'}\hat H_1 \ket{n_{\rm ex}=0,\zeta=1}}{n\omega_m} \ket{n_{\rm ex}=n,\zeta'} \, , \\
&\hat H_1 = \sum_{\nu, j} g_{\rm R}^{(\nu, j)} \hat{\tilde \sigma}_y^{(j)} (\hat b_\nu + \hat b_\nu^\dagger) - g_{\rm R}/\sqrt{N} (\hat{\tilde S}_+ \hat b_1 + \hat{\tilde S}_- \hat b_1^\dagger)
\, .
\end{align}
The perturbative coupling $\hat H_1$ for $\nu = 1$ is indicated by the purple dashed line in Fig.~\ref{fig:collective-sketches}.

We assume that all terms $\propto \hat{\tilde \sigma}_-^{(j)}$ and $\propto \hat b_\nu$ vanish, i.e.~that all ions are in the dark state and all phonon modes are in the vacuum state up to higher order corrections.
This results in state admixtures $\sim \eta_z \mathcal K_{\nu,j} \hat{\tilde \sigma}_+^{(j)} \hat b_{\nu}^\dagger \ket{n_{\rm ex}=0, \zeta=1}$ (see Appendix~\ref{app:collective} for exact expression).

As before, we want a purely dissipative master equation which we can transform into a rate equation for the populations, only.
To do this, we make an ansatz for the steady state density matrix: 

\begin{align}
    & \hat \rho^{(1)}(t\rightarrow \infty) =
    \nonumber \\
    &\qquad P_{\rm ex}(0,t\rightarrow \infty) \ket{n_{\rm ex}=0, \zeta=1}^{(1)}\bra{n_{\rm ex}=0, \zeta=1}^{(1)} \nonumber \\
    &\quad + \sum_{n_{\rm ex}>0,\zeta} \frac{P_{\rm ex}(n_{\rm ex},t\rightarrow\infty)}{\mathcal N_{N,n_{\rm ex}}} \ket{n_{\rm ex},\zeta}\bra{n_{\rm ex},\zeta} + \mathcal O(\eta_z^3) \, ,
    \\
    &\quad P_{\rm ex}(0,t) \equiv \bra{n_{\rm ex}=0,\zeta=1}^{(1)} \hat \rho^{(1)} \ket{n_{\rm ex}=0, \zeta=1}^{(1)}
    \\
    &\quad P_{\rm ex}(n_{\rm ex}>1,t) \equiv \bra{n_{\rm ex},\zeta} \hat \rho^{(1)} \ket{n_{\rm ex}, \zeta}
\end{align}
As for $N=1$, this ansatz absorbs the fast coherent dynamics in the basis transformation $\ket{n_{\rm ex}=0,\zeta=1} \rightarrow \ket{n_{\rm ex}=0,\zeta=1}^{(1)}$, and the dynamics of $\hat \rho^{(1)}$ are described by $P_{\rm ex}(n_{\rm ex},t)$, only, which follow a simple linear ordinary differential equation.
Consequently, states with $n_{\rm ex} > 0$ remain unmodified.
However, the dressed ground state $\ket{n_{\rm ex}=0,\zeta=1}^{(1)}$ is not perfectly dark anymore.
For a consistent expansion, we include both the corrections to this state and higher-order terms in the Lamb-Dicke expansion of the effective jump operators.
With these, we compute 

\begin{align}
    \gamma_{n_{\rm ex},\rm p} = \sum_{\zeta',\hat L} \abs{\bra{n_{\rm ex},\zeta'} \hat L \ket{n_{\rm ex}=0,\zeta=1}^{(1)}}^2
\end{align}

up to second order in $\eta_z$, where $\hat L$ sums over all effective jump operators expanded out to first order in $\eta_z$.
These processes pump excitations at rate $\gamma_{1,\mathrm{p}}$ into $n_\mathrm{ex}=1$ and  at rate $\gamma_{2,\mathrm{p}}$ into $n_\mathrm{ex}=2$, which are evaluated in Appendix~\ref{app:collective}.

The matrix $\mathbf M$ describing the dynamics of $\vec p_{\rm ex}$ following Eq.~\eqref{eq:dpex_dt} becomes
\begin{align}
    \mathbf M
    &=
    \begin{pmatrix}
        -\gamma_{1,\rm p} - \gamma_{2,\rm p} & \gamma_{N,\rm sc}^{(1)} & 0 & 0 & \cdots \\
        \gamma_{1,\rm p} & - \gamma_{N,\rm sc}^{(1)} & \gamma_{N,\rm sc}^{(2)} & 0 & \cdots \\
        \gamma_{2,\rm p} & 0 & -\gamma_{N,\rm sc}^{(2)} & \gamma_{N,\rm sc}^{(3)} & \ddots \\
        0 & 0 & 0 & -\gamma_{N,\rm sc}^{(3)} & \ddots \\
        \vdots & & & & \ddots
    \end{pmatrix}
    \, .
\end{align}
Its steady state has $P_{\rm ex}(n_{\rm ex},t\rightarrow\infty) = 0$ for $n_{\rm ex}>2$, and the normalized populations for $n_{\rm ex}\leq 2$ are
\begin{align}
    \vec p_{\rm ex}(t\rightarrow\infty)
    &=
    \frac{
    \qty[
        \gamma_{N,\rm sc}^{(1)} \gamma_{N,\rm sc}^{(2)},\ 
        \qty(\gamma_{1,p} + \gamma_{2,p} ) \gamma_{N,\rm sc}^{(2)},\ 
        \gamma_{2,p} \gamma_{N,\rm sc}^{(1)},0,
        \dots
    ]^T
    }{
    \qty(\gamma_{N,\rm sc}^{(1)} + \gamma_{1,p} + \gamma_{2,p} ) \gamma_{N,\rm sc}^{(2)}
    +
    \gamma_{2,p} \gamma_{N,\rm sc}^{(1)}
    }
    \, .
\end{align}

We now define vectors $\vec O$ which allow us to compute the observables as $\langle\hat O\rangle = \vec O \cdot \vec p_{\rm ex}$.
Due to the basis choice for $n_{\rm ex}=0$, the vectors associated with the observables take the form (see Appendix~\ref{app:collective})
\begin{align}
    \vec n_{\rm ex} &=
    \qty(
    \sum_{\nu}\frac{\eta_z^2 \Omega_e^2 \Omega_g^2 \omega_m^2}{\qty(\Omega_e^2 + \Omega_g^2)^2 \qty(\omega_m + \omega_\nu)^2},
    1,
    2,
    \dots
    )^T
    \, ,
    \\
    \vec n
    &=
    \qty(
    \frac{\eta_z^2\Omega_e^2\Omega_g^2}{4\qty(\Omega_e^2 + \Omega_g^2)^2},
    \frac{1}{N+1},
    \frac{4+2N}{2+N(N+1)},
    \dots
    )^T
    \, ,
    \label{eq:phonon_number_vector}
    \\
    \vec p_{\rm gs} &=
    \qty(
    1 - \sum_{\nu}\frac{\eta_z^2 \Omega_e^2 \Omega_g^2 \omega_m^2}{\qty(\Omega_e^2 + \Omega_g^2)^2 \qty(\omega_m + \omega_\nu)^2},
    0,
    0,
    \dots
    )^T
    \, .
    \label{eq:ground_state}
\end{align}
Here, we define the ground state probability $\hat p_{\rm gs} \equiv \ket{n_{\rm ex}=0,\zeta=1}\bra{n_{\rm ex}=0,\zeta=1}$.
The dots in Eq.~\eqref{eq:phonon_number_vector} need to be evaluated by state counting as $n_{\rm ex} - \bar N_+(n_{\rm ex},N)$.
The full expressions, used to evaluate dynamics, and closed forms for the steady state are given in Appendix~\ref{app:collective}.

Fig.~\ref{Fig:fig6ab}(d) shows good qualitative agreement between the analytic dynamics and the corresponding final ground state population, and the numerical solution of the full dynamics for $N=1$.
Interestingly, in the strong coupling regime, the final temperature depends on the asymmetry between $\Omega_g$ and $\Omega_e$, and is hottest for $\Omega_g \approx \Omega_e$ i.e. when $\alpha\approx \pi/2$.
However, there is a trade-off between decreasing the final temperature by changing the ratio $\sin(\alpha) \rightarrow 0$ since it also slows down  the cooling rate.

Beyond the steady-state, we can also estimate the decay rate.
The equation of motion for $\bar n_{\rm ex}(t) =  (\vec n_{\rm ex} \cdot \vec p_{\rm ex}(t))$ can be computed as
\begin{align}
    \dot{\bar n}_{\rm ex}
    &=
    \vec n_{\rm ex}^T \mathbf{M} \vec p_{\rm ex}(t)
    \label{eq:nex_dot}
    \\
    &\approx
    \qty[\sum_{n_{\rm ex}> 0} \frac{\gamma_{N,\rm sc}^{(n_{\rm ex})}}{\bar n_{\rm ex}(t)} P_{\rm ex}^{(0)}(n_{\rm ex},t) ] \times \bar n_{\rm ex}(t)
    \, ,
\end{align}
where in the second step we only keep leading-order terms in $\eta_z$.
Thus, we can identify the term in the square brackets as an effective decay rate, which depends on $P^{(0)}_{\rm ex}$.

For $n_{\rm ex} \geq N$, $\gamma_{N, \rm sc}^{(n_{\rm ex})}$ does not depend on $n_{\rm ex}$, and we can pull it out of the sum to define an effective instantaneous decay rate $\gamma_{N,\rm sc}^{(n_{\rm ex})}/\bar{n}_{\rm ex}$, as for $N=1$.


\subsubsection{Multimode effects}
While so far we have focused on cooling a single mode, one of the primary advantages of EIT cooling is its ability to simultaneously cool many modes.
For the multimode case, Eq.~\eqref{eq:taviscummingsa} becomes (after rotating out the $\omega_m(\hat b^\dagger \hat b + \hat{\tilde S}_z)$-term)
\begin{align}
    \hat H_{\rm mm} &=
    \qty(\sum_{\nu, j}\frac{g_{\rm R} \mathcal K_{\nu,j} \sqrt{\omega_1}}{\sqrt{\omega_\nu}}
    \hat{\tilde \sigma}_+^{(j)} \hat b_\nu + h.c.)
    + \sum_\nu (\omega_\nu - \omega_m) \hat b_\nu^\dagger \hat b_\nu
    \, .
\end{align}
Then, only modes with $\abs{\omega_\nu - \omega_s} \lesssim g_{\rm R}$ can exchange interactions between spin and motion, and cool.
In Appendix~\ref{Ap:strong-multimode}, we numerically confirm for $N=2$ that indeed modes up to $\abs{\omega_\nu - \omega_s} \lesssim 4g_{\rm R}$ can be cooled.
For each mode, ion-dependent couplings modify the eigenstates and thus the individual decay rates $\kappa_{\nu,-,j,l}^{(n_\mathrm{ex},\zeta',\zeta)}$ become mode dependent. However, their sum $\gamma_{N,\rm sc}^{(n_{\rm ex})}$ will not depend on the mode because $\sum_\nu \abs{\mathcal K_{\nu,j}}^2 = 1$.
Thus, all modes with $\abs{\omega_\nu - \omega_s} \ll g_{\rm R}$ cool with similar rate.

However, when multiple modes are cooled simultaneously, they share the same ions for cooling, leading to cross-talk between the modes.
In particular, the rate-limiting step for excitation removal is the photon scattering rate of the bright state $\gamma_{N,M,\rm mm}^{(n_{\rm ex})}$.
In the multi-ion case, we need to replace $n_{\rm ex}$ with the total number of excitations $\hat n_{\rm ex} = \sum_\nu \hat b_\nu^\dagger \hat b_\nu + \hat N_+$.
We can again assume that the states quickly average over all possible internal states with fixed excitation number
\begin{align}
    \gamma_{N,M,\rm mm}^{(n_{\rm ex})} =& 
    2 \overline{N_{+,\rm mm}}(n_{\rm ex}, N, M) \times \gamma_{N=1,\rm sc} 
    \label{eq:definition_gamma_N_nex_multimode}
    \\
    \overline{N_{+,\rm mm}}(n_{\rm ex}, N, M) =&
    \frac{\sum_{m=0}^{\min(n_{\rm ex},N)} m \binom{N}{m} \binom{M - 1 + n_{\rm ex} - m}{M - 1}}{\sum_{m=0}^{\min(n_{\rm ex},N)} \binom{N}{m} \binom{M - 1 + n_{\rm ex} - m}{M - 1}}
    \, .
    \label{eq:nplus_mm}
\end{align}
where the additional binomial coefficient accounts for the number of ways to distribute $n_{\rm ex} - m$ excitations across $M$ modes.
We plot $\overline{N_{+,\rm mm}}(n_{\rm ex},N,M)$ in Fig.~\ref{Fig:fig6ab}(d) for different $M$.
We find that for $M\ll N$, $\overline{N_{+,\rm mm}}$ closely resembles the single-mode $M=1$ case.
For $n_{\rm ex} \gg N$, we recover $\overline{N_{+,\rm mm}}(n_{\rm ex},N,M) \to N/2$, while for $n_{\rm ex} = 1 \ll N$, we can evaluate Eq.~\eqref{eq:nplus_mm} to find $\overline{N_{+,\rm mm}}(n_{\rm ex},N,M) = N / (N+M)$.

As in Eq.~\eqref{eq:nex_dot}, we can define an effective instantaneous cooling rate $\gamma_{N,M,\rm mm}^{(n_{\rm ex})} / \bar n_{\rm ex}$.
Assuming we want to cool $M \ll N$ modes, at sufficiently large $n_{\rm ex}$, these quantities scale as $\gamma_{N,M,\rm mm}^{(n_{\rm ex})} \propto N$ and $\bar n_{\rm ex} \propto M$, leading to a scaling of the cooling rate for $M$ modes with $N$ ions as $N/M$.
In Appendix~\ref{Ap:strong-multimode}, we confirm this intuition for $N=2$, and find that as we tune the second mode out of resonance, we smoothly interpolate between $N/M = 1$ and $N/M = 2$.

To estimate the final temperature, for $n_{\rm ex} \ll N$, instead we have $\gamma_{N,M,\rm mm}^{(n_{\rm ex})} \approx 2\gamma_{N=1,\rm sc} \frac{N}{N+M} n_{\rm ex}$.
Thus, for $M \ll N$, increasing the number of modes and $n_{\rm ex} \propto M$ together with it leaves the instantaneous cooling rate $\gamma_{N,M,\rm mm}^{(n_{\rm ex})} / \bar n_{\rm ex}$ unchanged.
In Appendix~\ref{Ap:strong-multimode}, we furthermore evaluate the reheating rates and the resulting steady state.
The reheating only slightly increases for small $M$.
As a consequence, for $M \ll N$, the final total excitation number (summed over all modes) remains approximately constant.
For $M = N$, the final total excitation number increases by approximately 2.5, but does not scale with $M$.
It may seem counterintuitive that the total excitation number is not proportional to the number of modes.
However, most excitations are stored in the ions internal state, which is independent of the number of modes.
Consequently, we expect efficient cooling even in the presence of many modes.

\begin{table*}[]
    \centering
    \renewcommand{\arraystretch}{1.4}
    \setlength{\extrarowheight}{4pt}
    
    \begin{tabular}{r|c|c}
        & Cooling rate & Cooling limit \\
        \hline
        Weak coupling ($2\gamma/\Delta_{\rm R} \gg \eta_z, g_{\rm O}/\omega_m$) &
        $\frac{\Delta_{\rm R}\omega_m}{\gamma} \qty(\frac{\eta_z^2}{4} + \frac{4g_{\rm O}^2}{\omega_m^2})$ &
        $1 - P(0,t\rightarrow\infty) \approx \frac{\gamma^2}{16\Delta_{\rm R}^2}$
        \\[6pt]
        \hline
        Strong coupling ($2\gamma/\Delta_{\rm R} \ll \eta_z$), $n_{\rm ex} \ll N$ &
        $\frac{\Delta_{\rm R}\omega_m}{\gamma}\frac{\gamma^2}{4\Delta_{\rm R}^2} \frac{2N}{N+1}$ &
        \multirow{2}{*}{%
        $1 - \langle \hat p_{\rm gs}\rangle (t\rightarrow\infty)
        \approx
        \eta_z^2 \frac{\qty[\sum_\nu\frac{8N(N+1)\omega_\nu^2 + 8N^2\omega_m^2}{(\omega_m + \omega_\nu)^2}]+3N^2+3N+2}{32N^2} $
        }
        \\[6pt]
        \cline{1-2}
        Strong coupling ($2\gamma/\Delta_{\rm R} \ll \eta_z$), $n_{\rm ex} \geq N$ &
        $\frac{\Delta_{\rm R}\omega_m}{\gamma}\frac{\gamma^2}{4\Delta_{\rm R}^2}\frac{N}{n_{\rm ex}}$ &
    \end{tabular}
    \caption{Cooling parameters for $\Delta_g = \Delta_e = \Delta_{\rm R}$, $\Omega_g = \Omega_e = \sqrt{2\Delta_{\rm R} \omega_m}$, such that $\omega_s = (\Omega_g^2 + \Omega_e^2)/4\Delta_{\rm R} = \omega_m$, and far detuned $\gamma,\Omega \ll \Delta_{\rm R}$, $\gamma = \gamma_g + \gamma_e$, in terms of the bare Hamiltonian parameters, and for $\gamma_m = 0$. Note that $\Omega_g$ is determined by $\Omega_g^2 = 2\Delta_{\rm R}\omega_m$. We also set $g_{\rm O}=0$ for the strong coupling case.
    $\omega_\nu$ are the axial eigenfrequencies of the crystal.
    We define the cooling rate here such that it locally fits an exponential decay.}
    \label{tab:analytic_results}
\end{table*}

\section{Spontaneous Emission Recoil}\label{sec:IV}
We now discuss how the recoil of the spontaneously emitted photons affects the dynamics.
In contrast to the laser photons, whose $\vec k$-vectors are controlled, spontaneously emitted photons are emitted into a random direction.
To account for that, we need to integrate over all possible emission angles weighted by their polarization-dependent emission pattern.
For circular polarization, the recoil along $Z$ is described by $W(\mu_l)=(3/8)(1+\mu_l^{2})$ with $\mu_{l}= (\vec{k}_{l}\cdot e_{Z})/|\vec{k}_{l}| $ and $e_{Z}$ the unit vector along $Z$~\cite{RevModPhys.58.699}.
The master equation then becomes
\begin{align}
    \partial_t \hat \rho 
    &=
    -i \qty[\hat H, \hat \rho] + \sum_{l=e,g} \int_{-1}^1 d\mu_l \mathcal D_{\hat L_{l,\mu_l}}\qty[\hat \rho]
    \, , 
    \label{eq:MaterEqWithRecoil}
    \\
    \hat{L}_{l,\mu_l}
    &=
    \sqrt{\gamma_{l}W(\mu_l)}\hat{\sigma}_{lr}e^{-i\mu_{l} |\vec{k}_{l}|\hat{Z}}
    \label{eq:JumpsWithRecoil}
\end{align}

\begin{figure}[t!]
	\includegraphics[width=0.495
    \columnwidth]{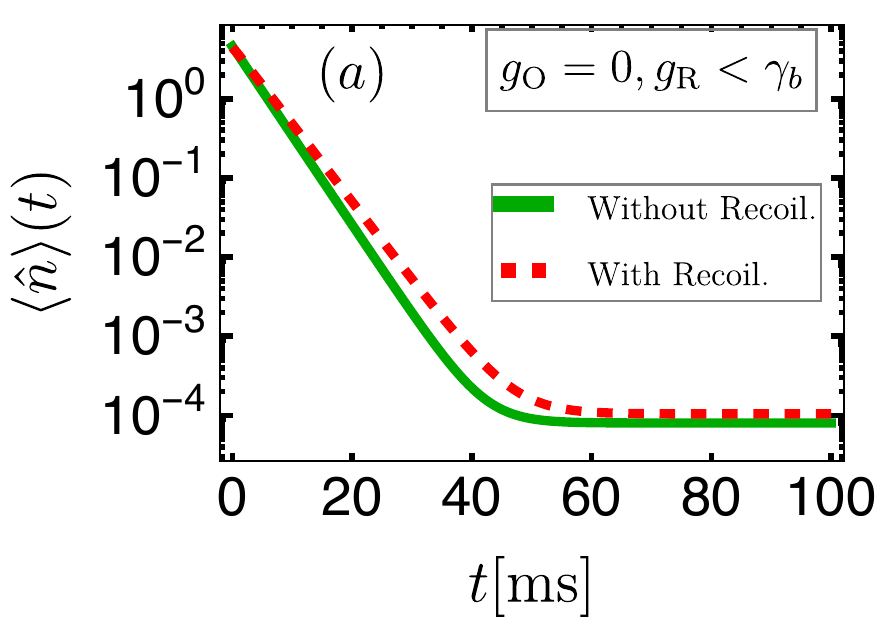}
	\includegraphics[width=0.495\columnwidth]{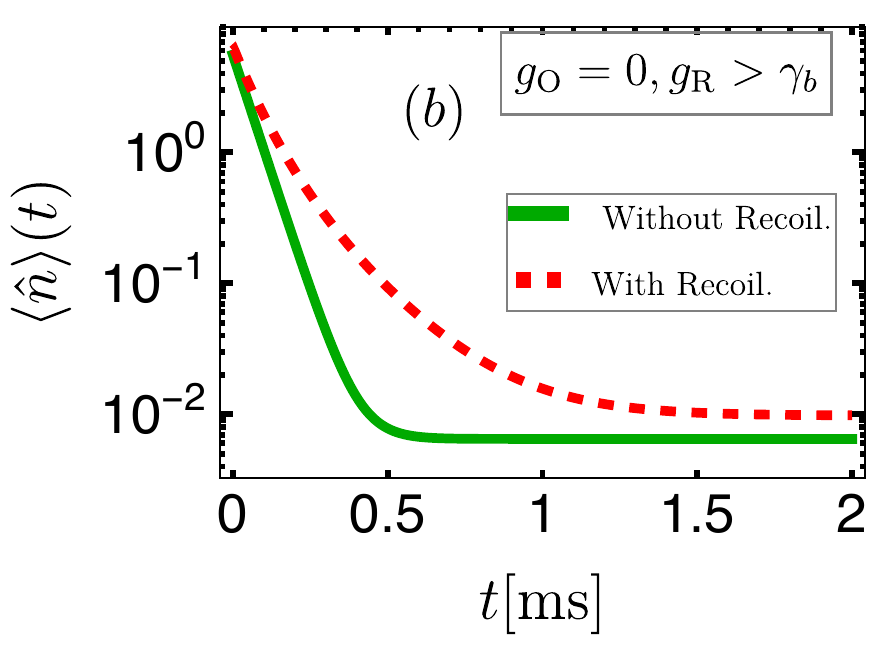}
\caption{Inelastic recoil effects on the cooling dynamics in weak (a) and strong (b) coupling regimes. We set $\gamma_{m}=0$  and $N=1$ for both cases. The other parameters are the same as previously stated for the corresponding cases.
}
\label{Fig:InelesticRecoil}
\end{figure}

 We numerically study the full quantum dynamics for $N=1$.
Naively, one might assume that this introduces major corrections to the final ground state population for $\abs{\vec k_l} \gg \Delta k_{z}$, which is especially true in the weak coupling regime.
However, the numerical simulations in Fig.~\ref{Fig:InelesticRecoil} show that at short times with $\langle \hat n \rangle \gtrsim 1$, the dynamics barely changes; at intermediate times, the cooling rate slows down, especially in the strong-coupling case; at late times, the final ground state fractions increase slightly ($\sim 25-50\%$) compared to ignoring the recoil of the emitted photon.
We can understand this by considering that the final population is determined by the competition of processes that pump population into or out of the state dark state $\ket{-,0}$.
However, the leading order contribution in the Lamb-Dicke parameter remains unchanged for both processes.
This is because the dark state is dark due to the destructive interference of the Raman lasers, irrespective of the recoil of the emitted photons.
Only a momentum kick by at least one of the Raman lasers can break this destructive interference, which is proportional to the Raman Lamb-Dicke parameter (see Appendix~\ref{app:collective}), and thus reduced due to the small incidence angles.
Finally, these corrections are much more prominent for the strong coupling regime, where the corrections due to the Raman Lamb-Dicke parameter are much larger compared to the weak coupling regime.

To estimate the effect for $N>1$ in the strong coupling regime, we compute the steady-state photon scattering rate, which determines how often a recoil is applied to each ion.
In the strong coupling regime, in first-order perturbation theory, and for balanced Raman Rabi frequencies, the steady-state photon scattering rate is given by
\begin{align}
    \gamma_{\rm scat} =& \sum_{j,l} \int d\mu_l \Tr[\hat \rho^{(1)}(t \rightarrow \infty) \hat L_{l,\mu_l}^{(j)\dagger} \hat L_{l,\mu_l}^{(j)}]
    \nonumber \\
    =& \frac{\Omega^2\gamma}{\Delta_{\rm R}^2} \eta_z^2 \qty[\frac{1}{4} + \sum_{\nu=1}^N \frac{\omega_\nu^2}{\qty(\omega_m + \omega_\nu)^2}]
    \, .
\end{align}
This term does increase with $N$ as $\sum_\nu \frac{\omega_\nu^2}{\qty(\omega_m + \omega_\nu)^2}$ in exactly the same way as $\gamma_{1,p} + 2\gamma_{2,p}$, which is the reheating rate for large $N$ without recoil.
If we assume that photon recoil simply introduces an additional heating rate proportional to $\gamma_{\rm scat}$, the change in relative importance will follow $\gamma_{\rm scat}/(\gamma_{1,p} + 2\gamma_{2,p})$, which evaluates to $\gamma_{\rm scat}/(\gamma_{1,p} + 2\gamma_{2,p}) = 32/3$ for $N=1$ and $\gamma_{\rm scat}/(\gamma_{1,p} + 2\gamma_{2,p}) = 16$ for $N\to \infty$.
Thus, we expect the relative effect of photon recoil in the strong coupling regime to increase by about 50\% as we increase $N$.
This is more than offset by the substantial speed-up in the early cooling dynamics that scales linearly with $N$.

Furthermore, we emphasize that the large and non-perturbative recoil effects are a consequence of the low mass of $^9$Be$^+$ ions, leading to a large single-photon Lamb-Dicke parameter $| \vec k_l|Z_{zpf} = 0.38$.
Most trapped-ion experiments operate at much lower single-photon Lamb-Dicke parameters.
E.g.~Lechner \textit{et al.}~demonstrated EIT cooling for radial modes in an 18-ion linear Paul trap with single-photon Lamb-Dicke parameter of $0.11$ for radial trap frequencies of $\omega_m = 2\pi \times 2.7$ MHz~\cite{PhysRevA.93.053401}.

\section{Optimal cooling parameters}
\label{sec:optimal-parameters}
\begin{figure}[t!]
	\includegraphics[width=0.51
    \columnwidth]{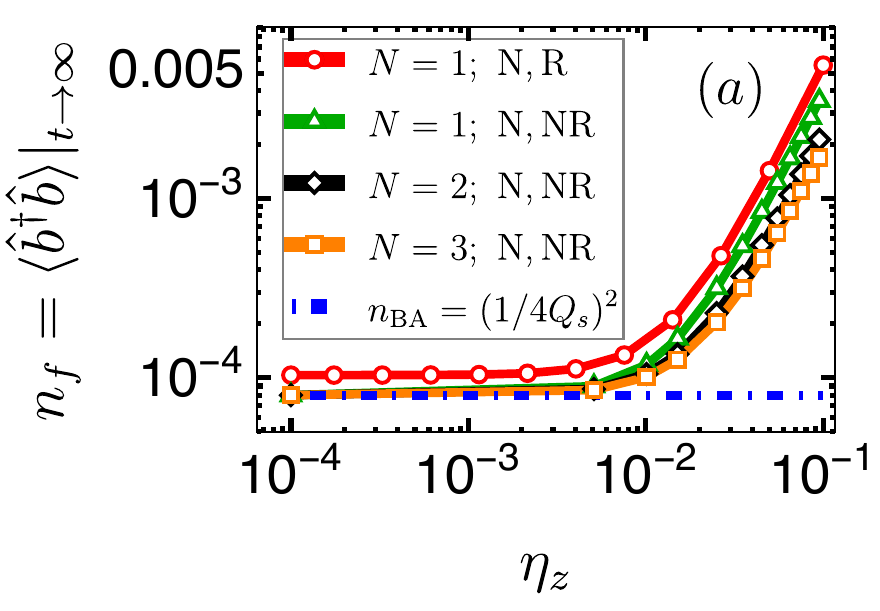}
    \includegraphics[width=0.475\columnwidth]{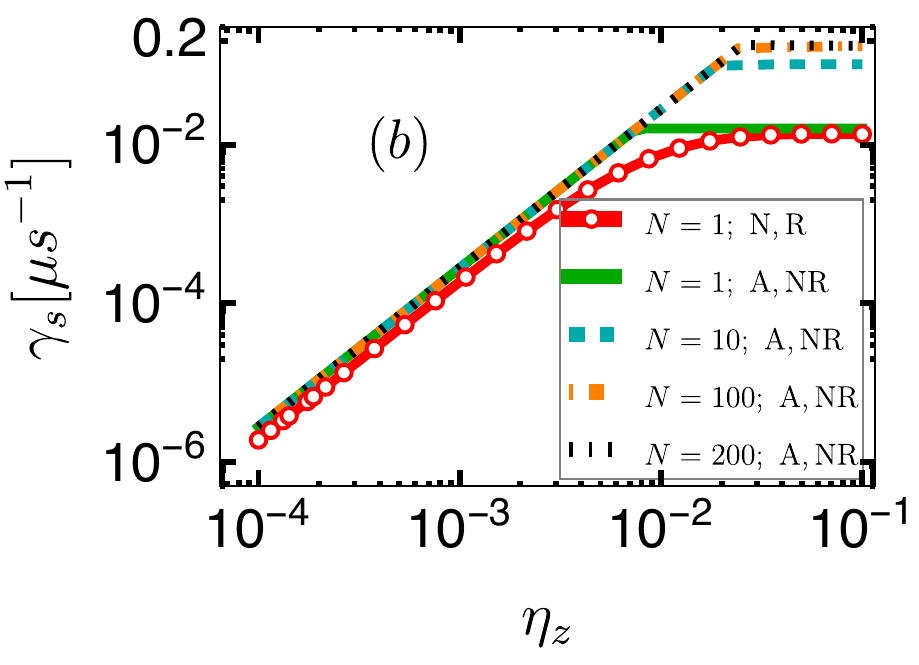}

\includegraphics[width=0.5\columnwidth]{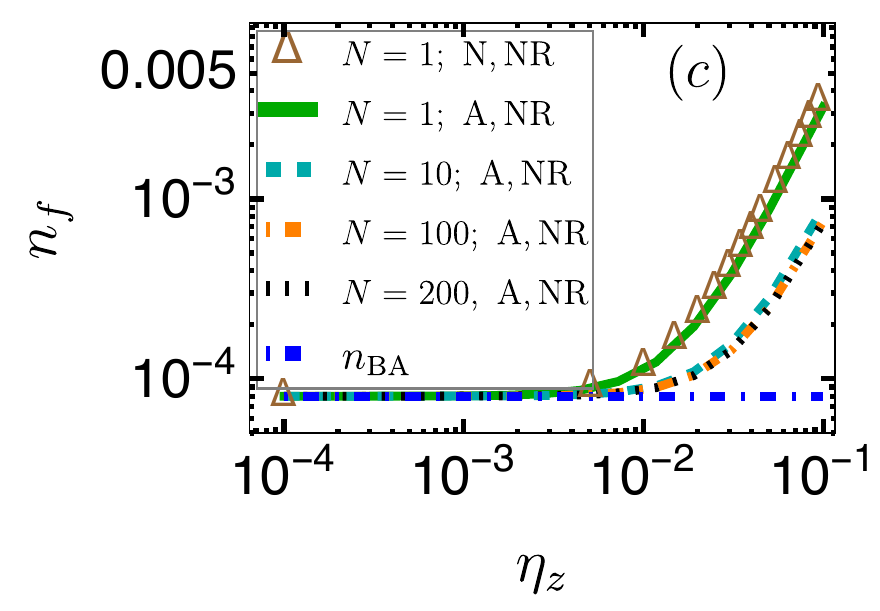}
\includegraphics[width=0.48\columnwidth]{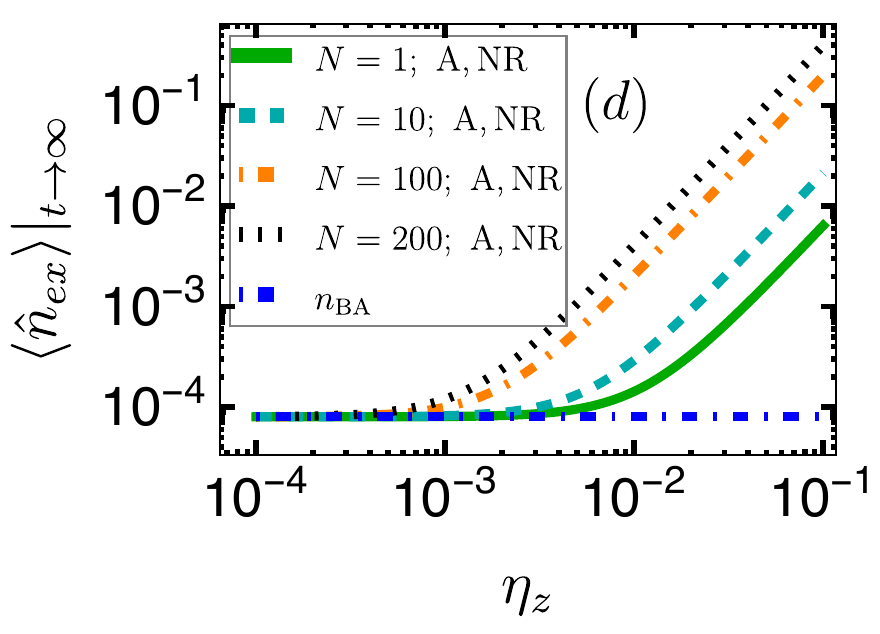}
\caption{
Cooling across different parameter regimes, computed numerically (N) or following analytical results (A). The numerical results further distinguish between including photon recoil (R) or not (NR).
(a) Steady-state phonon number as a function of $\eta_z$.
The red curve with dots includes emitted-photon recoil as a numerical solution of Eqs.~\eqref{eq:MaterEqWithRecoil} and \eqref{eq:JumpsWithRecoil}, the other curves do not include emitted-photon recoil following Eq.~\eqref{eq:MasterEqDressedstate1OB}. The horizontal blue dashdotted line represents the fundamental quantum backaction limit.
(b) Cooling rate during the first 50 \textmu s as a function of $\eta_z$. The red curve with dots is a numerical solution of Eqs.~\eqref{eq:MaterEqWithRecoil} and \eqref{eq:JumpsWithRecoil}, the curves without symbols represent our analytic results in Eq.~\eqref{eq:gamma_s_analytic}.
(c) Analytic steady-state mean phonon number as a function of $\eta_z$ for different $N$ given by Eq.~\eqref{eq:n_f_analytic}.
The triangles are the numeric solution plotted as a green curve with triangles in panel (a).
(d) Analytic steady-state total excitation number as function of $\eta_z$ computed as the sum of Eqs.~\eqref{Eq:rates} and \eqref{eq:analytic_excitations}.
Parameters: $\Omega_g = \Omega_e = 2\pi \times 40$ MHz, $\Delta_{\rm R} = 2\pi \times 512$ MHz, $\omega_m = 2\pi \times 1.59$ MHz, $n_{\rm th} = 4.6$, $\gamma_g = 2\pi \times 6$ \textmu s$^{-1}$, $\gamma_e = 2\pi \times 12$ \textmu s$^{-1}$.
}
\label{Fig:optimality}
\end{figure}

We summarize our results in Table~\ref{tab:analytic_results} and Fig.~\ref{Fig:optimality} to identify the optimal cooling parameters.
For $N=1$, we numerically compute the cooling dynamics and the steady state for different two-photon Lamb-Dicke parameters $\eta_z$ by tuning the angles $\theta_g$ and $\theta_e$ in Fig.~\ref{Fig:Fig1}(b).
Fig.~\ref{Fig:optimality}(a) shows the steady-state phonon number, extracted by computing the dynamics until saturation.
In the weak coupling regime $\eta_z \ll 10^{-2}$, the final phonon number $n_f \approx n_{\rm BA}$ remains essentially independent of $\eta_z$ at the backaction limit, with a small correction due to the recoil of the spontaneously emitted photon.
As $\eta_z$ increases beyond $10^{-2}$, the final phonon number increases as well, with identical trends when including and excluding the recoil.
Without recoil, increasing the ion number $N$ slightly suppresses the final phonon number.
These two regimes can be combined by summing the contributions from weak and strong coupling~\cite{PhysRevA.104.013117} as
\begin{align}
    n_f \approx \gamma^2/16\Delta_{\rm R}^2 + n_{f,\rm sc}
    \, ,
    \label{eq:n_f_analytic}
\end{align}
where $n_{f,\rm sc}$ is given by Eq.~\eqref{eq:analytic_phonons} in Appendix \ref{app:collective}.
This matches the numerical simulations as shown in Fig.~\ref{Fig:optimality}(c).
This is consistent with our analytical predictions for $g_O=0$ and $N=1$, in Table~\ref{tab:analytic_results}.

The cooling rate is shown in Fig.~\ref{Fig:optimality}(b).
We extract the initial cooling rate numerically from the 0-to-50 \textmu s dynamics as $\gamma_s = -\ln[\langle \hat n  (50 \text{ \textmu s})\rangle /\langle \hat n(0) \rangle]/(50 \text{ \textmu s})$ (red curve).
For $\eta_z \ll 10^{-2}$, the cooling rate increases as $\eta_z^2$, while it settles to a constant for $\eta_z \gg 10^{-2}$.
Both regimes are well captured by our analytic predictions in Table~\ref{tab:analytic_results} (green curves).
We combine both regimes as 
\begin{align}
    \gamma_s \approx \min(\eta_z^2\Delta_{\rm R}\omega_m/4\gamma ,\gamma_{N, \rm sc})
    \, ,
    \label{eq:gamma_s_analytic}
\end{align}
where $\gamma_{N,\rm sc}$ is computed from Eq.~\eqref{eq:nex_dot} by assuming a thermal distribution $(\vec p_{\rm ex})_n \propto \exp[-2n\times \mathrm{coth}(2n_{\rm th}+1)]$.
The analytical results that combine the steady state and cooling rate of weak and strong coupling can now be extended to $N \gg 1$.

While the weak-coupling cooling rate remains identical for changing $N$, the maximal cooling rate at large $\eta_z$ increases for large $N$ (black, orange, and blue curves in Fig.~\ref{Fig:optimality}(b)).
Consequently, the optimal $\eta_z$ shifts to larger values of $\eta_z$ with increasing $N$.
To compute the steady-state phonon number, we use $\omega_\nu$, which are here computed numerically for a Penning trap with $N$ ions and fixed axial center-of-mass frequency $\omega_m/(2\pi) = 1.59$ MHz~\cite{PhysRevX.14.031030}.
We find that the final phonon number behaves similar to the cooling rate: The weak coupling is independent of $N$, the transition point shifts to larger $\eta_z$ for larger $N$, and the cooling rate in the strong coupling regime improves (Fig.~\ref{Fig:optimality}(c)).

However, in the strong coupling regime, many ions can be in the bright state, and the entropy in the system is captured by $\langle \hat n_{\rm ex} \rangle > n_f$.
This contrasts the weak coupling regime, where the number of ions in the bright state vanishes because the bright-state decay is much faster than the pumping into the bright state, such that $\langle \hat n_{ex} \rangle \approx \langle \hat b^\dagger \hat b\rangle$.
Fig.~\ref{Fig:optimality}(d) shows the final excitation number computed by summing weak-coupling and strong coupling contributions.
While $N=1$ barely changes, the trend with increasing $N$ reverts and the crossover point shifts to smaller values of $\eta_z$:
The bright ion number increases with the total ion number, because additional ions introduce new phonon modes which can participate in the off-resonant creation of pairs of bright ions and non-center-of-mass phonons.
In summary, when optimizing to cool center-of-mass phonons, an ideal $N$-dependent value of $\eta_z$ optimizes both cooling rate and final temperature.
In contrast, when optimizing to remove all entropy, there is a range of good $\eta_z$, here between $10^{-3} \lesssim \eta_z \lesssim 2\times 10^{-2}$, which trade faster cooling for increased final temperature.
If $\eta_z$ can be tuned dynamically (e.g.~by changing the trap tightness), it might thus be optimal to initially work at large $\eta_z$ and then dynamically decrease $\eta_z$ to cool quickly and minimize the entropy.
This dynamical tuning could also be achieved by initially introducing an optical dipole force,  which can be understood as an effective increase in $\eta_z$ due to the equivalence of $g_{\rm R}$ and $g_{\rm O}$ observed in Eqs.~\eqref{eq:jaynescummings} and Eq.~\eqref{eq:SpinhamoEffMMa}. Then, turning the optical dipole force off after a steady state is reached reduces the effective $\eta_z$ and can provide a fast and efficient two-stage cooling process.

Now constraining the analysis to the crossover regime $\eta_z \sim \gamma/\Delta_{\rm R}$, the cooling rate scales as $\gamma_s(\eta_{z,\rm opt}) \propto \omega_m \gamma /\Delta_{\rm R}$ and the cooling limit scales as $n_f(\eta_{z,\rm opt}) \propto \gamma^2/\Delta_{\rm R}^2$.
This will provide a setup-specific trade-off, where increasing the detuning improves the cooling limit, but slows down the cooling.

\section{Conclusion} 
\label{sec:conclusion}
We have formulated a two-photon dark-state laser cooling scheme for trapped-ion arrays.
Our scheme directly generalizes EIT cooling to arbitrary Raman Rabi frequency ratios, an additional optical dipole force, and strong spin-motion coupling.
All of these can enhance the cooling rate under the right conditions.
The two-level description also provides a unified intuitive framework valid throughout all three regimes.
This has enabled us to unveil the nature of a collective cooling speed-up, which we attribute to an effective parallelization of cooling across multiple ions, and is unique to the strong coupling regime.

In addition to Penning traps, which we used to parametrize the simulations presented in this work,  our results can directly optimize current quantum simulation, quantum computing~\cite{PhysRevLett.125.053001}, and ion clock~\cite{dawel2025a} experiments in Paul traps, which often use EIT cooling as part of their toolbox. 

Paul traps are now also pushing increasingly large ion crystals, either in the form of 1D ion strings or even 2D crystals.
EIT cooling has already been demonstrated in ion strings~\cite{PhysRevA.93.053401} and 2D crystals~\cite{qiao2021double,kiesenhofer2023controlling,guo2024a}.
Since our scheme has similar requirements to EIT cooling, this opens the collective effects observed here as an interesting avenue to enhance cooling in these settings.
Especially in 2D, increasing micromotion with increasing $N$ broadens the cooling resonances. While this broadening poses a challenge to most protocols, including the ones presented here, often one can focus on cooling the transverse (i.e.~axial) direction, which can be made perpendicular to the micromotion~\cite{qiao2021double,kiesenhofer2023controlling}.
Furthermore, the enhanced cooling bandwidth $\propto g_{\rm R}$ in the strong coupling regime might help to overcome broadening.
In this context, it will be interesting to explore if our theory can be combined with existing methods to optimize EIT cooling~\cite{PhysRevLett.125.053001,fouka2025multilevel}.
It will also be interesting to explore extensions towards 3D ion crystals~\cite{PhysRevX.14.031030}.

Going beyond trapped ions, two-photon Raman cooling is also common in neutral atom~\cite{PhysRevX.2.041014,fd15-4g1n} and molecule tweezer arrays~\cite{park2023extended, PhysRevX.14.031002, Lu2024}.
In this direction, it will be promising to combine our work with recent work on EIT cooling beyond $\Lambda$-level structures~\cite{fouka2025multilevel}.

Furthermore, it might be interesting to relate our scheme to $\Lambda$-enhanced gray molasses when no trap is present~\cite{PhysRevLett.127.263201,phatak2024generalized}.
Finally, while here motion is typically well isolated, it will be interesting to study if dipolar interactions may produce similar collective enhancements~\cite{wang2023enhanced,rubies2025collectively}.


{\it Acknowledgments:} 
We thank  Deviprasath Palani and  Steve Pampel for  feedback on the manuscript and Samarth Hawaldar for providing the Penning crystal mode frequencies. We would like to express our sincere gratitude to Peter Zoller for his insightful discussions on this project during his visits to JILA. MMK thanks Sahin K. Özdemir for the support at Saint Louis University. This material is based upon work supported by the  the U.S. Department of Energy, Office of Science, National Quantum Information Science Research Centers, Quantum Systems Accelerator. We also acknowledge funding support from  ARO grant  W911NF24-1-0128,  FA9550-25-1-0080, the NSF JILA-PFC PHY-2317149  and  NIST. MMK acknowledges the support of Air Force Office of Scientific Research (AFOSR) Multi-disciplinary University Research Initiative (MURI) award on Programmable Systems with Non-Hermitian Quantum Dynamics (Award No. FA9550-21-1-0202).
DW acknowledges additional support by the German Federal Ministry of Research and Technology (BMFTR) project MUNIQC-ATOMS (Grant No.~13N16070). A.S. acknowledges support by the Department of Science and Technology, Govt. of India through the INSPIRE Faculty Award, by the Anusandhan National Research Foundation (ANRF), Govt. of India through the Prime Minister's Early Career Research Grant (PMECRG), and by IIT Madras through the New Faculty Initiation Grant (NFIG).

$^\circ$Now at Rigetti Computing, 775 Heinz Avenue, Berkeley, California 94710, USA.

\appendix
\section{Lasers induced effective spin dynamics}\label{Ap:AppendixA}
\subsection{Effective Spin Hamiltonian }

Following the schematics shown in Fig.~\ref{Fig:Fig1} (b), we derive the  effective spin dynamics of the ground levels after adiabatic elimination of the excited state in the  Raman driven three level system discussed in the main text.  We use the notation $\ket{l}\bra{k}=\hat{\sigma}_{lk}$ interchangeably. We start by formally writing the Raman laser  interaction as
\begin{align}
\hat{H}_{\rm RLI}=&-\omega_{gr}\hat{\sigma}_{gg}-\omega_{er}\hat{\sigma}_{ee}+\frac{\Omega_{g}}{2}\pr{\hat{\sigma}_{rg} e^{i \pr{\vec{k}_{g}.\vec{\hat{\bm{R}}}-\omega_{g}^{L}t}}+ {\rm H.C.}}\nonumber\\
&+\frac{\Omega_{e}}{2}\pr{\hat{\sigma}_{re}e^{i \pr{\vec{k}_{e}.\vec{\hat{\bm{R}}}-\omega_{e}^{L}t}}+ {\rm H.C.}}
\end{align}
We further consider that the first Raman laser  $ ~(\sigma-{\rm polarized})$ with a wave vector $\bm{k}_{g}$ drives the transition $\ket{g}\rightarrow\ket{r}$. This propagates at an angle $\theta_{g}$ with respect to the $X-Y$ plane, yet it has no component along the $Y$-axis. Therefore $\vec{k}_{g}=k_{gx}~\bm{e_{x}}+k_{gz}~\bm{e_{z}}$ with $\bm{e_{i}}$ being the unit vector along the  $\bm{i}_{\rm th}$ direction. The second Raman beam with wave vector $\bm{k}_{g}$ drives the transition $\ket{e}\rightarrow\ket{r} ~(\pi-{\rm polarized})$ and considered to be propagating in a way such that $\vec{k}_{e}=k_{ex}~\bm{e_{x}}+k_{ez}~\bm{e_{z}}$, with $e_{x,y,z}$ are the unit vectors. Moreover, the position vector of the ion  is $\vec{R}= X~\bm{e_{x}}+Y~\bm{e_{y}}+Z~\bm{e_{z}}$. The Hamiltonian then reads
\begin{align}
\hat{H}_{\rm RLP}=&-\omega_{gr}\hat{\sigma}_{gg}-\omega_{er}\hat{\sigma}_{ee}+\frac{\Omega_{g}}{2}\pr{\hat{\sigma}_{rg} e^{i \pr{k_{gx}\hat{X}+k_{gz}\hat{Z}-\omega_{g}^{L}t}}+ {\rm H.C.}}\nonumber\\
&+\frac{\Omega_{e}}{2}\pr{\hat{\sigma}_{re} e^{i \pr{k_{ex}\hat{X}+k_{ez}\hat{Z}-\omega_{e}^{L}t}}+ {\rm H.C.}}.
\end{align}
Here $\Omega_{g}$ and $\Omega_{e}$ are the single-photon Rabi frequencies as induced by two Raman lasers.

In order to derive the effective Hamiltonian and effective jump operator for the two lower spin states in the case of highly detuned Raman lasers, i.e. $\{\Delta_{g},\Delta_{e}\}\gg \{\Omega_{g,e}, \gamma_{e,g}\}$, we use the the effective operator formalism \cite{PhysRevA.85.032111} to adiabatically eliminate the optically excited state $\ket{r}$. Assigning the excited state $\ket{r}$ to have at zero energy, the excited state Hamiltonian is $\hat{H}_{\ket{r}}=0$. Therefore the non-Hermitian $\textstyle{\hat{H}_{\rm NH}\equiv\hat{H}_{\ket{r}}-(i/2)\sum_{\hat{L}_{k}}\hat{L}_{k}^{\dagger} \hat{L}_{k}}$ operator reads
\begin{align}
\hat{H}_{\rm NH}=-i\frac{\gamma}{2}\hat{\sigma}_{rr}~, \qquad\gamma\equiv\gamma_{g}+\gamma_{e}.
\end{align}
The inverse operator corresponding to each of the ground state are given by
\begin{align}
&\pr{\hat{H}^{g}_{\rm NH}}^{-1}=\pr{\omega_{gr}-\omega_{g}^{L}-i\pc{\gamma/2}}^{-1}\hat{\sigma}_{rr}\nonumber\\
&\pr{\hat{H}^{e}_{\rm NH}}^{-1}=\pr{\omega_{er}-\omega_{e}^{L}-i\pc{\gamma/2}}^{-1}\hat{\sigma}_{rr}.
\end{align}
The propagators that connect the lower spin state $\ket{l}({\rm with} ~l=g,e)$, to the optically excited state $\ket{r}$ are given by
\begin{align}
&\hat{V}_{+}^{g}\pc{t}=\frac{\Omega_{g}}{2}e^{-i\omega_{g}^{L}t} e^{ik_{gz}\hat{Z}}\hat{\sigma}_{rg},\quad  \hat{V}_{-}^{g}\pc{t}\equiv\quad \hat{V}_{+}^{g \dagger}\pc{t},\nonumber\\
&\hat{V}_{+}^{e}\pc{t}=\frac{\Omega_{e}}{2}e^{-i\omega_{e}^{L}t}e^{ik_{ez}\hat{Z}}\hat{\sigma}_{re},\quad \hat{V}_{-}^{e}\pc{t}\equiv\quad \hat{V}_{+}^{e\dagger}\pc{t}.
\end{align}
The total excitation and de-excitation operator are respectively given by
\begin{align}
\hat{V}_{+}\pc{t}=\sum_{l=g,e} \hat{V}_{+}^{l}\pc{t},\quad  \hat{V}_{-}\pc{t}=\hat{V}_{+}^{\dagger}\pc{t}.
\end{align}
The effective Hamiltonian is constructed through the expression,
\begin{align}
\hat{H}_{\rm eff}=&-\frac{1}{2}\pr{\hat{V}_{-}\pc{t}\sum_{l=g,e}\pr{\hat{H}^{l}_{\rm NH}}^{-1}\hat{V}_{+}^{l}\pc{t}+ {\rm H.C.}}+\hat{H}_{\rm LM}\\
&\quad {\rm where},\quad   \hat{H}_{\rm LM}=-\omega_{gr}\hat{\sigma}_{gg}-\omega_{er}\hat{\sigma}_{ee}.  
\end{align}
In the following, we also add the ODF Hamiltonian $\hat{H}_{\rm ODF}=g_{\rm O} (\hat{\sigma}_{ee}-\hat{\sigma}_{gg})(\hat{b}+\hat{b}^{\dagger})$ which is diagonal in the spin basis.

\begin{align}
\hat{H}_{\rm eff} =-\frac{1}{2}&\biggl[\biggl(\frac{\Omega_{g}}{2}e^{i\omega_{g}^{L}t}e^{-i(k_{gz}\hat{Z}+k_{gx}\hat{X})}\hat{\sigma}_{gr}+\frac{\Omega_{e}}{2}e^{i\omega_{e}^{L}t}e^{-i(k_{ez}\hat{Z}+k_{ex}\hat{X})}\hat{\sigma}_{er}\biggr)\nonumber \\
&~\biggl(\frac{\Omega_{g}e^{-i\omega_{g}^{L}t}e^{i(k_{gz}\hat{Z}+k_{gx}\hat{X})}}{2\pr{\Delta_{g}-i\pc{\gamma/2}}}\hat{\sigma}_{rg}+\frac{\Omega_{e}e^{-i\omega_{e}^{L}t}e^{i(k_{ez}\hat{Z}+k_{ex}\hat{X})}}{2\pr{\Delta_{e}-i\pc{\gamma/2}}}\hat{\sigma}_{re}\biggr)\nonumber\\
&+{\rm H.C.}\biggr]+\hat{H}_{\rm LM}+ \hat{H}_{\rm ODF},
\end{align}
Here we have defined $\Delta_{l}\equiv \omega_{lr}-\omega_{l}^{L} $ for $l=\{g,e\}$, which are the corresponding detunings of the RL. We combine the transition terms $\hat{\sigma}_{gg},\hat{\sigma}_{ge},\hat{\sigma}_{eg},\hat{\sigma}_{ee} $ with their corresponding part from the hermitian conjugate terms to get the effective Hamiltonian
\begin{align}
\hat{H}_{\rm eff} =&-\frac{1}{2}\biggl[\frac{\Omega_{g}^2}{4}\px{\frac{1}{\Delta_{g}-i\pc{\gamma/2}}+\frac{1}{\Delta_{g}+i\pc{\gamma/2}}}\hat{\sigma}_{gg}\nonumber\\
&+\frac{\Omega_{e}^2}{4}\px{\frac{1}{\Delta_{e}-i\pc{\gamma/2}}+\frac{1}{\Delta_{e}+i\pc{\gamma/2}}}\hat{\sigma}_{ee}+\nonumber\\
&\biggl(\frac{\Omega_{g}\Omega_{e}}{4}\px{\frac{1}{\Delta_{e}-i\pc{\gamma/2}}+\frac{1}{\Delta_{g}+i\pc{\gamma/2}}}e^{i\pc{\Delta k_{x}\hat{X}-\omega^{\rm d} t}} e^{i{\Delta k_z}\hat{Z}} \hat{\sigma}_{ge}\nonumber\\
&+{\rm H.C.}\biggr)\biggr]+\hat{H}_{\rm LM}+ \hat{H}_{\rm ODF},
\end{align}
where $\omega^{\rm d}\equiv \omega_{e}^{L}-\omega_{g}^{L}$, $\Delta k_{x}\equiv \md{k_{ex}}-\md{k_{gx}}$ and $\Delta k_{z}\equiv \md{k_{ez}}-\md{k_{gz}}$. To disregard the in plane momentum kicks, it is required that $\Delta k_{x}=0$. This is set  by choosing angles $\theta_{g}$ and $\theta_{e}$ such that $k_{x}^{\rm d}=0$, while ${\Delta k_z}$ gives suitable wave vector difference in the transverse direction to establish the spin motion coupling (cf. see Fig.~\ref{Fig:Fig1}(b)).This leads to the following effective Hamiltonian:
\begin{align}
\hat{H}_{\rm eff}=&\biggl[-\frac{\Delta_{g}\Omega_{g}^{2}}{4\Delta_{g}^{2}+\gamma^{2}}\hat{\sigma}_{gg}-\frac{\Delta_{e}\Omega_{e}^{2}}{4\Delta_{e}^{2}+\gamma^{2}}\hat{\sigma}_{ee}\nonumber\\
&+\biggl(\frac{\Omega_{g}\Omega_{e}\pc{\Delta_{g}+\Delta_{e}}}{8\pc{\Delta_{g}-i\gamma/2}\pc{\Delta_{e}+i\gamma/2}}e^{i\pc{{\Delta k_z}\hat{Z}-\omega^{\rm d}t}}\hat{\sigma}_{ge}+{\rm H.C.}\biggr)\biggr]\nonumber\\
&-\omega_{gr}\hat{\sigma}_{gg}-\omega_{er}\hat{\sigma}_{ee}+\hat{H}_{\rm ODF}.
\end{align}
We now move into the rotating frame of the RL drives through the transformation $U=e^{-i\pc{\omega_{g}^{L}\hat{\sigma}_{gg}+\omega_{e}^{L}\hat{\sigma}_{ee}}t}$
\begin{align}
\hat{H}_{\rm eff}=&\pc{\Delta_{g}-\frac{\Delta_{g}\Omega_{g}^{2}}{4\Delta_{g}^{2}+\gamma^{2}}}\hat{\sigma}_{gg}+\pc{\Delta_{e}-\frac{\Delta_{e}\Omega_{e}^{2}}{4\Delta_{e}^{2}+\gamma^{2}}}\hat{\sigma}_{ee}\nonumber\\
&+\pc{\frac{\Omega_{g}\Omega_{e}\pc{\Delta_{g}+\Delta_{e}}}{8\pc{\Delta_{g}-i\gamma/2}\pc{\Delta_{e}+i\gamma/2}}e^{i{\Delta k_z}\hat{Z}}\hat{\sigma}_{ge}+{\rm H.C.}}\nonumber\\
&+\hat{H}_{\rm ODF}.
\end{align}
$\hat{H}_{\rm ODF}$ is unaffected under such transformation. This is the effective Hamiltonian for the lower spin states as stated for the Eq.~\ref{eq:SpinhamoEffRaman} of the main text.
\subsection{Effective Jump operators}
The state $\ket{r}$ decays to $\ket{g}$ and $\ket{e}$ with their corresponding jump operator to be $\hat{L}_{g}=\sqrt{\gamma_{g}W(\mu_g)}\hat{\sigma}_{gr}e^{-i\mu_{g} |\vec{k}_{g}|\hat{Z}}$ and $\hat{L}_{e}=\sqrt{\gamma_{e}W(\mu_e)}\hat{\sigma}_{er}e^{-i\mu_{e} |\vec{k}_{e}|\hat{Z}}$ respectively. We keep these jump operator more general and take into account the scattered photons recoil effects. Here $W(\mu_{l})=(3/8)(1+\mu_{l}^{2})$ describes the angular distribution of the dipole radiation pattern and has to be integrated over the variable $\mu_g, \mu_e$ in the interval $[-1,1]$. According to the effective operator formalism, the effective jump operators for the lower spin states manifold is given by the expression
 \begin{align}
\hat{L}^{\rm eff}_{q}=\hat{L}_{q}\sum_{l=g,e}\pr{\hat{H}^{l}_{\rm NH}}^{-1}\hat{V}_{+}^{l}\pc{t}, \quad {\rm where}\quad q=\px{g,e}
\end{align} 
Inserting the required constitute of the above expression and following the step of previous section we get
\begin{align}
\hat{L}^{\rm eff}_{g}&=\frac{\sqrt{\gamma_{g}W(\mu_g)}\Omega_{g}e^{-i\omega_{g}^{L}t}e^{-i\mu_{g} |\vec{k}_{g}|\hat{Z}}e^{ik_{gz}\hat{Z}}}{2\pr{\Delta_{g}-i\pc{\gamma/2}}}\hat{\sigma}_{gg}\nonumber\\
&+\frac{\sqrt{\gamma_{g}W(\mu_g)}\Omega_{e}e^{-i\omega_{e}^{L}t}e^{-i\mu_{g} |\vec{k}_{g}|\hat{Z}}e^{ik_{ez}\hat{Z}}}{2\pr{\Delta_{e}-i\pc{\gamma/2}}}\hat{\sigma}_{ge} 
\end{align} 
\begin{align}
\hat{L}^{\rm eff}_{e}&=\frac{\sqrt{\gamma_{e}W(\mu_e)}\Omega_{g}e^{-i\omega_{g}^{L}t}e^{-i\mu_{e} |\vec{k}_{e}|\hat{Z}}e^{ik_{gz}\hat{Z}}}{2\pr{\Delta_{g}-i\pc{\gamma/2}}}\hat{\sigma}_{eg}\nonumber\\
&+\frac{\sqrt{\gamma_{e}W(\mu_e)}\Omega_{e}e^{-i\omega_{e}^{L}t}e^{-i\mu_{e} |\vec{k}_{e}|\hat{Z}}e^{ik_{ez}\hat{Z}}}{2\pr{\Delta_{e}-i\pc{\gamma/2}}}\hat{\sigma}_{ee} 
\end{align}
As done before for the Hamiltonian part, we move into the rotating frame of the Raman laser frequencies $\omega_{g}^{L}$ and $\omega_{e}^{L}$ using the transformation $U=e^{-i\pc{\omega_{g}^{L}\hat{\sigma}_{gg}+\omega_{e}^{L}\hat{\sigma}_{ee}}t}$ we get,
\begin{align}
\hat{L}^{\rm eff}_{g}&=\frac{\sqrt{\gamma_{g}W(\mu_g)}\Omega_{g}e^{-i\omega_{g}^{L}t}e^{-i\mu_{g} |\vec{k}_{g}|\hat{Z}}e^{ik_{gz}\hat{Z}}}{2\pr{\Delta_{g}-i\pc{\gamma/2}}}\hat{\sigma}_{gg}\nonumber\\
&+\frac{\sqrt{\gamma_{g}W(\mu_g)}e^{-i\omega_{g}^{L}t}\Omega_{e}e^{-i\mu_{g} |\vec{k}_{g}|\hat{Z}}e^{ik_{ez}\hat{Z}}}{2\pr{\Delta_{e}-i\pc{\gamma/2}}}\hat{\sigma}_{ge} 
\end{align} 
\begin{align}
\hat{L}^{\rm eff}_{e}&=\frac{\sqrt{\gamma_{e}W(\mu_e)}\Omega_{g}e^{-i\omega_{e}^{L}t}e^{-i\mu_{e} |\vec{k}_{e}|\hat{Z}}e^{ik_{gz}\hat{Z}}}{2\pr{\Delta_{g}-i\pc{\gamma/2}}}\hat{\sigma}_{eg}\nonumber\\
&+\frac{\sqrt{\gamma_{e}W(\mu_e)}\Omega_{e}e^{-i\omega_{e}^{L}t}e^{-i\mu_{e} |\vec{k}_{e}|\hat{Z}}e^{ik_{ez}\hat{Z}}}{2\pr{\Delta_{e}-i\pc{\gamma/2}}}\hat{\sigma}_{ee} 
\end{align}
The common phase factor $e^{-i\omega_{g}^{L}t}$ and $e^{-i\omega_{e}^{L}t}$ in the first and second jump operators respectively is disregarded since it cancels out while writing the corresponding master equation. This results in: 
\begin{align}\label{Eqsupp:recoiljump1}
\hat{L}^{\rm eff}_{g}&=\frac{\sqrt{\gamma_{g}W(\mu_g)}\Omega_{g}e^{-i\mu_{g} |\vec{k}_{g}|\hat{Z}}e^{ik_{gz}\hat{Z}}}{2\pr{\Delta_{g}-i\pc{\gamma/2}}}\hat{\sigma}_{gg}\nonumber\\
&+\frac{\sqrt{\gamma_{g}W(\mu_g)}\Omega_{e}e^{-i\mu_{g} |\vec{k}_{g}|\hat{Z}}e^{ik_{ez}\hat{Z}}}{2\pr{\Delta_{e}-i\pc{\gamma/2}}}\hat{\sigma}_{ge},
\end{align} 
\begin{align}\label{Eqsupp:recoiljump2}
\hat{L}^{\rm eff}_{e}&=\frac{\sqrt{\gamma_{e}W(\mu_e)}\Omega_{g}e^{-i\mu_{e} |\vec{k}_{e}|\hat{Z}}e^{ik_{gz}\hat{Z}}}{2\pr{\Delta_{g}-i\pc{\gamma/2}}}\hat{\sigma}_{eg}\nonumber\\
&+\frac{\sqrt{\gamma_{e}W(\mu_e)}\Omega_{e}e^{-i\mu_{e} |\vec{k}_{e}|\hat{Z}}e^{ik_{ez}\hat{Z}}}{2\pr{\Delta_{e}-i\pc{\gamma/2}}}\hat{\sigma}_{ee}.
\end{align}

\subsubsection{Effective jump operators without  scattered photon recoil effects}
Each term in these jump operators describes the momentum transfer to the particle when a corresponding spin conserving or spin flip decay process occurs. Momentarily, we keep only the terms which are zeroth order in the paramters $\eta_{ggg}(\mu_g)\equiv(k_{gz}-\mu_g |\vec{k}_{g}|)Z_{zpf}$, $\eta_{egg}(\mu_g)\equiv(k_{ez}-\mu_g |\vec{k}_{g}|)Z_{zpf}$ , $\eta_{gee}(\mu_e)\equiv(k_{gz}-\mu_e |\vec{k}_{e}|)Z_{zpf}$ and $\eta_{eee}(\mu_e)\equiv(k_{ez}-\mu_e |\vec{k}_{e}|)Z_{zpf}$, where $Z_{zpf}$ is the zero point fluctuation of the particle transverse coordinate. For strong confinement, such as the case  in trapped ions settings, higher order terms are suppressed  and the underlying master equation results into the jump operators as given by
\begin{align}
\hat{L}^{\rm eff}_{g}=\frac{\sqrt{\gamma_{g} W(\mu_g)}\Omega_{g}}{2\Delta_{g}-i\gamma}\hat{\sigma}_{gg}+\frac{\sqrt{\gamma_{g}W(\mu_g)}\Omega_{e}}{2\Delta_{e}-i\gamma}\hat{\sigma}_{ge} ,
\end{align} 
\begin{align}
\hat{L}^{\rm eff}_{e}=\frac{\sqrt{\gamma_{e}W(\mu_e)}\Omega_{g}}{2\Delta_{g}-i\gamma}\hat{\sigma}_{eg}+\frac{\sqrt{\gamma_{e}W(\mu_e)}\Omega_{e}}{2\Delta_{e}-i\gamma}\hat{\sigma}_{ee}.
\end{align}
Integrating the function $W(\mu_l),~l=\{g,e\}$ which is common to both terms in each jump operator, we extract the final form of the jump operators from the master equation which are given by
\begin{align}
\hat{L}^{\rm eff}_{g}=\frac{\sqrt{\gamma_{g}}\Omega_{g}}{2\Delta_{g}-i\gamma}\hat{\sigma}_{gg}+\frac{\sqrt{\gamma_{g}}\Omega_{e}}{2\Delta_{e}-i\gamma}\hat{\sigma}_{ge} ,
\end{align} 
\begin{align}
\hat{L}^{\rm eff}_{e}=\frac{\sqrt{\gamma_{e}}\Omega_{g}}{2\Delta_{g}-i\gamma}\hat{\sigma}_{eg}+\frac{\sqrt{\gamma_{e}}\Omega_{e}}{2\Delta_{e}-i\gamma}\hat{\sigma}_{ee}.
\end{align}
These jump operators define the effective spin decay channels and reported in Eq. \eqref{eq:SpinJumpEff} of the main text. These equations can also be rewritten into dressed dark- and bright-state basis as:
\begin{align} \label{eq:DissipatinJumpDressedframe1a} 
\hat{\tilde{L}}_{\rm 1,\alpha}&=\tilde{\alpha}_{++}^{s}\ket{+}\bra{+}+\tilde{\alpha}_{+-}^{s}\ket{+}\bra{-}+\tilde{\alpha}_{-+}^{s}\ket{-}\bra{+}\nonumber\\
&+\tilde{\alpha}_{--}^{s}\ket{-}\bra{-}
\end{align}
\begin{align} \label{eq:DissipatinJumpDressedframe1b} 
\hat{\tilde{L}}_{\rm 1,\beta}&=\tilde{\beta}_{++}^{s}\ket{+}\bra{+}+\tilde{\beta}_{+-}^{s}\ket{+}\bra{-}+\tilde{\beta}_{-+}^{s}\ket{-}\bra{+}\nonumber\\
&+\tilde{\beta}_{--}^{s}\ket{-}\bra{-},
\end{align}
where the coefficient are determined by

\begin{align}
&\tilde{\alpha}_{++}^{s}=\frac{i\sqrt{W(\mu_g)\gamma_{g}}\Omega_{g}\px{(\gamma+2 i \Delta_{\rm g})\Omega_{e}^2+(\gamma+2 i \Delta_{\rm e})\Omega_{g}^2}}{(\gamma+2 i \Delta_{\rm e})(\gamma+2 i \Delta_{\rm g})(\Omega_{g}^2+\Omega_{e}^2)}\nonumber\\
&\tilde{\alpha}_{+-}^{s}=\frac{2\sqrt{W(\mu_g)\gamma_{g}}\Omega_{e}\Omega_{g}^2\px{\Delta_{g}-\Delta_{e}}}{(\gamma+2 i \Delta_{\rm e})(\gamma+2 i \Delta_{\rm g})(\Omega_{g}^2+\Omega_{e}^2)}\nonumber\\
&\tilde{\alpha}_{-+}^{s}=\frac{i\sqrt{W(\mu_g)\gamma_{g}}\Omega_{e}\px{(\gamma+2 i \Delta_{\rm g})\Omega_{e}^2+(\gamma+2 i \Delta_{\rm e})\Omega_{g}^2}}{(\gamma+2 i \Delta_{\rm e})(\gamma+2 i \Delta_{\rm g})(\Omega_{g}^2+\Omega_{e}^2)}\nonumber\\
&\tilde{\alpha}_{--}^{s}=\frac{2\sqrt{W(\mu_g)\gamma_{g}}\Omega_{g}\Omega_{e}^2\px{\Delta_{g}-\Delta_{e}}}{(\gamma+2 i \Delta_{\rm e})(\gamma+2 i \Delta_{\rm g})(\Omega_{g}^2+\Omega_{e}^2)}
\end{align}

\begin{align}
&\tilde{\beta}_{++}^{s}=\frac{i\sqrt{W(\mu_e)\gamma_{e}}\Omega_{e}\px{(\gamma+2 i \Delta_{\rm g})\Omega_{e}^2+(\gamma+2 i \Delta_{\rm e})\Omega_{g}^2}}{(\gamma+2 i \Delta_{\rm e})(\gamma+2 i \Delta_{\rm g})(\Omega_{g}^2+\Omega_{e}^2)}\nonumber\\
&\tilde{\beta}_{+-}^{s}=\frac{2\sqrt{W(\mu_e)\gamma_{e}}\Omega_{g}\Omega_{e}^2\px{\Delta_{g}-\Delta_{e}}}{(\gamma+2 i \Delta_{\rm e})(\gamma+2 i \Delta_{\rm g})(\Omega_{g}^2+\Omega_{e}^2)}\nonumber\\
&\tilde{\beta}_{-+}^{s}=\frac{-i\sqrt{W(\mu_e)\gamma_{e}}\Omega_{g}\px{(\gamma+2 i \Delta_{\rm g})\Omega_{e}^2+(\gamma+2 i \Delta_{\rm e})\Omega_{g}^2}}{(\gamma+2 i \Delta_{\rm e})(\gamma+2 i \Delta_{\rm g})(\Omega_{g}^2+\Omega_{e}^2)}\nonumber\\
&\tilde{\beta}_{--}^{s}=\frac{2\sqrt{W(\mu_e)\gamma_{e}}\Omega_{e}\Omega_{g}^2\px{\Delta_{g}-\Delta_{e}}}{(\gamma+2 i \Delta_{\rm e})(\gamma+2 i \Delta_{\rm g})(\Omega_{g}^2+\Omega_{e}^2)}
\end{align}
Under the two photon Raman resonance condition $\Delta_{g}=\Delta_{e}$, the coefficient $\alpha_{+-}^s=\alpha_{--}^s=\beta_{+-}^s=\beta_{--}^s=0$ for zeroth order in the parameters $\eta_{ggg}(\mu_g)$, $\eta_{egg}(\mu_g)$ , $\eta_{gee}(\mu_e)$ and $\eta_{eee}(\mu_e)$. Thus, the dark state does not decay and the steady state of the spin is a perfect dark state. In Eqs.~\eqref{eq:DissipatinJumpDressedframe12}-\eqref{eq:DissipatinJumpDressedframe13} of the main text, we use these spin jump operators (Though the coefficient are relabeled therein for convenience ) written in the dressed dark and bright state basis along with two-photon Raman resonance condition to perform the cooling analysis in the weak spin-boson coupling limit.

\section{Analysis for the dissipative dynamics of the motion  }\label{Ap:AppendixB}
\subsection{Adiabatic elimination of spin degrees of freedom}
We derive the master equation of the  external mechanical motion (consider as system of interest) based on the fact that spin (consider as the bath) relaxes to its steady state quite fast as compared to the time scale of the joint spin-motion system evolution. We follow the so-called projection operator method \cite{Breuer2007}, originally developed by the Nakajima and Zwanzig \cite{Nakajima1958, PhysRev.124.983}. Here the basic idea is to derive the master equation for the motion of some slow collective variables. Let us keep the discussion quite general momentarily, with the  density operator of the system is $\hat{\rho}_{S}$ and of the total system is $\hat{\rho}$. In order to derive the reduce density operator of the system, it is convenient to define the projector super operator $\hat{\mathcal{P}}$ such that the following properties hold:
\begin{align}
\hat{\rho}\rightarrow\hat{\mathcal{P}}\hat{\rho}=\Tr_{B}\pr{\hat{\rho}}\otimes\hat{\rho}_{B}\equiv\hat{\rho}_{S}\otimes\hat{\rho}_{B},
\end{align}
where $\hat{\rho}_{B}$ is some fixed steady state of the bath. This super-operator projects on the relevant part of the total density matrix $\hat{\rho}$. The quantity $\hat{\mathcal{P}}\hat{\rho}$ gives the complete information required to construct the reduced density matrix $\hat{\rho}_{S}$ of the system. Accordingly, a complementary super-operator $\hat{\mathcal{Q}}$ is introduced such that:
\begin{align}
\mathcal{Q\hat{\rho}}=\hat{\rho}-\hat{\mathcal{P}}\hat{\rho},
\end{align}
projects on the irrelevant part of the density matrix. The super-operators $\hat{\mathcal{P}}$ and $\hat{\mathcal{Q}}$ are maps in the state space of the combined system defined by the total Hilbert space $\mathcal{H}=\mathcal{H}_{S}\otimes\mathcal{H}_{B}$. As such, the following properties are extracted from these representation: $\hat{\mathcal{P}}+\hat{\mathcal{Q}}=\mathbf{I},~
\hat{\mathcal{P}}^{2}=\hat{\mathcal{P}},~
\hat{\mathcal{Q}}^{2}=\hat{\mathcal{Q}},~
\hat{\mathcal{P}}\hat{\mathcal{Q}}=\hat{\mathcal{Q}}\hat{\mathcal{P}}=0$. The density matrix $\hat{\rho}_{B}$ used above is an operator in $\mathcal{H}_{B}$. 

In order to derive an effective master equation for the system of interest, i.e. the mechanical mode, we adiabatically eliminate the spin degrees of freedom. Let us consider the separation of the time scales such that $\{g_{\rm R},g_{\rm O}\}< \gamma_{ b}$, meaning that all the spin relaxations time scales are shorter than the time scale for the mode dynamics under the effects of spin-phonon coupled interaction. The spin state is hardly affected by the interaction on short time scale and therefore it can be adiabatically eliminated. We also point out the different energy scale $\gamma_{m} \bar{n}_{\rm th}\ll\gamma_{ b}$ such that these dissipation channel has a separation of time scales. We define the small parameters $\epsilon_{1}=\{g_{\rm R}/\gamma_{ b}, g_{\rm O}/\gamma_{ b}\}$ and $\epsilon_{2}=\gamma_{m}n_{\rm th}/\gamma_{ b}$ such that the spin-motion coupled dynamics, and also the intrinsic dissipation to the resonator, are considered as perturbations. Using Eqs.~\eqref{eq:SpinJumpEff}-~\eqref{eq:EffIntarctionhamTerm} of the main text, we therefore  write the master equation of the whole setup in the Liouvillian form as given by
\begin{align}
\dot{\hat{\rho}}=\mathcal{L}_{0}^{m}\pc{\hat{\rho}}+\mathcal{L}_{0}^{s}\pc{\hat{\rho}}+\mathcal{L}_{1}\pc{\hat{\rho}}+\mathcal{L}_{2}\pc{\hat{\rho}}.
\end{align}
Here the zeorth-order Liouvillian terms describes the unitary dynamics of the mechanical resonator and the total (unitary and dissipative) spin dynamics, which are respectively given by
\begin{align}
\mathcal{L}_{0}^{m}\pc{\hat{\rho}}\equiv-i\pr{\hat{H}_{m},\hat{\rho}},
\end{align}
\begin{align}
\mathcal{L}_{0}^{s}\pc{\hat{\rho}}&=-i\pr{\hat{H}_{s},\hat{\rho}}+\sum\nolimits_{\hat{L}_{s}}\mathcal{D}_{\hat{L}_{s}}\pr{\hat{\rho}}\nonumber\\
&\equiv-i\pr{\hat{H}_{ s},\hat{\rho}}+\mathcal{L}_{s,d}\pr{\hat{\rho}}.
\end{align}
where $\hat{L}_{s}=\{\hat{L}_{g}^{{\rm eff}},\hat{L}_{e}^{{\rm eff}}\}$. The first order terms in $\epsilon_{1}$ are described  as
\begin{align}
\mathcal{L}_{1}\pc{\hat{\rho}}=-i\pr{\hat{H}_{int},\hat{\rho}},
\end{align}
and the terms first order in $\epsilon_{2}$ is given by the coupling of the resonator with its environment such that,
\begin{align}
\mathcal{L}_{2}\pc{\hat{\rho}}\equiv \sum\nolimits_{\hat{L}_{p}}\mathcal{D}_{\hat{L}_{p}}\pr{\hat{\rho}}, \quad{\rm where}\quad \hat{L}_{p}=\{\hat{L}^{m}_{\hat{b}},\hat{L}^{m}_{\hat{b}^{\dagger}}\}.
\end{align}
In this form, the fast spin degrees are considered as effective bath to system of interest which is the mechanical oscillator. Under the assumption that $\{g_{\rm O},g_{\rm R}\}\ll\gamma_{ b}$, the spin goes to its steady state $\hat{\rho}_{ss}$ on a time scale which is much faster than the time scale of the interaction between mechanical motion and the spin. It is therefore possible to to project the master equation in the form $\hat{\rho}(t)=\hat{\rho}_{ss}\otimes\Tr_{s}\pr{\hat{\rho}(t)}\equiv\hat{\rho}_{ss}\otimes\hat{\rho}_{m}(t)$. In order to remove the action of coupling term on the projected state, we first make use of the displacement operator $\hat{D}$ such that it performs a coherent shift on the mechanical mode only, such that:
\begin{align}
\hat{D}\hat{b} \hat {D}^{\dagger}=\hat{b}+\beta. 
\end{align}
In such shifted frame, the new density operator is given by $\bar{\hat{\rho}}=\hat{D} \hat{\rho} \hat{D}^{\dagger}$. Thus the master equation in the shifted frame takes the form
\begin{align}
\dot{\bar{\hat{\rho}}}=\mathcal{L}_{0}^{s}\bar{\hat{\rho}}+\hat{D}\pc{\mathcal{L}_{0}^{m}\hat{\rho}+\mathcal{L}_{1}\hat{\rho}+\mathcal{L}_{2}\hat{\rho} }\hat{D}^{\dagger} \quad \text{with}
\end{align}
\begin{align}
\hat{D}\mathcal{L}_{0}^{m}\hat{\rho} \hat{D}^{\dagger}=\mathcal{L}_{0}^{m}\bar{\hat{\rho}}-i\omega_{m}\pr{\beta \hat{b}^{\dagger}+ \beta^{*}\hat{b},\bar{\hat{\rho}}},\\
\hat{D}\mathcal{L}_{1}\hat{\rho} \hat{D}^{\dagger}=\mathcal{L}_{1}\bar{\hat{\rho}}-i(\beta+\beta^{*})\pr{\hat{F},\bar{\hat{\rho}}},\\
\hat{D}\mathcal{L}_{2}\hat{\rho} \hat{D}^{\dagger}=\mathcal{L}_{2}\bar{\hat{\rho}}+\gamma_{m}\pr{\beta^{*}\hat{b}-\beta \hat{b}^{\dagger},\bar{\hat{\rho}}}.
\end{align}
Here the shift acquired by the operator $\hat{D} \mathcal{L}_{1} \hat{D}^{\dagger}$ acts only on the spin degree and it is included in the spins dynamics by defining the new Liouvillian 
\begin{align}
\tilde{\mathcal{L}}_{0}^{s}\bar{\hat{\rho}}\equiv\mathcal{L}_{0}^{s}\bar{\hat{\rho}}-i(\beta+\beta^{*})\pr{\hat{F},\bar{\hat{\rho}}}
\end{align}
Under such transformations, the master equation in the shifted frame reads
\begin{align}
\dot{\bar{\hat{\rho}}}&=\pc{\mathcal{L}_{0}^{m}+\tilde{\mathcal{L}}_{0}^{s}+\mathcal{L}_{1}+\mathcal{L}_{2}}\bar{\hat{\rho}}\nonumber\\
&+\pr{\pc{\gamma_{m}-i\omega_{m}}\beta^{*}\hat{b}-\pc{\gamma_{m}+i\omega_{m}}\beta \hat{b}^{\dagger},\bar{\hat{\rho}}}.
\label{eq:ae1}
\end{align}
We now project the master equation onto the factorized form such that 
\begin{align}
\hat{\mathcal{P}}\bar{\hat{\rho}}=\hat{\rho}_{ss}\otimes\bar{\hat{\rho}}_{m}, \quad \hat{\mathcal{P}}\tilde{\mathcal{L}}_{0}^{s}\bar{\hat{\rho}}=0, \quad \text {and}\quad  \hat{\mathcal{Q}}\bar{\hat{\rho}}=\bar{\hat{\rho}}-\hat{\mathcal{P}}\bar{\hat{\rho}}.
\end{align}
The amount of the displacement $\beta$ is determined in a way such that the projection of $\hat{\mathcal{P}}$ on the last term of the Eq. \eqref{eq:ae1} cancels the coupling term $\hat{\mathcal{P}}\mathcal{L}_{1}\hat{\mathcal{P}}\bar{\hat{\rho}}=-i \hat{\rho}_{ss}\otimes\langle\hat{F}\rangle\pr{ \hat{b}+\hat{b}^{\dagger},\hat{\rho}_{m}}$. Therefore we have
\begin{align}
\beta=-\frac{\langle\hat{F}\rangle}{\omega_{m}-i\gamma_{m}}, \quad \beta^{*}=-\frac{\langle\hat{F}\rangle}{\omega_{m}+i\gamma_{m}}.
\label{eq:alphaeq}
\end{align}
By realizing that $\pr{\hat{\mathcal{P}},\mathcal{L}_{0}^{m}}=0$ and $\pr{\hat{\mathcal{P}},\mathcal{L}_{2}}=0$, the master equation in the $\hat{\mathcal{P}}$ space takes the form
\begin{align}
\hat{\mathcal{P}}\dot{\bar{\hat{\rho}}}=\hat{\mathcal{P}}\mathcal{L}_{0}^{m}\hat{\mathcal{P}}\bar{\hat{\rho}}+\hat{\mathcal{P}}\mathcal{L}_{1}\hat{\mathcal{Q}}\bar{\hat{\rho}}+\hat{\mathcal{P}}\mathcal{L}_{2}\hat{\mathcal{P}}\bar{\hat{\rho}}.
\label{eq:ae2}
\end{align}
In order to eliminate the terms containing the orthogonal space operators in the last equation, we write the master equation for the orthogonal $\hat{\mathcal{Q}}-$ space. By employing the properties of the projectors, we have $\hat{\mathcal{Q}}\mathcal{L}_{0}^{m}\hat{\mathcal{P}}\bar{\hat{\rho}}=0$, $\hat{\mathcal{Q}}\mathcal{L}_{0}^{s}\hat{\mathcal{P}}\bar{\hat{\rho}}=0$ and $\hat{\mathcal{Q}}\mathcal{L}_{2}\hat{\mathcal{P}}\bar{\hat{\rho}}=0$. Therefore, the equation for the orthogonal space reads
\begin{align}
\hat{\mathcal{Q}}\dot{\bar{\hat{\rho}}}&=\hat{\mathcal{Q}}\pc{\mathcal{L}_{0}^{m}+\tilde{\mathcal{L}}_{0}^{s}+\mathcal{L}_{2}}\hat{\mathcal{Q}}\bar{\hat{\rho}}+\hat{\mathcal{Q}}\mathcal{L}_{1}\hat{\mathcal{P}}\bar{\hat{\rho}}\nonumber\\
&-i\hat{\mathcal{Q}}\pr{(\hat{F}-\langle\hat{F}\rangle)\pc{\hat{b}+\hat{b}^{\dagger}},\hat{\mathcal{Q}}\bar{\hat{\rho}}}.
\end{align}
We only keep those terms which effectively lead to an equation that is second order in the perturbation parameters $\epsilon_{1,2}$. In effect, the formal solution to last equation (with higher order terms disregarded) reads
\begin{align}
\hat{\mathcal{Q}}\bar{\hat{\rho}}\pc{t}&={\rm e}^{\hat{\mathcal{Q}}\pc{\mathcal{L}_{0}^{m}+\tilde{\mathcal{L}}_{0}^{s}}(t-t_{1})}\hat{\mathcal{Q}}\bar{\hat{\rho}}\pc{t_{1}}\nonumber\\
&+\int_{t_{1}}^{t}d\tau{\rm e}^{\hat{\mathcal{Q}}\pc{\mathcal{L}_{0}^{m}+\tilde{\mathcal{L}}_{0}^{s}}(t-\tau)}\hat{\mathcal{Q}}\mathcal{L}_{1}\hat{\mathcal{P}}\bar{\hat{\rho}}\pc{\tau}
\label{eq:ae3}
\end{align}
From Eq. \eqref{eq:ae2}, we take the solution $\hat{\mathcal{P}}\bar{\hat{\rho}}\pc{t}$ that is on the order $\mathcal{O}\pc{\epsilon^{0}_{1,2}}$. This is once again with the desire that the final equation is of the second order in coupling parameters. Thus the obtained solution reduces to
\begin{align}
\hat{\mathcal{P}}\bar{\hat{\rho}}(t)={\rm e}^{{P}\mathcal{L}_{0}^{m}(t-t_{0})}\hat{\mathcal{P}}\hat{\rho}(t_{0}).
\end{align}
By inserting this solution in Eq.~\eqref{eq:ae3}, we get:
\begin{align}
\hat{\mathcal{Q}}\bar{\hat{\rho}}\pc{t}=\int_{0}^{\infty}d\tau{\rm e}^{\hat{\mathcal{Q}}\pc{\mathcal{L}_{0}^{m}+\tilde{\mathcal{L}}_{0}^{s}}\tau}\hat{\mathcal{Q}}\mathcal{L}_{1}{\rm e}^{-\hat{\mathcal{P}}\mathcal{L}_{0}^{m}\tau}\hat{\mathcal{P}}\bar{\hat{\rho}}(t).
\label{eq:ae4}
\end{align}
Here we have rewritten this reduced form of the Eq.~\eqref{eq:ae3} based on the following grounds: Since we are interested in time scale larger than $\gamma_{b}^{-1}$ for which $\tilde{\mathcal{L}}_{0}^{s}$ decays, the first term in Eq.~\eqref{eq:ae3} will not contribute for such dynamics. Moreover the integrand of the equation decays on the time scale $\gamma_{b}^{-1}$. This allow us to make the Markov approximation, where we extend the lower limit of integration to $t_{1}\rightarrow-\infty$ and further replace $ \hat{\rho}(t_{0})$ by $  \hat{\rho}(t)$ in Eq.~\eqref{eq:ae3}. 

In order to get the dynamical equation for the system subspace only, we eliminate the orthogonal subspace. To do so, we insert the obtained solution from Eq. \eqref{eq:ae4} into Eq. \eqref{eq:ae2}. This results into effective equation which is contained by $\hat{\mathcal{P}}\bar{\hat{\rho}}$ terms only, representing the dynamics of the system subspace only. Afterwards, we project the master equation onto the factorized form and trace over the spin degree of freedom. The resultant master equation for the mechanical mode is given by
\begin{align}
\dot{\bar{\hat{\rho}}}_{m}&=\pc{\mathcal{L}_{0}^{m}+\mathcal{L}_{2}}\bar{\hat{\rho}}_{m}\nonumber\\
&+\Tr_{s}\pr{\int_{0}^{\infty}d\tau\mathcal{L}_{1}\hat{\mathcal{Q}}{\rm e}^{\pc{\mathcal{L}_{0}^{m}+\tilde{\mathcal{L}}_{0}^{s}} \tau } \mathcal{L}_{1} \hat{\rho}_{ss}\otimes{\rm e}^{-\mathcal{L}_{0}^{m}\tau} \bar{\hat{\rho}}}.
\end{align}
We further evaluate the second term of the master equation. We define the fluctuations of the operator $\widehat{\delta F}=\hat{F}-\av{{F}}_{ss}$ around its steady state mean field value $\av{{F}}_{ss}$- and further use the definition: $\Tr_{s}\pr{\widehat{\delta F} {\rm e}^{\tilde{\mathcal{L}}_{0}^{s}\tau}\widehat{\delta F} \hat{\rho}_{ss}}=\av{\widehat{\delta F}\pc{\tau} \widehat{\delta F}}_{ss}$, $\Tr_{s}\pr{\widehat{\delta F} {\rm e}^{\tilde{\mathcal{L}}_{0}^{s}\tau}\hat{\rho}_{ss}\widehat{\delta F} }=\av{ \widehat{\delta F}\widehat{\delta F}\pc{\tau}}_{ss}$ and ${\rm e}^{\mathcal{L}_{0}^{m}\tau}b=b(-\tau)$. Introducing these relations, we obtain an equation where the spin degrees are completely traced out and the mechanical mode master equation reads
\begin{align}
&\dot{\bar{\hat{\rho}}}_{m}=\pc{\mathcal{L}_{0}^{m}+\mathcal{L}_{2}}\bar{\hat{\rho}}_{m}\nonumber\\
&-\int_{0}^{\infty}d\tau \av{\widehat{\delta F}(\tau) \widehat{\delta F}}_{ss}\pr{ \hat{b}+\hat{b}^{\dagger},\pr{\hat{b}\pc{-\tau}+\hat{b}^{\dagger}\pc{-\tau},\hat{\rho}_{m} }}\nonumber\\
&-\int_{0}^{\infty}d\tau \av{\pr{\widehat{\delta F}(\tau), \widehat{\delta F}}}_{ss}\pr{ \hat{b}+\hat{b}^{\dagger}, \hat{\rho}_{m}\pc{\hat{b}\pc{-\tau}+\hat{b}^{\dagger}\pc{-\tau}}}
\end{align}
We further define the complex spectral function $\mathcal{S}\pc{\omega}$ which is the Laplace transform of the correlation function $\av{\widehat{\delta F}(\tau) \widehat{\delta{F}}}_{ss}$ of the spin steady state and written as
 \begin{align}
\mathcal{S}(\omega)=\mathcal{S}_{\rm R}\pc{\omega}+i \mathcal{S}_{\rm I}\pc{\omega}\equiv \int_{0}^{\infty}d\tau e^{i\omega \tau}\av{\widehat{\delta F}(\tau) \widehat{\delta{F}}}_{ss}
 \end{align}
The complex nature of the spectral function arises due to the complex correlation function. In principal, the spin steady state correlation should be calculated by the Liouvillian $\tilde{\mathcal{L}}_{0}^{s}$, which is given by 
\begin{align}
\tilde{\mathcal{L}}_{0}^{s}\bar{\hat{\rho}}=\mathcal{L}_{0}^{s}\bar{\hat{\rho}}+\frac{2 i \omega_{m}\langle\hat{F}\rangle}{\omega_{m}^2+\gamma_{m}^2}\pr{\hat{F},\bar{\hat{\rho}}},
\end{align}
where we have invoked the value of $\beta$ from Eq. \eqref{eq:alphaeq}. However the shifts determined by the second term gives a small correction, i.e. $\mathcal{O}\pc{\{g_{\rm O}/\omega_{m},g_{\rm R}/\omega_{m}\}\ll 1}$, therefore we calculate the steady state correlation with respect to $\mathcal{L}_{0}^{s}$. We thus have the effective master equation for the mechanical motion as given by
\begin{align}
\dot{\bar{\hat{\rho}}}_{m}=&\pc{\mathcal{L}_{0}^{m}+\mathcal{L}_{2}}\bar{\hat{\rho}}_{m}-\biggl\{\mathcal{S}\pc{\omega_{m}}\pr{\hat{b}+\hat{b}^{\dagger},\pr{\hat{b},\bar{\hat{\rho}}_{m} }} \nonumber\\
&+\mathcal{S}\pc{-\omega_{m}}\pr{\hat{b}+\hat{b}^{\dagger},\pr{\hat{b}^{\dagger},\bar{\hat{\rho}}_{m} } }\biggr\}\nonumber\\
&-\px{\mathcal{S}\pc{\omega_{m}}-\mathcal{S}^{*}\pc{-\omega_{m}} }\pr{\hat{b}+\hat{b}^{\dagger},\bar{\hat{\rho}}_{m}\hat{b}}\nonumber\\
&-\px{\mathcal{S}\pc{-\omega_{m}}-\mathcal{S}^{*}\pc{\omega_{m}} }\pr{\hat{b}+\hat{b}^{\dagger},\bar{\hat{\rho}}_{m}\hat{b}^{\dagger}}
\end{align}
We further apply the rotating wave approximation and neglect fast oscillating terms. This lead to equation of motion of the mechanical resonator
\begin{align}
\dot{\bar{\hat{\rho}}}_{m}=&-i\omega_{m}\pr{\hat{b}^{\dagger}\hat{b},\bar{\hat{\rho}}_{m}}\nonumber\\
&-2i\px{\mathcal{S}_{\rm I}\pc{\omega_{m}}\pr{\hat{b}^{\dagger}\hat{b},\bar{\hat{\rho}}_{m}}+\mathcal{S}_{\rm I}\pc{-\omega_{m}}\pr{\hat{b} \hat{b}^{\dagger},\bar{\hat{\rho}}_{m}} }\nonumber\\
&+\px{2\mathcal{S}_{\rm R}\pc{\omega_{m}}+\gamma_{m}\pc{n_{\rm th}+1}}\mathcal{D}_{\hat{b}}\pr{\bar{\hat{\rho}}_{m}}\nonumber\\
&+\px{ 2\mathcal{S}_{\rm R}\pc{-\omega_{m}}+\gamma_{m}n_{\rm th}}\mathcal{D}_{\hat{b}^{\dagger}}\pr{\bar{\hat{\rho}}_{m}}.
\end{align}
In this equation, the terms containing $\mathcal{S}_{\rm I}$ cause a small frequency shift of the oscillator and they don't contribute in the population change of the oscillator. We thus ignore such terms and further remove the subscript for real part, that is, we define $S\pc{\omega_{m}} \equiv 2\mathcal{S}_{\rm R}\pc{\omega_{m}}$.  Moreover, in the un-displaced frame, the terms proportional to $\sim \beta^{*} b, \beta b^{\dagger}$ will appear which also don't contribute to the cooling processes, therefore these are neglected. The final form of the equation is then given by
\begin{align}
\dot{\hat{\rho}}_{m}=&-i\omega_{m}\pr{b^{\dagger}b,\hat{\rho}_{m}}+\pr{S\pc{\omega_{m}}+\gamma_{m}\pc{n_{\rm th}+1}}\mathcal{D}_{\hat{b}}\pr{\hat{\rho}_{m}}\nonumber\\
&+\pr{ S\pc{-\omega_{m}}+\gamma_{m}n_{\rm th}}\mathcal{D}_{\hat{b}^{\dagger}}\pr{\hat{\rho}_{m}}.
\label{eq:eafinal}
\end{align}
This is the effective master equation of the mode and corresponding jump operator are written in the Eq.~\eqref{eq:Spin-Absorption} of the main text, where we write it in terms of the effective mechanical mode jump operators. 
\subsection{Contribution by different dissipative channels in cooling dynamics}\label{Ap:AppendixB2}
\begin{figure}[t!]
\includegraphics[width=0.38\textwidth]{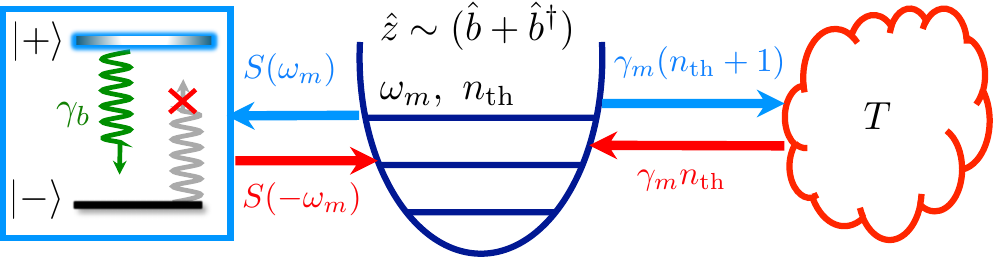}
\caption{Cooling and heating due to different dissipative channels.}
\label{Fig:ApFig3not}
\end{figure}
We now analyze the cooling and heating rates from various dissipating channels and rewrite the master Eq.~\eqref{eq:eafinal} as
\begin{align}
\dot{\hat{\rho}}_{m}=&-i\omega_{m}\pr{\hat{b}^{\dagger}\hat{b},\hat{\rho}_{m}}
+\mathcal{D}_{\hat{L}_{\hat{b}}^{m}}\pr{\hat{\rho}_{m}}+
\mathcal{D}_{\hat{L}_{\hat{b}^{\dagger}}^{m}}\pr{\hat{\rho}_{m}}\nonumber\\
&+\mathcal{D}_{\hat{L}_{\hat{b}}^{m,{ s}}}\pr{\hat{\rho}_{m}}+\mathcal{D}_{\hat{L}_{\hat{b}^{\dagger}}^{m,s}}\pr{\hat{\rho}_{m}},
\label{eq:mevhnicaldensiysepchannels}
\end{align}
where, $\textstyle{\hat{L}_{\hat{b}}^{m,s}=
\sqrt{\gamma_s\pc{n_{s}+1}}~\hat{b},\quad \hat{L}_{\hat{b}^{\dagger}}^{m,s}=\sqrt{\gamma_s n_{s}}~\hat{b}^{\dagger}}$. In these expressions, the damping channels with rate $\gamma_{m}$, as written in the first line of Eq.~\eqref{eq:mevhnicaldensiysepchannels}, are due to the intrinsic-uncontrolled heat bath. Moreover from the second line, we obtain the RL-controlled damping rate $\gamma_s=S\pc{\omega_{m}}-S\pc{-\omega_{m}}$ and the occupation number $n_s=S\pc{-\omega_{m}}/[S\pc{\omega_{m}}-S\pc{-\omega_{m}}]$, which are both purely due to internal spin states. All these channels are schematically shown in Fig. \ref{Fig:ApFig3not}. The net cooling rate and steady-state final occupation number are respectively rewritten in terms of these quantities as
\begin{align} 
\Gamma_{c}=	\gamma_s+\gamma_{m}, \quad  n_{f}=\frac{\gamma_{s}n_s+\gamma_{m}n_{\rm th}}{\gamma_s+\gamma_{m}}.
\label{eq:netopticalcooling&steadystatephonon}
\end{align}
Furthermore, for resonance condition $\omega_{m}=\omega_{s}$, we get the expression for the optimal cooling rate $\Gamma_{c,{\rm opt}}=	\gamma_{s,{\rm opt}}+\gamma_{m}$, and the final minimum occupation number takes the expression
\begin{align} \label{eq:Shiftedupbound}
n_{f, {\rm min}}=\frac{\gamma_{s,{\rm opt}}n_{s,{\rm  min}}+\gamma_{m}n_{\rm th}}{\gamma_{s,{\rm opt}}+\gamma_{m}}.
\end{align}
To explicitly calculate the optimal spin-induced expressions $\gamma_{s, {\rm opt}}$ and $n_{\rm s, min}$, we set $\gamma_{m}=0$. We thus get the expression for spin-induced optimal cooling rate as given by 
\begin{align}
\gamma_{s,{\rm opt}}= \frac{64 \Delta_{\rm R}^2(\gamma^{2}+4\Delta_{\rm R}^2) \left(4 g_{\rm O}^2 \Omega_{g}^2\Omega_{e}^2 +g_{\rm R}^2\left(\Omega_{g}^2+\Omega_{e}^2\right)^2\right)}{\gamma (\gamma^{2}+16\Delta_{\rm R}^2)(\Omega_{g}^2+\Omega_{e}^2)^3},
\label{eq:SpininducecoolingRateopt}
\end{align}
This cooling rate is further maximized for equal Rabi frequencies $\Omega_{g}=\Omega_{e}$.  In addition, a straightforward calculation results into quantum back action limit 
\begin{align}
n_{\rm BA}\equiv n_{f, {\rm min}}|_{\gamma_{m}\rightarrow 0} =n_{\rm s, min}=\pc{\frac{\gamma}{4 \Delta_{\rm R}}}^2,
\label{eq:Spininduceminoccupation}
\end{align}
which is the fundamental limit on the minimum occupation. Such a limit arises due to having non-zero bright state linewidth as activated by the RL induced decay channels such as those stated in Eq.~\eqref{eq:DissipatinJumpDressedframe12} of the main text. 

From the Lorentzian dark-bright spin absorption spectrum plotted in Fig.~\ref{Fig:Fig3} (a), it can be inferred that the ratio of the half of the bright state width $\gamma_{b}/2$ to the first available side-band transition, which occurs at $2\omega_{s}$, is given by $\gamma_{b}/4\omega_{s}=(1/4Q_{s})$. Here $Q_{s}$ is the quality factor of the Lorentzian absorption spectrum of the dark steady state. Therefore the QBL on minimum possible occupation number can be written as
\begin{align}
n_{\rm BA}=(1/4Q_{s})^2.
\label{eq:SpininduceminoccupationQF}
\end{align}
This expression establishes a relation between the QBL and the quality factor of the proposed Lorentzian spin absorption spectrum of Eq.~\eqref{eq:Lorentzian} of the main text. 

\subsection{Spin absorption spectrum }\label{spectrumcal}
In this appendix, we calculate the spin absorption spectrum as determined by the the corresponding force fluctuation operators. First we take the definition from the above subsections which has shown to take the form
\begin{align}
S\pc{\omega}=2~{\rm Re}\pr{\int_{0}^{\infty}d\tau e^{i\omega \tau}\langle\widehat{\delta F}\pc{\tau} \widehat{\delta F}\rangle_{ss}}.
\end{align} 
Th operator $\hat{F}=\vec{c}~\vec{\hat{\sigma}}$ is parametrized with coupling row vector $\textstyle{\vec {c}=[0,g_{\rm R},g_{\rm O}]}$, wherein the vector of the Pauli operators is $\textstyle{\vec{\hat{\sigma}}=[\hat{\sigma}_{x},\hat{\sigma}_{y},\hat{\sigma}_{z}]^T}$. Therefore the spectrum is rewritten as
\begin{align}
S\pc{\omega}=2~{\rm Re}\pr{\vec{c}\int_{0}^{\infty}d\tau e^{i\omega \tau}\langle\vec{\widehat{\delta\sigma}}\pc{\tau} \widehat{\delta F}\rangle_{ss}}.
\end{align}
The integral term can be identified as the Laplace transform (with Laplace domain variable $s=i\omega$ ) of the fluctuation correlation function, therefore the spectrum is given by
\begin{align}
\textstyle{S\pc{\omega}=2~{\rm Re}[\lim_{s\rightarrow i\omega}\vec{c}~\langle\vec{\widehat{\delta\sigma}}\pc{s} \widehat{\delta F}\rangle_{ss}]}.
\label{eq:SpecLaplacedomaindef}
\end{align}
Using Eq.~\eqref{eq:SpinJumpEff} and Eq.~\eqref{eq:Hamiltonian4} of the main text, the optical Bloch equations as dictated by the spin only part (spin Hamiltonian and its dissipation) takes the form
\begin{align}
\partial_{t}\langle\vec{\hat{\sigma}}\rangle=\vec{\vec{A}}\langle\vec{\hat{\sigma}}\rangle+\vec{\Gamma}.
\end{align}
Here the dynamics is determined by the kernel matrix $\vec{\vec{A}}$ and the dissipation vector $\vec{\Gamma}$, which are respectively given by
\begin{align}
&\vec{\vec{A}}=\left(
\begin{array}{ccc}
 -\frac{\gamma_{b}}{2} &-\omega_{\rm LS} & 0 \\
\omega_{\rm LS} &  -\frac{\gamma_{b}}{2} & -\Omega_{\rm R} \\
 \frac{\Omega_{e} \Omega_{g} (\gamma_{e}-\gamma_{g})}{\gamma^2+4 \Delta_{\rm R}^2} & \Omega_{\rm R} & -\frac{\gamma_{e} \Omega_{g}^2+\gamma_{g}\Omega_{e}^2}{\gamma^2+4 \Delta_{\rm R}^2} \\
\end{array}
\right),\nonumber\\
&\vec{\Gamma}=\left(
\begin{array}{ccc}
\frac{\Omega_{e}\Omega_{g} \gamma}{\gamma^2+4 \Delta_{\rm R}^2}\\
0\\
\frac{\gamma_{g} \Omega_{e}^2-\gamma_{e} \Omega_{g}^2}{\gamma^2+4 \Delta_{\rm R}^2}
\end{array}
\right).
\end{align}
The optical Bloch equations further allow to calculate the steady state solution as given by $\langle\vec{\hat{\sigma}}\rangle_{ss}=-\vec{\vec{A}}^{-1}\vec{\Gamma}$. Hence the equation for motion for the spin fluctuation (around the steady state mean field value) is given by
\begin{align}
\partial_{t}\langle\vec{\widehat{\delta\sigma}}(t)\rangle=\vec{\vec{A}}\langle\vec{\widehat{\delta\sigma}}(t)\rangle.
\end{align}
In the Laplace domain, this equation takes the form
\begin{align}
\langle\vec{\widehat{\delta\sigma}}(s)\rangle=(s \mathbb{I} -\vec{\vec{A}})^{-1}\langle\vec{\widehat{\delta\sigma}}\rangle.
\end{align}
According to quantum regression theorem, the dynamics for the correlation $\langle\vec{\widehat{\delta\sigma}}(s) \widehat{\delta F}\rangle$ follows the same equation as above. Therefore from Eq.~\eqref{eq:SpecLaplacedomaindef}, we get the spectrum taking the form
\begin{align}
S\pc{\omega}=2~{\rm Re}\pr{\lim_{s\rightarrow -i\omega}\vec{c}~(s  \mathbb{I}-\vec{\vec{A}})^{-1}\langle\vec{\widehat{\delta\sigma}} \widehat{\delta F}\rangle_{ss}}.
\end{align}
This equation is rewritten into the steady state expectation values of the Pauli operators as given by
\begin{align}
S\pc{\omega}=2~{\rm Re}\pr{\lim_{s\rightarrow -i\omega}\vec{c}~(s  \mathbb{I}-\vec{\vec{A}})^{-1}\pr{\langle\vec{\hat{\sigma}} \hat{F}\rangle_{ss}-\langle\vec{\hat{\sigma}}\rangle_{ss}\langle\hat{F}\rangle_{ss}}}.
\end{align}
The final analytical form of the spectrum is a Lorentzian function which is given by Eq. \eqref{eq:Lorentzian} of the main text.
\section{Analysis for the full three level (3L) dynamics}\label{Ap:AppendixC}
In order to benchmark the analytically formulated two-level dark state cooling theory as stated in the main text, we numerically simulate the master equation by taking into account the internal full three-levels $\{\ket{g},\ket{e},\ket{r}\}$ of the particle and its external motion. For this purpose, we take the $3L$ Hamiltonian in Eq.~ \eqref{eq:hamoltoninaSystemLabeled} of the main text and move into interaction picture with respect to RL frequencies $\omega_{g}^{L},~\omega_{e}^{L}$ through the transformation $U=e^{-i\pc{\omega_{g}^{L}\hat{\sigma}_{gg}+\omega_{e}^{L}\hat{\sigma}_{ee}}t}$. In effect we get the $3L$ Hamiltonian which reads
\begin{align}
\hat{H}_{3L}=&\Delta_{g}\hat{\sigma}_{gg}+\Delta_{e}\hat{\sigma}_{ee}+\frac{\Omega_{g}}{2}\pr{\hat{\sigma}_{rg} e^{i k_{gz}\hat{Z}}+ {\rm H.C.}}\nonumber\\
&+\frac{\Omega_{e}}{2}\pr{\hat{\sigma}_{re} e^{i k_{ez}\hat{Z}}+ {\rm H.C.}}+\hat{H}_{\rm ODF}+\omega_{m}\hat{b}^{\dagger}\hat{b}, 
\end{align}
\noindent where $\Delta_{g}=\omega_{g}^{L}-\omega_{gr}$ and $\Delta_{e}=\omega_{e}^{L}-\omega_{er}$ are detunings of the RL. Note that $\hat{H}_{\rm ODF}$ remained unchanged under this transformation. Similarly, the optical decay jump operator would accumulate an unimportant phase factor which is canceled while writing the master equation. Therefore jump operator for the internal three-level system would be the same as stated in Eq.~\eqref{eq:SystemBasicJumpoperators} of the main text. Performing a Lamb-Dicke expansion for the transverse $z$ coordinate and canonically quantizing the motion, we get the Hamiltonian
\begin{align}
\hat{H}_{\rm 3L}=&\hat{H}_{\rm ODF}+ \sum_{l=g,e}[\Delta_{l}\hat{\sigma}_{ll}+\frac{\Omega_{l}}{2}\pr{\hat{\sigma}_{rl}+ {\rm H.C.}}]\nonumber\\
&+\frac{\Omega_{g}\eta_{gz}}{2}\pr{i\hat{\sigma}_{rg} (\hat{b}+\hat{b}^{\dagger})+ {\rm H.C.}}\nonumber\\
&+\frac{\Omega_{e}\eta_{ez}}{2}\pr{i\hat{\sigma}_{re} (\hat{b}+\hat{b}^{\dagger})+ {\rm H.C.}}+\omega_{m}\hat{b}^{\dagger}\hat{b}.
\label{Eq:threelevelLambDickehamoltonian}
\end{align}
\begin{figure}[t]
\includegraphics[width=0.24\textwidth]{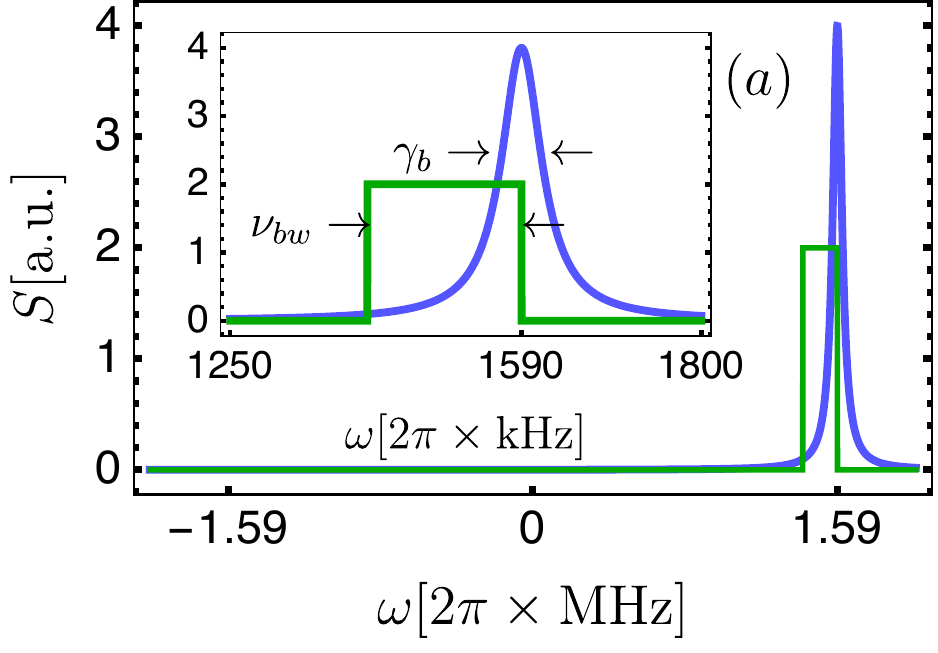}
\includegraphics[width=0.23\textwidth]{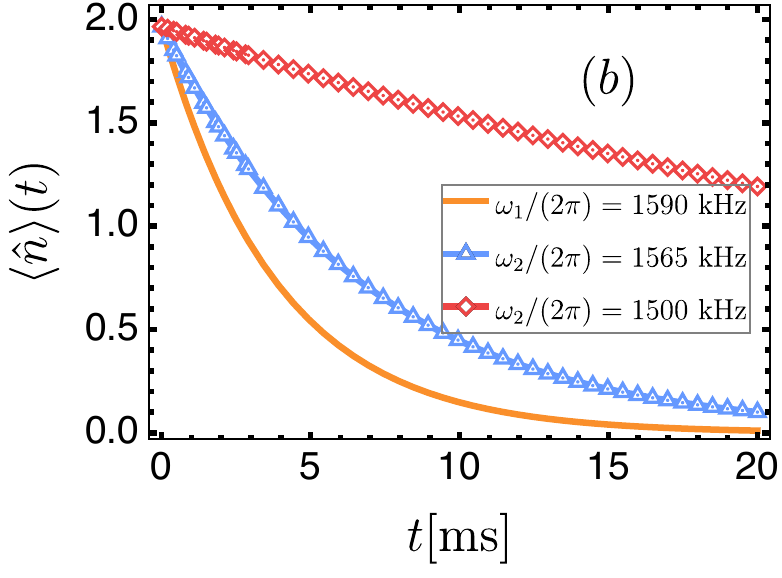}
\caption{(a): Spin absorption spectrum with a band of transverse modes frequencies $\nu_{bw}$ are put in resonance within the bright state linewidth $\gamma_{b}$ of the spectrum. This would allow to cool the many-modes simultaneously. (b): Multi-mode cooling: When $|\omega_{2}-\omega_{1}|<\gamma_{b}$, the pair of motional modes ${\omega_1,\omega_2}$ is cooled simultaneously, as evidenced by the dynamics of their mean phonon occupations and illustrated by the orange--blue curves. In contrast, simultaneous cooling becomes inefficient when $|\omega_{2}-\omega_{1}|>\gamma_{b}$, as shown by the orange--red curves. The parameters are identical to those of the weak-coupling regime [cf. Fig.~\ref{Fig:fig3abcd}], with $\gamma_{b}/(2\pi)=56~{\rm kHz}$.}
\label{Fig:ApFig2}
\end{figure}
Here $\eta_{gz}=k_{gz}\sqrt{\hbar/2m\omega_{m}}$ and $\eta_{ez}=k_{ez}\sqrt{\hbar/2m\omega_{m}}$ are the Lamb Dicke parameters associated with the corresponding RL and considered to be small $\{\eta_{gz},\eta_{ez}\}\ll 1$. The full dynamics of ion's internal three levels $({\rm 3L})$ and external degrees of freedom is then described by the Hamiltonian in Eq.~\eqref{Eq:threelevelLambDickehamoltonian} of the main text along with ion's spontaneous decay jump operators  $\hat{L}_{d}=\{\hat{L}_{g}, \hat{L}_{e}\}$ in Eq.~\eqref{eq:SystemBasicJumpoperators} of the main text (with ignoring the external mechanical effects in the incoherent decay processes, i.e. valid for deep traps), and coupling of motion to the thermal bath described by jump operators $\hat{L}_{p}=\{\hat{L}^{m}_{\hat{b}},\hat{L}^{m}_{\hat{b}^{\dagger}}\} $ in Eq.~\eqref{eq:JumThermalPhonon1} of the main text such that the three-level $(3L)$ master equation is
\begin{align}
\partial_{t}\hat{\rho}_{e,g,r}=&-i[\hat{H}_{\rm 3L},\hat{\rho}_{e,g,r}]+\sum\nolimits_{\hat{L}_{d}}\mathcal{D}_{\hat{L}_{d}}\pr{\hat{\rho}_{e,g,r}}\nonumber\\
&+\sum\nolimits_{\hat{L}_{p}}\mathcal{D}_{\hat{L}_{p}}\pr{\hat{\rho}_{e,g,r}}.
\label{eq:MasterEq3L}
\end{align}
We simulate such a $3L$ master equation numerically by exact diagonalization (ED) method. The steady-state dynamics is determined by the zero eigenvalue component of the Liouvillian super operators of the dynamics. For the steady-state regime, we obtain numerically the mean phonon number as a function of system parameters, phonon Fock state distribution function. Note that the optical excited state scattering rate, that is shown in Fig.~\ref{Fig:Fig3} (b) of the main text, is obtained via the optical Bloch dynamics of the three-level density matrix $\hat{\rho}_{e,g,r}$ in the limit where we ignore the mechanical effects on the three-level system. Therefore, it is obtained by setting $(\{\eta_{gz},\eta_{ez},g_{0}\}\rightarrow 0, \gamma_{m}=0)$ in Eq.~\eqref{eq:MasterEq3L}.
\section{Multi-mode simultaneous cooling in the weak coupling regime} \label{Ap:AppendixD}
We now focus on the simultaneous cooling of many modes in the weak coupling regime.
As an example, in Fig \ref{Fig:ApFig2} (a), we show a band of frequency $\nu_{bw}/(2\pi)= 180 ~{\rm kHz}$ for $\gamma_{b}/(2\pi)=56~{\rm kHz}$ while the highest frequency mode is made resonant with the dark-bright spin transition frequency. This figure shows that the lower modes falling within the bandwidth $\gamma_{b}/2$ are sufficiently close to resonance to cool simultaneously with the highest frequency mode. Tuning the bright state linewidth then allows one, in principle, to cover the entire band and cool all modes. It should be also noted that both $\omega_{s}$ and $\gamma_{b}$ are not independent, therefore one should find optimal tuning where most of the bandwidth find resonance condition while falling withing the bright state line width. Additionally, due to different couplings for the different modes in Eq.~\eqref{eq:SpinhamoEffMMa} would result in different cooling rates for each mode.  As an illustrative example, we quantify these effects using the dressed-frame multi-ion, multi-mode formalism described by Eqs. \eqref{eq:SpinhamoEffMMa} and \eqref{eq:DissipatinJumpDressedframe12}, where the latter is generalized to the multi-ion case. For a system of two ions supporting two motional modes $\{\omega_1,\omega_2\}$, we perform the unitary transformation $\hat{U}=e^{-i\
(\omega_{1}\sum_\nu\hat{b}_\nu^{\dagger}\hat{b}_\nu+(\omega_s/2)\sum_j\hat{\tilde \sigma}_{z}^{(j)})t}$ which transforms the system into the interaction picture, yielding the Hamiltonian
\begin{align}
\hat{H}_{\rm int}
&= (\omega_{2}-\omega_{1}) \hat{b}_{2}^{\dagger}\hat{b}_{2}-i g_{\rm R}
\sum_{j=1}^{2}
\sum_{\nu=1}^{2}
\mathcal{K}_{\nu,j}
\sqrt{\frac{\omega_{1}}{\omega_{\nu}}}
\left(
\hat{b}_{\nu}\hat{\tilde{\sigma}}_{+}^{(j)}
-
\hat{b}_{\nu}^{\dagger}\hat{\tilde{\sigma}}_{-}^{(j)}
\right).
\end{align}
Here we have applied the rotating-wave approximation (RWA), retaining only the resonant contributions while neglecting all rapidly oscillating terms in the Hamiltonian. This is achieved by imposing the resonance condition $\omega_s=\omega_1$. In the interaction picture, the spin jump operators likewise decompose into four independent jump operators, given by:
\begin{align}
\hat{L}_{1}^{(j)} &= \tilde{\alpha}_{gg}\,\ket{+}\!\bra{+}^{(j)}, \\
\hat{L}_{2} ^{(j)}&= \tilde{\alpha}_{ee}\,\ket{+}\!\bra{+}^{(j)}, \\
\hat{L}_{3}^{(j)} &= \tilde{\alpha}_{ge}\,\ket{-}\!\bra{+}^{(j)}, \\
\hat{L}_{4}^{(j)} &= \tilde{\alpha}_{eg}\,\ket{-}\!\bra{+}^{(j)}.
\end{align}
We numerically simulate these equations, and the resulting weak-coupling multimode cooling dynamics are shown in Fig. \ref{Fig:ApFig2} (b). The two motional modes, ${\omega_1,\omega_2}$, are cooled simultaneously when the condition $|\omega_2-\omega_1|<\gamma_b/2$ is satisfied. In contrast, for $|\omega_2-\omega_1|>\gamma_b/2$, simultaneous cooling of both modes becomes inefficient (see figure caption).
\section{Collective dynamics in the strong coupling regime}
\label{app:collective}
In this section, we evaluate Eq.~\eqref{eq:gamma_N_def} as well as the corrections $\gamma_{1,p}$ and $\gamma_{2,p}$.

We first compute the spontaneous emission operators.
The effective jump operators are Eq.~\eqref{Eqsupp:recoiljump1} and \eqref{Eqsupp:recoiljump2}, with the replacements $\hat Z \rightarrow \hat Z_j$ and $\hat \sigma_{kl} \rightarrow \hat \sigma_{kl}^{(j)}$.
As stated in the main text, $\hat Z_j = \sum_\nu \mathcal K_{\nu, j} \sqrt{\hbar / 2m_I\omega_\nu} \qty(\hat b_\nu + \hat b_\nu^\dagger)$.
To allow comparisons of parameters to other regimes, we use the single-particle Lamb-Dicke parameter $\eta_z$.

We can then compute the matrix elements of the effective Lindblad operators. Since each Lindblad operator acts only on the state (both internal and motion) of one ion, we suppress the other internal states from our notation.
\begin{widetext}
\begin{align}
    \hat L_{l,\mu_l; \rm eff}^{(j)}
    &=
    \sum_{\alpha,\beta = \pm}
    \bra{\alpha} \hat L_{l,\mu_l; \rm eff}^{(j)} \ket{\beta}
    \ket{\alpha}\bra{\beta}
    \, , \label{first}\\
    \bra{+} \hat L_{l,\mu_l; \rm eff}^{(j)} \ket{+}
    &=
    \frac{1}{\Omega_g^2 + \Omega_e^2} \frac{\sqrt{\gamma_l}}{2\Delta_R - \mi \gamma} \Omega_l \sum_{l'=g,e} \Omega_{l'}^2 
    \exp[\qty(\eta_{l',z} - \mu_l\eta_{l}) \sum_\nu \mathcal K_{\nu,j} \qty(\hat b_\nu + \hat b_\nu^\dagger)] 
    \, , \\
    \bra{-} \hat L_{l,\mu_l; \rm eff}^{(j)} \ket{+}
    &=
    \frac{1}{\Omega_g^2 + \Omega_e^2} \frac{\sqrt{\gamma_l}}{2\Delta_R - \mi \gamma} (-1)^l \Omega_{\bar l} \sum_{l' = g,e} \Omega_{l'}^2 
    \exp[\qty(\eta_{l',z} - \mu_l\eta_{l}) \sum_\nu \mathcal K_{\nu,j} \qty(\hat b_\nu + \hat b_\nu^\dagger)] 
    \, , \\
    \bra{+} \hat L_{l,\mu_l; \rm eff}^{(j)} \ket{-}
    &=
    \frac{1}{\Omega_g^2 + \Omega_e^2} \frac{\sqrt{\gamma_l}}{2\Delta_R - \mi \gamma} \Omega_l \sum_{l'=g,e} \Omega_{l'} \Omega_{\bar l'} (-1)^{l'} 
    \exp[\qty(\eta_{l',z} - \mu_l\eta_{l}) \sum_\nu \mathcal K_{\nu,j} \qty(\hat b_\nu + \hat b_\nu^\dagger)] 
    \, , 
    \end{align}
    \begin{align}
    \bra{-} \hat L_{l,\mu_l; \rm eff}^{(j)} \ket{-}
    &=
    \frac{1}{\Omega_g^2 + \Omega_e^2} \frac{\sqrt{\gamma_l}}{2\Delta_R - \mi \gamma} \Omega_{\bar l} \sum_{l'} \Omega_{l'} \Omega_{\bar l'} (-1)^{l'}
    \exp[\qty(\eta_{l',z} - \mu_l\eta_{l}) \sum_\nu \mathcal K_{\nu,j} \qty(\hat b_\nu + \hat b_\nu^\dagger)] 
    \, ,
\end{align}
\end{widetext}
with $\bar e = g$ and $\bar g = e$ and $(-1)^g \equiv -1$, $(-1)^e \equiv 1$.

We now evaluate the jump operator to zeroth order in $\eta_{l=e,g,z}$.
Note that $\sum_l \Omega_l \Omega_{\bar l} (-1)^l = \Omega_g \Omega_e \sum_l (-1)^l = 0 $, so that $\bra{\pm, n} \hat L_{l,\mu_l;\rm eff} \ket{-,n} = 0$ as expected for our definition of dark state.
Then,
\begin{align}
    \hat L_{l,\mu_l;\rm eff}^{(j)}
    &=
    \frac{\sqrt{\gamma_l}}{2\Delta_R - \mi \gamma} [\Omega_l \hat{\tilde \sigma}_{++}^{(j)} + (-1)^l \Omega_{\bar l} \hat{\tilde \sigma}_{-+}^{(j)}
    ]
    +
    \mathcal O(\eta_{z})
    \, .
    \label{eq:Leff_collective_eta0}
\end{align}
 This is independent of $\mu_l$ at $\mathcal O(\eta_z^0)$,  and we drop $\mu_l$ from the notation.

By replacing  Eq.~\eqref{eq:collective_decay_definition} into Eq.~\eqref{eq:gamma_N_def}, we find
\begin{align}
    \gamma_{N, \rm sc}^{(n_\mathrm{ex})}
    &=
    \frac{1}{\mathcal N_{N,n_\mathrm{ex}}} \sum_{\zeta,\zeta',j,l} \kappa_{-,j,l}^{(n_\mathrm{ex},\zeta',\zeta)} \, ,
    \label{eq:definition_gamma_N_sc_nex}
\end{align}
with
\begin{align}
    \kappa_{-,j,l}^{(n_\mathrm{ex},\zeta',\zeta)}
    &=
    \abs{\bra{n_\mathrm{ex}-1,\zeta'} \hat L_{l;\rm eff}^{(j)} \ket{n_\mathrm{ex},\zeta}}^2
    \, .
    \label{eq:definition_kappa_minus_manybody}
\end{align}
We plug this term  in Eq.~\eqref{eq:Leff_collective_eta0} to find
\begin{align}
    \sum_{\zeta',j,l}
    \kappa_{-,j,l}^{n_\mathrm{ex},\zeta',\zeta}
    &=
    \frac{\gamma_g \Omega_e^2 + \gamma_e \Omega_g^2}{4\Delta_{\rm R}^2 + \gamma^2}
    \bra{n_\mathrm{ex}, \zeta} \hat N_+ \ket{n_\mathrm{ex}, \zeta}
    \\
    &=
    2\gamma_{N=1,\rm sc}
    \bra{n_\mathrm{ex}, \zeta} \hat N_+ \ket{n_\mathrm{ex}, \zeta}
    \, ,
\end{align}
which is Eq.~\eqref{eq:gamma_nsc} in the main text.

We will need two specific values below:
\begin{align}
    \gamma_{N,\rm sc}^{(1)}
    &=
    \frac{2N}{N+1}\gamma_{N=1,\rm sc}
    \\
    \gamma_{N,\rm sc}^{(2)}
    &=
    \frac{4N^2}{N^2+N+2}\gamma_{N=1,\rm sc}
    \, .
\end{align}

To estimate the ground state fraction, we need to include processes up to $\mathcal O(\eta_{l,z}^2)$ for the state with $n_{\rm{ex}}=0$.
Since the Hamiltonian in Eq.~\eqref{eq:SpinhamoEffMMa} also off-resonantly drives blue sidebands, and $g_\mathrm{R}/\omega_m \sim \eta_z$, we need to compute the eigenstates as dressed by the blue sideband.
We find that the dark state becomes
\begin{align}
    &\ket{n_{\rm ex}=0,\zeta=1}^{(1)}
    =
    \nonumber \\
    &\quad
    \qty(1 - \sum_\nu \frac{\eta_z^2 \Omega_e^2 \Omega_g^2 \omega_m^2}{ 2\qty(\Omega_e^2 + \Omega_g^2)^2\qty(\omega_m + \omega_\nu)^2}) \ket{n_{\rm ex}=0,\zeta=1}
    \nonumber
    \\
    &
    -
    \sum_{\nu,j}
    \frac{\mi \eta_z\Omega_e\Omega_g\omega_m}{(\Omega_e^2 + \Omega_g^2)(\omega_m+\omega_\nu)}
    \hat b_\nu^\dagger \mathcal K_{\nu,j}\hat{\tilde \sigma}_{+-}^{(j)} \ket{n_{\rm ex}=0, \zeta=1}
    \, .
    \label{eq:n_ex_1}
\end{align}

We then find up to order $\eta_z$, we can write
\begin{align}
    \hat L_{l,\mu_l;\rm eff}^{(j)} \ket{n_{\rm ex}=0,\zeta=1}^{(1)}
    &=
    \frac{\mi\eta_z\Omega_g\Omega_e\sqrt{\gamma_l}}{2\Delta_{\rm R} (\Omega_g^2 + \Omega_e^2)}
    \sum_\nu
    \frac{\omega_\nu}{\omega_m + \omega_\nu}
    \mathcal K_{\nu,j}\hat b_\nu^\dagger
    \nonumber
    \\
    &\qquad
    \qty(
    \Omega_{\bar l}
    +
    \Omega_l
    \hat{\tilde \sigma}^{(j)}_{+-}
    )
    \ket{n_{\rm ex}=0,\zeta=1}
    \, .
    \label{eq:app_jump_nex0}
\end{align}
This includes both the $\mathcal O(\eta_z)$ contributions of $\hat L_{\rm eff}$ and of $\ket{n_{\rm ex}=0,\zeta=1}^{(1)}$.

We now want to count all $\ket +$ and $\hat b_{1}^\dagger$ (center of mass) excitation.
We thus count excitations created by $\hat b_\nu^\dagger \hat{\tilde \sigma}_{+-}^{(j)}$ as $n_{\rm ex} = 2$ for $\nu = 1$, and as $n_{\rm ex} = 1$ otherwise.
We find
\begin{align}
    \gamma_{1,p}
    &=
    \sum_{\zeta,j,l}
    \abs{
    \bra{n_{\rm ex}=1,\zeta} \hat L_{l,\mu_l;\rm eff}^{(j)} \ket{n_{\rm ex}=0,\zeta=1}^{(1)}
    }^2
    \nonumber
    \\
    &=
    \qty[\frac{\Omega_g\Omega_e\eta_z}{4\Delta_{\rm R}\qty(\Omega_g^2 + \Omega_e^2)}]^2
    \nonumber
    \\
    &\quad
    \qty[
    \qty(\sum_{\nu\neq 0} \frac{4\omega_\nu^2 }{(\omega_m + \omega_\nu)^2})
    \qty(\Omega_g^2\gamma_g + \Omega_e^2\gamma_e)
    +
    \Omega_g^2\gamma_e + \Omega_e^2\gamma_g
    ]
    \label{eq:definition_gamma_1_p}
    \\
    &\overset{\Omega_g=\Omega_e}{=}
    \frac{\eta_z^2}{8}\gamma_{N=1,\rm sc} \qty[\sum_{\nu} \frac{4\omega_\nu^2}{\qty(\omega_m + \omega_\nu)^2}]
    \, ,
    \\
    \gamma_{2,p}
    &=
    \qty[\frac{\Omega_g\Omega_e\eta_z}{4\Delta_{\rm R}\qty(\Omega_g^2 + \Omega_e^2)}]^2
    \qty(\Omega_g^2\gamma_g + \Omega_e^2\gamma_e)
    \\
    &\overset{\Omega_g=\Omega_e}{=}
    \frac{\eta_z^2}{8}\gamma_{N=1,\rm sc}
    \, .
\end{align}
We used $\sum_j \mathcal K_{\nu,j}^2 = 1$.

\subsection{Dressed density matrix}
We now provide some details on the vectorization of the operators.
Since the density matrix is diagonal at order $\eta_z^0$ in the basis $\ket{n_{\rm ex},\zeta}$, in first-order perturbation theory in $\eta_z$, we can write the density matrix as
\begin{align}
    \hat \rho^{(1)}(t) =&
    \sum_{n_{\rm ex},\zeta} \frac{P_{\rm ex}^{(1)}(n_{\rm ex}, t)}{\mathcal N_{N,n_{\rm ex}}} \ket{n_{\rm ex}, \zeta}^{(1)}\bra{n_{\rm ex}, \zeta}^{(1)} 
    \, ,
\end{align}
with $P_{\rm ex}^{(1)}(n_{\rm ex},t) = P_{\rm ex}^{(0)}(n_{\rm ex},t) + \mathcal O(\eta_z^2)$ and $\ket{n_{\rm ex},\zeta}^{(1)} = \ket{n_{\rm ex},\zeta} + \mathcal O(\eta_z)$.

Using $P_{\rm ex}^{(1)}(n_{\rm ex},t\rightarrow\infty) = \mathcal O(\eta_z^2)$ for $n_{\rm ex} > 0$, this simplifies to
\begin{align}
    \hat \rho^{(1)}(t\rightarrow \infty) =&
    P_{\rm ex}(0,t\rightarrow \infty) \ket{n_{\rm ex}=0, \zeta=1}^{(1)}\bra{n_{\rm ex}=0, \zeta=1}^{(1)} 
    \nonumber
    \\
    &+ \sum_{n_{\rm ex}>0,\zeta} \frac{P_{\rm ex}(n_{\rm ex},t\rightarrow\infty)}{\mathcal N_{N,n_{\rm ex}}} \ket{n_{\rm ex},\zeta}\bra{n_{\rm ex},\zeta} 
    \nonumber
    \\
    &+ \mathcal O(\eta_z^3)
    \, ,
    \label{eq:rho1_steadystate_appendix}
\end{align}
where $P_{\rm ex}(n_{\rm ex},t)$ only accounts for the ground-state corrections, as discussed in the main text.

\subsection{Vectorization of operators}
\label{ap_sec:vectorization_collective}

To define $\vec O$, we now enforce the condition $\vec O \cdot \vec p_{\rm ex}(t\rightarrow\infty) = \Tr[\hat O \hat \rho^{(1)}(t\rightarrow\infty)]$ to find
\begin{align}
    &\sum_j \qty(\vec O)_j P_{\rm ex}(j,t\rightarrow\infty)
    =
    \nonumber \\
    &\quad P_{\rm ex}(0,t\rightarrow\infty) 
    \bra{n_{\rm ex}=0,\zeta=1}^{(1)} \hat O \ket{n_{\rm ex}=0,\zeta=1}^{(1)}
    \nonumber \\
    &\qquad
    + \sum_{n_{\rm ex},\zeta} \frac{P(n_{\rm ex},t\rightarrow\infty)}{\mathcal N_{N,n_{\rm ex}}} \bra{n_{\rm ex},\zeta} \hat O \ket{n_{\rm ex},\zeta}
    \, .
\end{align}
By matching both sides of the equation, we find
\begin{align}
    \qty(\vec O)_0 &= \bra{n_{\rm ex}=0,\zeta=1}^{(1)} \hat O \ket{n_{\rm ex}=0,\zeta=1}^{(1)}
    \, ,
    \\
    \qty(\vec O)_j &= \frac{\sum_\zeta \bra{n_{\rm ex}=j,\zeta} \hat O \ket{n_{\rm ex}=j,\zeta}}{\mathcal N_{N,j}} \qquad
    \text{for $j>0$.}
\end{align}

For $j>2$, the observables are then given by
\begin{align}
    \qty(\vec n_{\rm ex})_j &= j 
    \, , \\
    \qty(\vec n)_j &= j - \overline{N_+}(j,N) 
    \, , \\
    \qty(\vec p_{\rm gs})_j &= 0
    \, .
\end{align}

\subsection{Closed form analytic steady state}
We evaluate the steady state expectation values as $\langle \hat O \rangle = \vec O \cdot \vec p_{\rm ex}(t\rightarrow\infty)$, and find
\begin{align}
    \langle \hat p_\mathrm{gs}\rangle (t \rightarrow \infty) &=
    \frac{
    \gamma_{N,\mathrm{sc}}^{(1)}\gamma_{N,\mathrm{sc}}^{(2)}
    }{
    \gamma_{N,\mathrm{sc}}^{(1)}\gamma_{N,\mathrm{sc}}^{(2)} +
    \gamma_{1,\mathrm{p}}\gamma_{N,\mathrm{sc}}^{(2)} +
    \gamma_{2,\mathrm{p}}\gamma_{N,\mathrm{sc}}^{(2)} +
    \gamma_{2,\mathrm{p}}\gamma_{N,\mathrm{sc}}^{(1)}
    }
    \nonumber \\
    &\quad
    -
    \sum_\nu \frac{\eta_z^2\Omega_e^2\Omega_g^2\omega_m^2}{(\Omega_e^2 + \Omega_g^2)^2(\omega_m + \omega_\nu)^2}
    \, ,
    \label{eq:analytic_groundstate}
    \\
    \langle \hat n_\mathrm{ex} \rangle (t \rightarrow \infty) &=
    \frac{
    \gamma_{1,\mathrm{p}}\gamma_{N,\mathrm{sc}}^{(2)} +
    \gamma_{2,\mathrm{p}}\gamma_{N,\mathrm{sc}}^{(2)} +
    2\gamma_{2,\mathrm{p}}\gamma_{N,\mathrm{sc}}^{(1)}
    }{
    \gamma_{N,\mathrm{sc}}^{(1)}\gamma_{N,\mathrm{sc}}^{(2)} +
    \gamma_{1,\mathrm{p}}\gamma_{N,\mathrm{sc}}^{(2)} +
    \gamma_{2,\mathrm{p}}\gamma_{N,\mathrm{sc}}^{(2)} +
    \gamma_{2,\mathrm{p}}\gamma_{N,\mathrm{sc}}^{(1)}
    }
    \nonumber \\
    &\quad
    +
    \sum_{\nu\neq0} \frac{\eta_z^2\Omega_e^2\Omega_g^2\omega_m^2}{(\Omega_e^2 + \Omega_g^2)^2(\omega_m + \omega_\nu)^2}
    +
    \frac{\eta_z^2\Omega_e^2\Omega_g^2}{2(\Omega_e^2 + \Omega_g^2)^2}
    \, ,
    \label{eq:analytic_excitations}
    \\
    \langle \hat n \rangle (t \rightarrow \infty) &=
    \frac{
    \frac{\gamma_{1,\mathrm{p}}\gamma_{N,\mathrm{sc}}^{(2)}}{N+1} +
    \frac{\gamma_{2,\mathrm{p}}\gamma_{N,\mathrm{sc}}^{(2)}}{N+1} +
    \frac{\gamma_{2,\mathrm{p}}\gamma_{N,\mathrm{sc}}^{(1)} (2N+4)}{N^2 + N + 2}
    }{
    \gamma_{N,\mathrm{sc}}^{(1)}\gamma_{N,\mathrm{sc}}^{(2)} +
    \gamma_{1,\mathrm{p}}\gamma_{N,\mathrm{sc}}^{(2)} +
    \gamma_{2,\mathrm{p}}\gamma_{N,\mathrm{sc}}^{(2)} +
    \gamma_{2,\mathrm{p}}\gamma_{N,\mathrm{sc}}^{(1)}
    }
    \nonumber \\
    &\quad
    +
    \frac{\eta_z^2\Omega_e^2\Omega_g^2}{4(\Omega_e^2 + \Omega_g^2)^2}
    \, .
    \label{eq:analytic_phonons}
\end{align}

\subsection{Photon Scattering Rate}
Following Sec.~\ref{ap_sec:vectorization_collective}, we can decompose the steady-state photon scattering rate as
\begin{align}
    \gamma_{\rm scat} =& \sum_{j,l} \int d\mu_l \Tr[\hat \rho^{(1)}(t \rightarrow \infty) \hat L_{l,\mu_l}^{(j)\dagger} \hat L_{l,\mu_l}^{(j)}]
    \\
    =& \vec R \cdot \vec p_{\rm ex}(t \to \infty)
    \label{eq:definition_gamma_scat}
\end{align}
by vectorizing $\hat R = \sum_{j,l} \int d\mu_l \hat L_{l,\mu_l}^{(j)\dagger} \hat L_{l,\mu_l}^{(j)}$.
We find from Eqs.~\eqref{eq:Leff_collective_eta0} and \eqref{eq:app_jump_nex0}
\begin{align}
    \qty(\vec R)_0 =& \sum_{j,l} \int d\mu_l W(\mu_l) \times
    \nonumber \\
    &\quad ^{(1)}\hspace{-3pt}\bra{n_{\rm ex}=0,\zeta=1} \qty(\hat L_{l,\mu_l;\rm eff}^{(j)})^\dagger \hat L_{l,\mu_l;\rm eff}^{(j)} \ket{n_{\rm ex}=0,\zeta=1}^{(1)}
    \nonumber \\
    =& \gamma \frac{\Omega_g^2\Omega_e^2}{\qty(\Omega_g^2 + \Omega_e^2)\Delta_{\rm R}^2} \frac{\eta_z^2}{4} \sum_\nu \frac{\omega_\nu^2}{(\omega_m + \omega_\nu)^2}
    \, ,
    \\
    \qty(\vec R)_{n_{\rm ex}>0} =& 
    \sum_{j,l} \int d\mu_l W(\mu_l)
    \frac{\gamma_l}{4\Delta_{\rm R}^2 + \gamma^2} \qty(\Omega_l^2 + \Omega_{\bar l}^2) \langle \hat n_+^{(j)} \rangle + \mathcal O(\eta_z)
    \nonumber \\
    =& \frac{\qty(\Omega_g^2 + \Omega_e^2)\gamma}{4\Delta_{\rm R}^2 + \gamma^2} \overline{N_+}(n_{\rm ex}, N) + \mathcal O(\eta_z)
\end{align}

This gives for $\Omega_g = \Omega_e = \Omega$, and ignoring terms of order $\gamma^2/\Delta_{\rm R}^2$ and $\eta_z^3$
\begin{align}
    &\langle \hat R \rangle (t \to \infty) =
    \frac{1}{\qty(\gamma_{N,\rm sc}^{(1)} + \gamma_{1,p} + \gamma_{2,p})\gamma_{N,\rm sc}^{(2)} + \gamma_{2,p} \gamma_{N,\rm sc}^{(1)}}
    \times
    \nonumber \\
    &\quad \qty[
    \gamma_{N,\rm sc}^{(1)} \gamma_{N,\rm sc}^{(2)} R_0 +
    \qty(\gamma_{1,p} + \gamma_{2,p}) \gamma_{N,\rm sc}^{(2)} R_1 +
    \gamma_{2,p} \gamma_{N,\rm sc}^{(1)} R_2
    ]
    \nonumber \\
    &\qquad \qquad \approx
    \frac{\Omega^2\gamma}{\Delta_{\rm R}^2} \eta_z^2 \qty[\frac{1}{4} + \sum_\nu \frac{\omega_\nu^2}{\qty(\omega_m + \omega_\nu)^2}]
\end{align}

\section{Multimode Cooling in the strong coupling regime}
\label{Ap:strong-multimode}

\begin{figure}
    \centering
    \includegraphics[width=0.495\linewidth]{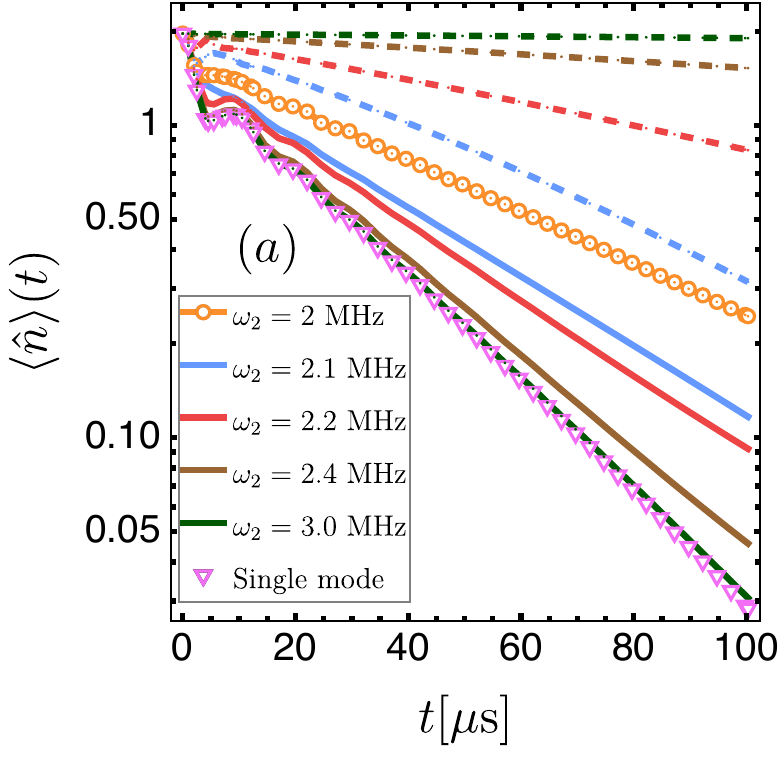}
    \includegraphics[width=0.495\linewidth]{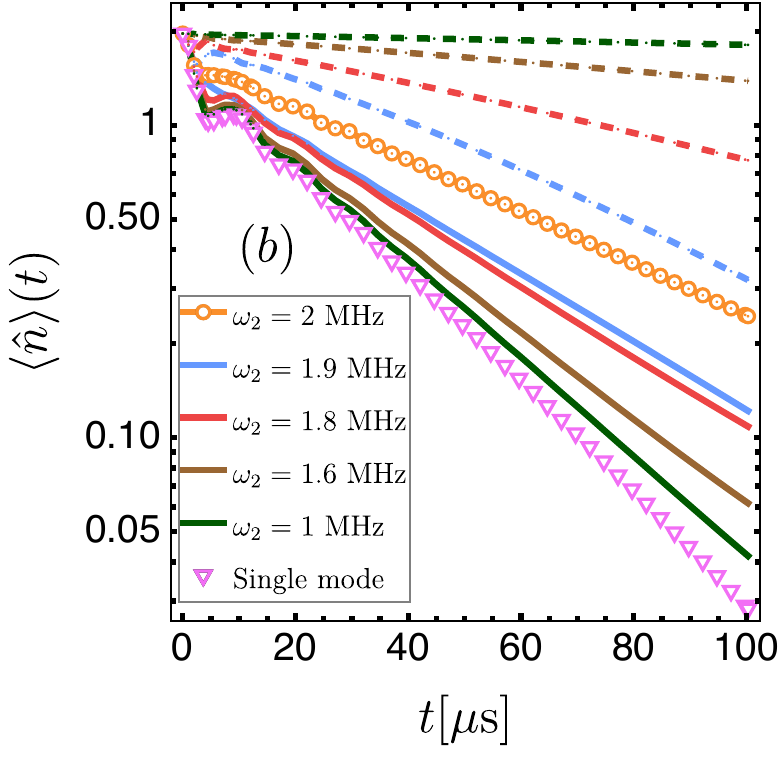}
    \caption{
    Two-mode cooling in the strong coupling regime.
    (a) Mean phonon number as a function of time for two ions computed from Eqs.~\eqref{eq:SpinhamoEffMMa} and \eqref{eq:SpinJumpEffm} after eliminating the excited state.
    The continuous lines are the mode occupations $\langle \hat b_1^\dagger \hat b_1 \rangle $ of center of mass phonon number which is cooled resonantly [$\omega_1/(2\pi) = \omega_{\rm s}/(2\pi) = 2$ MHz]
    The dash-dotted lines are the stretch mode occupation $\langle \hat b_2^\dagger \hat b_2\rangle$ with frequency $\omega_2$, where different colors correspond to different stretch-mode frequencies.
    The orange circles correspond to two fully decoupled ions, such that $\omega_1 = \omega_2$ and $\langle \hat b_1^\dagger \hat b_1\rangle = \langle \hat b_2^\dagger \hat b_2 \rangle$.
    The magenta triangles correspond to the single-mode solution in the main text.
    (b) Same as (a), but for $\omega_2 \leq \omega_1$, as expected for radial modes.
    For both cases, we assume an initial thermal occupation $p(n) \propto \exp(-n/n_{\rm th})$ with $n_{\rm th} = 2$ for both modes, truncated at $n_{\rm max} = 15$, giving $\langle \hat b_\nu^\dagger \hat b_\nu \rangle \approx 1.96$.
    We start with all ions initially in $\ket - $ and set $g_{\rm O} = 0$, $g_{\rm R} = 2\pi \times 50$ kHz, and $g_{\rm R} / \gamma_b = 1.4$, balanced Raman Rabi frequencies $\Omega_g = \Omega_e$.
    }
    \label{fig:strong-multimode}
\end{figure}

For $g_{\rm O} = 0$, the simultaneous cooling dynamics of multiple modes is described by
\begin{align}
    \hat H_{\rm eff} =& 
    \sum_{\nu=1}^N \omega_{\nu} \hat b_{\nu}^{\dagger} \hat b_{\nu}
    + \sum_{j=1}^{N}\frac{\omega_{s}}{2} \hat{\tilde \sigma}_{z}^{(j)}
    \nonumber \\
    &+ g_{\rm R}\sum_{j=1}^N\sum_{\nu=1}^N
    \mathcal K_{\nu,j}\sqrt{ \frac{\omega_{1}}{\omega_{\nu}} }
    \qty(\hat b_{\nu}+\hat b_{\nu}^{\dagger})\hat{\tilde \sigma}_{y}^{(j)}
    \, ,
\end{align}
and effective jump operators
\begin{align}
\hat{L}_{g}^{({\rm eff},j)}=\alpha_{gg}\hat{\sigma}_{gg}^{(j)}+\alpha_{ge}\hat{\sigma}_{ge}^{(j)},~ \hat{L}_{e}^{({\rm eff},{j})}=\alpha_{eg}\hat{\sigma}_{eg}^{(j)}+\alpha_{ee}\hat{\sigma}_{ee}^{(j)},
\end{align}

\subsection{$N = 2$}

Figure \ref{fig:strong-multimode} shows numerical simulations of the resulting dynamics for $N=2$, such that
\begin{align}
    \mathcal K = \frac{1}{\sqrt 2}\begin{pmatrix}
        1 & 1 \\ 1 & -1
    \end{pmatrix}
    \, .
\end{align}
For simplicity, we furthermore set $\alpha_{gg} = \alpha_{eg} = \alpha_{ge} = \alpha_{ee} = \gamma_b/2$.

The orange circles for $\omega_2/(2\pi) = 2$ MHz correspond to two decoupled ions and also directly overlap with the $N=1$ solution.
Over the full range of parameters analyzed, the center-of-mass mode cools faster than in this single-ion case and slower than for a fully isolated mode with $N=2$ ions.
It approaches the limit of having only the center-of-mass mode for far off-resonant $\omega_2 \gg \omega_1$ or $\omega_1 \gg \omega_2$, and approaches the single-ion case for $\omega_2 \rightarrow \omega_1$.

While the center-of-mass mode cooling speeds up, the stretch mode cooling slows down for increasing detuning.
Nevertheless, the stretch mode cools significantly up to $\abs{\omega_2 - \omega_1} \approx 4g_{\rm R}$.
We attribute this trade-off to the mechanism discussed in Fig.~\ref{fig:collective-sketches}.
The collective enhancement comes from the fact that the center-of-mass mode phonons have $N$ ions available to be cooled in parallel.
If multiple modes are cooled, the total excitation removal rate is distributed across both modes.
We see this as a trade-off where, as we tune $\omega_2$ out of resonance, the increase in cooling rate in mode 1 is directly accompanied by a decrease in cooling rate in mode 2.

Finally, we note that the cooling dynamics depend only weakly on whether $\omega_2 > \omega_1$ or $\omega_2 < \omega_1$, corresponding to axial or radial stretching, but mainly on the magnitude $\abs{\omega_2 - \omega_1}$.

\subsection{Large $N$}
We now estimate the reheating processes for large $N$.
We start from Eq.~\eqref{eq:app_jump_nex0}, but now we could all excitations created by $\hat b_\nu^\dagger \Omega_l \hat{\tilde \sigma}_{+-}^{(j)}$ as $n_{\rm ex} = 2$ for $1 \leq \nu \leq M$, and as $n_{\rm ex} = 1$, otherwise. Furthermore, we count $\hat b_\nu^\dagger \Omega_{\bar l}$ as pumping int $n_{\rm ex} = 1$ for $1 \leq \nu \leq M$.
We evaluate
\begin{align}
    \gamma_{1,\rm p}^{(M)} =& \sum_{\zeta,l,j} \abs{\bra{n_{\rm ex}=1,\zeta} \hat L_{l,\mu_l;\rm eff}^{(j)} \ket{n_{\rm ex}=0, \zeta = 1}^{(1)}}^2
    \nonumber \\
    =& \frac{\eta_z^2 \Omega_g^2 \Omega_e^2}{4 \Delta_{\rm R}^2 \qty(\Omega_g^2 + \Omega_e^2)^2} 
    \Bigg[
    \sum_{\nu=1}^M \frac{\omega_\nu^2\qty(\Omega_g^2 \gamma_e + \Omega_e^2 \gamma_g)}{(\omega_m + \omega_\nu)^2} 
    \nonumber \\
    &+
    \sum_{\nu=M+1}^N \frac{\omega_\nu^2\qty(\Omega_g^2 \gamma_g + \Omega_e^2 \gamma_e)}{(\omega_m + \omega_\nu)^2} 
    \Bigg]
    \nonumber \\
    &\overset{\Omega_g = \Omega_e}{=} \frac{\eta_z^2}{8}\gamma_{N=1,\rm sc} \qty[\sum_{\nu=1}^N \frac{4\omega_\nu^2}{(\omega_m + \omega_\nu)^2}] = \gamma_{1,\rm p},
    \label{eq:definition_gamma_1_p_manymode}
\end{align}
\begin{align}
    \gamma_{2,\rm p}^{(M)} =& \sum_{\zeta,l,j} \abs{\bra{n_{\rm ex}=2,\zeta} \hat L_{l,\mu_l;\rm eff}^{(j)} \ket{n_{\rm ex}=0, \zeta = 1}^{(1)}}^2
    \nonumber \\
    =& \frac{\eta_z^2 \Omega_g^2 \Omega_e^2 \qty(\Omega_g^2\gamma_g + \Omega_e^2 \gamma_e)}{4 \Delta_{\rm R}^2 \qty(\Omega_g^2 + \Omega_e^2)^2}\sum_{\nu=1}^M \frac{\omega_\nu^2}{(\omega_m + \omega_\nu)^2}
    \nonumber \\
    &\overset{\Omega_g=\Omega_e; \omega_\nu \approx \omega_m}{\approx}
    M\frac{\eta_z^2}{8}\gamma_{N=1,\rm sc} = M \gamma_{2,\rm p} \, .
\end{align}

Finally, we also need to evaluate the excitation number in the ground state from Eq.~\eqref{eq:n_ex_1}
\begin{align}
    n_{\rm ex,0} \equiv&^{(1)}\bra{n_{\rm ex}=1, \zeta=1} \hat N_+ + \sum_{\nu = 1}^M \hat b_\nu^\dagger \hat b_\nu \ket{n_{\rm ex}=1, \zeta=1}^{(1)}
    \nonumber \\
    =& \frac{\eta_z^2 \Omega_e^2 \Omega_g^2}{4(\Omega_e^2 + \Omega_g^2)^2} \qty[2\sum_{\nu=1}^M  \frac{4\omega_m^2}{(\omega_m + \omega_\nu)^2} + \sum_{\nu=M+1}^N  \frac{4\omega_m^2}{(\omega_m + \omega_\nu)^2}]
    \, .
\end{align}

With that, we can compute analogous to Eq.~\eqref{eq:analytic_excitations}
\begin{align}
    \langle \hat n_{\rm ex} \rangle (t \to \infty)
    \approx& \frac{\gamma_{1,\rm p}^{(M)} + 2\gamma_{2,\rm p}^{(M)}}{\gamma_{N,M,\rm mm}^{(1)}} + n_{\rm ex,0}
    \nonumber \\
    \overset{\Omega_g = \Omega_e; \omega_{\nu \leq M} \approx \omega_m}{\approx}& \frac{\eta_z^2}{8} \qty[\frac{M}{2} + \frac{MN}{N+M} + \frac{2N+M}{2N} \qty(\sum_{\nu = 1}^N \frac{4\omega_m^2}{(\omega_m+\omega_\nu)^2})]
    \, .
\end{align}
For $M \ll N$, this is still dominated by the last term, and thus only slightly changes with $N$.

\bibliographystyle{apsrev4-2}
\bibliography{References.bib}

@article{PhysRevLett.82.1971,
  title = {Quantum Computation with Ions in Thermal Motion},
  author = {S\o{}rensen, Anders and M\o{}lmer, Klaus},
  journal = {Phys. Rev. Lett.},
  volume = {82},
  issue = {9},
  pages = {1971--1974},
  numpages = {0},
  year = {1999},
  month = {Mar},
  publisher = {American Physical Society},
  doi = {10.1103/PhysRevLett.82.1971},
  url = {https://link.aps.org/doi/10.1103/PhysRevLett.82.1971}
}

@article{PhysRevA.62.022311,
  title = {Entanglement and quantum computation with ions in thermal motion},
  author = {S\o{}rensen, Anders and M\o{}lmer, Klaus},
  journal = {Phys. Rev. A},
  volume = {62},
  issue = {2},
  pages = {022311},
  numpages = {11},
  year = {2000},
  month = {Jul},
  publisher = {American Physical Society},
  doi = {10.1103/PhysRevA.62.022311},
  url = {https://link.aps.org/doi/10.1103/PhysRevA.62.022311}
}

@article{fd15-4g1n,
  title = {Loading and Imaging Atom Arrays via Electromagnetically Induced Transparency},
  author = {Qiu, Emily H. and \ifmmode \check{S}\else \v{S}\fi{}umarac, Tamara and Niu, Peiran and Tsesses, Shai and Wassaf, Fadi and Spierings, David C. and Chen, Meng-Wei and Uysal, Mehmet T. and Bartlett, Audrey and Menssen, Adrian J. and Lukin, Mikhail D. and Vuleti\ifmmode \acute{c}\else \'{c}\fi{}, Vladan},
  journal = {PRX Quantum},
  volume = {7},
  issue = {2},
  pages = {020358},
  numpages = {13},
  year = {2026},
  month = {Jun},
  publisher = {American Physical Society},
  doi = {10.1103/fd15-4g1n},
  url = {https://link.aps.org/doi/10.1103/fd15-4g1n}
}

@article{kiesenhofer2023controlling,
	author = {Kiesenhofer, Dominik and Hainzer, Helene and Zhdanov, Artem and Holz, Philip C. and Bock, Matthias and Ollikainen, Tuomas and Roos, Christian F.},
	title = {{Controlling Two-Dimensional Coulomb Crystals of More Than 100 Ions in a Monolithic Radio-Frequency Trap}},
	journal = {PRX Quantum},
	volume = {4},
	number = {2},
	pages = {020317},
	year = {2023},
	month = apr,
	publisher = {American Physical Society},
	doi = {10.1103/PRXQuantum.4.020317}
}

@article{guo2024a,
	author = {Guo, S.-A. and Wu, Y.-K. and Ye, J. and Zhang, L. and Lian, W.-Q. and Yao, R. and Wang, Y. and Yan, R.-Y. and Yi, Y.-J. and Xu, Y.-L. and Li, B.-W. and Hou, Y.-H. and Xu, Y.-Z. and Guo, W.-X. and Zhang, C. and Qi, B.-X. and Zhou, Z.-C. and He, L. and Duan, L.-M.},
	title = {{A site-resolved two-dimensional quantum simulator with hundreds of trapped ions}},
	journal = {Nature},
	volume = {630},
	pages = {613--618},
	year = {2024},
	month = jun,
	issn = {1476-4687},
	publisher = {Nature Publishing Group},
	doi = {10.1038/s41586-024-07459-0}
}

@article{PhysRevLett.73.2829,
  title = {Laser Cooling to a Single Quantum State in a Trap},
  author = {Dum, R. and Marte, P. and Pellizzari, T. and Zoller, P.},
  journal = {Phys. Rev. Lett.},
  volume = {73},
  issue = {21},
  pages = {2829--2832},
  numpages = {0},
  year = {1994},
  month = {Nov},
  publisher = {American Physical Society},
  doi = {10.1103/PhysRevLett.73.2829},
  url = {https://link.aps.org/doi/10.1103/PhysRevLett.73.2829}
}

@article{PhysRevA.102.033115,
  title = {Sideband ground-state cooling of graphene with Rydberg atoms via vacuum forces},
  author = {Khan, M. Miskeen and Ribeiro, S. and Mendon\ifmmode \mbox{\c{c}}\else \c{c}\fi{}a, J. T. and Ter\ifmmode \mbox{\c{c}}\else \c{c}\fi{}as, H.},
  journal = {Phys. Rev. A},
  volume = {102},
  issue = {3},
  pages = {033115},
  numpages = {9},
  year = {2020},
  month = {Sep},
  publisher = {American Physical Society},
  doi = {10.1103/PhysRevA.102.033115},
  url = {https://link.aps.org/doi/10.1103/PhysRevA.102.033115}
}

@article{wu2025electromagnetically,
	author = {Wu, Jenny J. and Hou, Pan-Yu and Erickson, Stephen D. and Brandt, Adam D. and Wan, Yong and Zarantonello, Giorgio and Cole, Daniel C. and Wilson, Andrew C. and Slichter, Daniel H. and Leibfried, Dietrich},
	title = {{Electromagnetically-induced-transparency cooling with a tripod structure in a hyperfine trapped ion with mixed-species crystals}},
	journal = {Phys. Rev. A},
	volume = {111},
	number = {4},
	pages = {043109},
	year = {2025},
	month = apr,
	publisher = {American Physical Society},
	doi = {10.1103/PhysRevA.111.043109}
}

@article{qiao2021double,
	author = {Qiao, Mu and Wang, Ye and Cai, Zhengyang and Du, Botao and Wang, Pengfei and Luan, Chunyang and Chen, Wentao and Noh, Heung-Ryoul and Kim, Kihwan},
	title = {{Double-Electromagnetically-Induced-Transparency Ground-State Cooling of Stationary Two-Dimensional Ion Crystals}},
	journal = {Phys. Rev. Lett.},
	volume = {126},
	number = {2},
	pages = {023604},
	year = {2021},
	month = jan,
	publisher = {American Physical Society},
	doi = {10.1103/PhysRevLett.126.023604}
}

@article{moses2023race,
  doi = {10.1103/PhysRevX.13.041052},
  title={A race-track trapped-ion quantum processor},
  author={Moses, Steven A and Baldwin, Charles H and Allman, Michael S and Ancona, R and Ascarrunz, L and Barnes, C and Bartolotta, J and Bjork, B and Blanchard, P and Bohn, M and others},
  journal={Physical Review X},
  volume={13},
  number={4},
  pages={041052},
  year={2023},
  publisher={APS},
}

@article{park2023extended,
	author = {Park, Annie J. and Picard, Lewis R. B. and Patenotte, Gabriel E. and Zhang, Jessie T. and Rosenband, Till and Ni, Kang-Kuen},
	title = {{Extended Rotational Coherence of Polar Molecules in an Elliptically Polarized Trap}},
	journal = {Phys. Rev. Lett.},
	volume = {131},
	number = {18},
	pages = {183401},
	year = {2023},
	month = oct,
	publisher = {American Physical Society},
	doi = {10.1103/PhysRevLett.131.183401}
}

@article{rubies2025collectively,
	author = {Rubies-Bigorda, Oriol and Holzinger, Raphael and Asenjo-Garcia, Ana and Romero-Isart, Oriol and Ritsch, Helmut and Ostermann, Stefan and Gonzalez-Ballestero, Carlos and Yelin, Susanne F. and Rusconi, Cosimo C.},
	title = {{Collectively enhanced ground-state cooling in subwavelength atomic arrays}},
	journal = {Phys. Rev. A},
	volume = {112},
	number = {2},
	pages = {023714},
	year = {2025},
	month = aug,
	publisher = {American Physical Society},
	doi = {10.1103/bhwv-ndtj}
}

@article{wang2023enhanced,
	author = {Wang, Chung-Hsien and Wang, Yi-Cheng and Chen, Chi-Chih and Wang, Chun-Che and Jen, H. H.},
	title = {{Enhanced dark-state sideband cooling in trapped atoms via photon-mediated dipole-dipole interactions}},
	journal = {Phys. Rev. A},
	volume = {107},
	number = {2},
	pages = {023117},
	year = {2023},
	month = feb,
	publisher = {American Physical Society},
	doi = {10.1103/PhysRevA.107.023117}
}

@article{zhang2022parallel,
	author = {Zhang, Jie and Zhang, Man-Chao and Xie, Yi and Wu, Chun-Wang and Ou, Bao-Quan and Chen, Ting and Bao, Wan-Su and Haljan, Paul and Wu, Wei and Zhang, Shuo and Chen, Ping-Xing},
	title = {{Parallel Electromagnetically Induced Transparency near Ground-State Cooling of a Trapped-Ion Crystal}},
	journal = {Phys. Rev. Appl.},
	volume = {18},
	number = {1},
	pages = {014022},
	year = {2022},
	month = jul,
	publisher = {American Physical Society},
	doi = {10.1103/PhysRevApplied.18.014022}
}

@article{sun2023sympathetic,
	author = {Sun, Chenglong and Cui, Kaifeng and Chao, Sijia and Wei, Yuanfei and Yuan, Jinbo and Cao, Jian and Shu, Hualin and Huang, Xueren},
	title = {{Sympathetic electromagnetically induced transparency ground state cooling of a 40Ca+{\textendash}27Al+ pair in an 27Al+ clock}},
	journal = {Chin. Phys. B},
	volume = {32},
	number = {5},
	pages = {050601},
	year = {2023},
	month = may,
	issn = {1674-1056},
	publisher = {Chinese Physical Society and IOP Publishing Ltd},
	doi = {10.1088/1674-1056/aca39d}
}

@article{dawel2025a,
	author = {Dawel, Fabian and Pelzer, Lennart and Dietze, Kai and Kramer, Johannes and Hild, Marek and King, Steven A. and Spethmann, Nicolas C. H. and Klose, Joshua and Stahl, Kilian and D{\ifmmode\ddot{o}\else\"{o}\fi}rscher, S{\ifmmode\ddot{o}\else\"{o}\fi}ren and Benkler, Erik and Lisdat, Christian and Porsev, Sergey G. and Safronova, Marianna S. and Schmidt, Piet O.},
	title = {{A high-stability optical clock based on a continuously ground-state cooled Al$^+$ ion without compromising its accuracy}},
	journal = {arXiv},
	year = {2025},
	month = sep,
	eprint = {2509.22525},
	doi = {10.48550/arXiv.2509.22525}
}

@article{scharnhorst2018experimental,
	author = {Scharnhorst, Nils and Cerrillo, Javier and Kramer, Johannes and Leroux, Ian D. and W{\ifmmode\ddot{u}\else\"{u}\fi}bbena, Jannes B. and Retzker, Alex and Schmidt, Piet O.},
	title = {{Experimental and theoretical investigation of a multimode cooling scheme using multiple electromagnetically-induced-transparency resonances}},
	journal = {Phys. Rev. A},
	volume = {98},
	number = {2},
	pages = {023424},
	year = {2018},
	month = aug,
	publisher = {American Physical Society},
	doi = {10.1103/PhysRevA.98.023424}
}

@article{huang2024electromagnetically,
	author = {Huang, Chuanxin and Wang, Chenxi and Zhang, Hongxuan and Hu, Hongyuan and Wang, Zuqing and Mao, Zhichao and Li, Shijiao and Hou, Panyu and Wu, Yukai and Zhou, Zichao and Duan, Luming},
	title = {{Electromagnetically Induced Transparency Cooling of High-Nuclear-Spin Ions}},
	journal = {Phys. Rev. Lett.},
	volume = {133},
	number = {11},
	pages = {113204},
	year = {2024},
	month = sep,
	publisher = {American Physical Society},
	doi = {10.1103/PhysRevLett.133.113204}
}

@article{bartolotta2024laser,
	author = {Bartolotta, John P. and Estey, Brian and Foss-Feig, Michael and Hayes, David and Gilbreth, Christopher N.},
	title = {{Laser cooling trapped-ion crystal modes beyond the Lamb-Dicke regime}},
	journal = {arXiv},
	year = {2024},
	month = nov,
	eprint = {2411.18818},
	doi = {10.48550/arXiv.2411.18818}
}

@article{fouka2025multilevel,
	author = {Fouka, Katya and Shankar, Athreya and Tan, Ting Rei and Safavi-Naini, Arghavan},
	title = {{Multilevel Electromagnetically Induced Transparency Cooling}},
	journal = {arXiv},
	year = {2025},
	month = jun,
	eprint = {2506.14546},
	doi = {10.48550/arXiv.2506.14546}
}

@book{sakurai2020modern,
  title={Modern quantum mechanics},
  author={Sakurai, Jun John and Napolitano, Jim},
  year={2020},
  publisher={Cambridge University Press}
}

@article{PhysRevX.14.031030,
  title = {Bilayer Crystals of Trapped Ions for Quantum Information Processing},
  author = {Hawaldar, Samarth and Shahi, Prakriti and Carter, Allison L. and Rey, Ana Maria and Bollinger, John J. and Shankar, Athreya},
  journal = {Phys. Rev. X},
  volume = {14},
  issue = {3},
  pages = {031030},
  numpages = {32},
  year = {2024},
  month = {Aug},
  publisher = {American Physical Society},
  doi = {10.1103/PhysRevX.14.031030},
  url = {https://link.aps.org/doi/10.1103/PhysRevX.14.031030}
}

@article{PhysRevX.14.031002,
  title = {Raman Sideband Cooling of Molecules in an Optical Tweezer Array to the 3D Motional Ground State},
  author = {Bao, Yicheng and Yu, Scarlett S. and You, Jiaqi and Anderegg, Lo\"{\i}c and Chae, Eunmi and Ketterle, Wolfgang and Ni, Kang-Kuen and Doyle, John M.},
  journal = {Phys. Rev. X},
  volume = {14},
  issue = {3},
  pages = {031002},
  numpages = {10},
  year = {2024},
  month = {Jul},
  publisher = {American Physical Society},
  doi = {10.1103/PhysRevX.14.031002},
  url = {https://link.aps.org/doi/10.1103/PhysRevX.14.031002}
}

@article{Lu2024,
  title = {Raman sideband cooling of molecules in an optical tweezer array},
  volume = {20},
  ISSN = {1745-2481},
  url = {http://dx.doi.org/10.1038/s41567-023-02346-3},
  DOI = {10.1038/s41567-023-02346-3},
  number = {3},
  journal = {Nature Physics},
  publisher = {Springer Science and Business Media LLC},
  author = {Lu,  Yukai and Li,  Samuel J. and Holland,  Connor M. and Cheuk,  Lawrence W.},
  year = {2024},
  month = jan,
  pages = {389–394}
}

@article{RevModPhys.82.1041,
  title = {Quantum interface between light and atomic ensembles},
  author = {Hammerer, Klemens and S\o{}rensen, Anders S. and Polzik, Eugene S.},
  journal = {Rev. Mod. Phys.},
  volume = {82},
  issue = {2},
  pages = {1041--1093},
  numpages = {0},
  year = {2010},
  month = {Apr},
  publisher = {American Physical Society},
  doi = {10.1103/RevModPhys.82.1041},
  url = {https://link.aps.org/doi/10.1103/RevModPhys.82.1041}
}

@article{Castelvecchi2023,
  title = {Quantum-computing approach uses single molecules as qubits for first time},
  ISSN = {1476-4687},
  url = {http://dx.doi.org/10.1038/d41586-023-03943-1},
  DOI = {10.1038/d41586-023-03943-1},
  journal = {Nature},
  publisher = {Springer Science and Business Media LLC},
  author = {Castelvecchi,  Davide},
  year = {2023},
  month = dec 
}

@article{RevModPhys.87.637,
  title = {Optical atomic clocks},
  author = {Ludlow, Andrew D. and Boyd, Martin M. and Ye, Jun and Peik, E. and Schmidt, P. O.},
  journal = {Rev. Mod. Phys.},
  volume = {87},
  issue = {2},
  pages = {637--701},
  numpages = {65},
  year = {2015},
  month = {Jun},
  publisher = {American Physical Society},
  doi = {10.1103/RevModPhys.87.637},
  url = {https://link.aps.org/doi/10.1103/RevModPhys.87.637}
}

@article{RevModPhys.89.035002,
  title = {Quantum sensing},
  author = {Degen, C. L. and Reinhard, F. and Cappellaro, P.},
  journal = {Rev. Mod. Phys.},
  volume = {89},
  issue = {3},
  pages = {035002},
  numpages = {39},
  year = {2017},
  month = {Jul},
  publisher = {American Physical Society},
  doi = {10.1103/RevModPhys.89.035002},
  url = {https://link.aps.org/doi/10.1103/RevModPhys.89.035002}
}

@article{PhysRevLett.99.093902,
  title = {Quantum Theory of Cavity-Assisted Sideband Cooling of Mechanical Motion},
  author = {Marquardt, Florian and Chen, Joe P. and Clerk, A. A. and Girvin, S. M.},
  journal = {Phys. Rev. Lett.},
  volume = {99},
  issue = {9},
  pages = {093902},
  numpages = {4},
  year = {2007},
  month = {Aug},
  publisher = {American Physical Society},
  doi = {10.1103/PhysRevLett.99.093902},
  url = {https://link.aps.org/doi/10.1103/PhysRevLett.99.093902}
}

@article{PhysRevLett.116.063601,
  title = {Laser Cooling of a Micromechanical Membrane to the Quantum Backaction Limit},
  author = {Peterson, R. W. and Purdy, T. P. and Kampel, N. S. and Andrews, R. W. and Yu, P.-L. and Lehnert, K. W. and Regal, C. A.},
  journal = {Phys. Rev. Lett.},
  volume = {116},
  issue = {6},
  pages = {063601},
  numpages = {6},
  year = {2016},
  month = {Feb},
  publisher = {American Physical Society},
  doi = {10.1103/PhysRevLett.116.063601},
  url = {https://link.aps.org/doi/10.1103/PhysRevLett.116.063601}
}

@article{RevModPhys.77.633,
  title = {Electromagnetically induced transparency: Optics in coherent media},
  author = {Fleischhauer, Michael and Imamoglu, Atac and Marangos, Jonathan P.},
  journal = {Rev. Mod. Phys.},
  volume = {77},
  issue = {2},
  pages = {633--673},
  numpages = {0},
  year = {2005},
  month = {Jul},
  publisher = {American Physical Society},
  doi = {10.1103/RevModPhys.77.633},
  url = {https://link.aps.org/doi/10.1103/RevModPhys.77.633}
}

@article{Roussy2023,
  title = {An improved bound on the electron’s electric dipole moment},
  volume = {381},
  ISSN = {1095-9203},
  url = {http://dx.doi.org/10.1126/science.adg4084},
  DOI = {10.1126/science.adg4084},
  number = {6653},
  journal = {Science},
  publisher = {American Association for the Advancement of Science (AAAS)},
  author = {Roussy,  Tanya S. and Caldwell,  Luke and Wright,  Trevor and Cairncross,  William B. and Shagam,  Yuval and Ng,  Kia Boon and Schlossberger,  Noah and Park,  Sun Yool and Wang,  Anzhou and Ye,  Jun and Cornell,  Eric A.},
  year = {2023},
  month = jul,
  pages = {46–50}
}

@article{Bluvstein2023,
  title = {Logical quantum processor based on reconfigurable atom arrays},
  volume = {626},
  ISSN = {1476-4687},
  url = {http://dx.doi.org/10.1038/s41586-023-06927-3},
  DOI = {10.1038/s41586-023-06927-3},
  number = {7997},
  journal = {Nature},
  publisher = {Springer Science and Business Media LLC},
  author = {Bluvstein,  Dolev and Evered,  Simon J. and Geim,  Alexandra A. and Li,  Sophie H. and Zhou,  Hengyun and Manovitz,  Tom and Ebadi,  Sepehr and Cain,  Madelyn and Kalinowski,  Marcin and Hangleiter,  Dominik and Bonilla Ataides,  J. Pablo and Maskara,  Nishad and Cong,  Iris and Gao,  Xun and Sales Rodriguez,  Pedro and Karolyshyn,  Thomas and Semeghini,  Giulia and Gullans,  Michael J. and Greiner,  Markus and Vuletić,  Vladan and Lukin,  Mikhail D.},
  year = {2023},
  month = dec,
  pages = {58–65}
}

@article{Egan2021,
  title = {Fault-tolerant control of an error-corrected qubit},
  volume = {598},
  ISSN = {1476-4687},
  url = {http://dx.doi.org/10.1038/s41586-021-03928-y},
  DOI = {10.1038/s41586-021-03928-y},
  number = {7880},
  journal = {Nature},
  publisher = {Springer Science and Business Media LLC},
  author = {Egan,  Laird and Debroy,  Dripto M. and Noel,  Crystal and Risinger,  Andrew and Zhu,  Daiwei and Biswas,  Debopriyo and Newman,  Michael and Li,  Muyuan and Brown,  Kenneth R. and Cetina,  Marko and Monroe,  Christopher},
  year = {2021},
  month = oct,
  pages = {281–286}
}

@article{Vilas2024,
  title = {An optical tweezer array of ultracold polyatomic molecules},
  volume = {628},
  ISSN = {1476-4687},
  url = {http://dx.doi.org/10.1038/s41586-024-07199-1},
  DOI = {10.1038/s41586-024-07199-1},
  number = {8007},
  journal = {Nature},
  publisher = {Springer Science and Business Media LLC},
  author = {Vilas,  Nathaniel B. and Robichaud,  Paige and Hallas,  Christian and Li,  Grace K. and Anderegg,  Loïc and Doyle,  John M.},
  year = {2024},
  month = apr,
  pages = {282–286}
}

@article{PhysRevLett.127.263201,
  title = {High Phase-Space Density of Laser-Cooled Molecules in an Optical Lattice},
  author = {Wu, Yewei and Burau, Justin J. and Mehling, Kameron and Ye, Jun and Ding, Shiqian},
  journal = {Phys. Rev. Lett.},
  volume = {127},
  issue = {26},
  pages = {263201},
  numpages = {6},
  year = {2021},
  month = {Dec},
  publisher = {American Physical Society},
  doi = {10.1103/PhysRevLett.127.263201},
  url = {https://link.aps.org/doi/10.1103/PhysRevLett.127.263201}
}

@book{PMeystre2007,
  title = {Elements of Quantum Optics},
  ISBN = {9783540742111},
  url = {http://dx.doi.org/10.1007/978-3-540-74211-1},
  author={Meystre, Pierre and Sargent, Murray},
  DOI = {10.1007/978-3-540-74211-1},
  publisher = {Springer Berlin Heidelberg},
  year = {2007}
}

@book{Breuer2007,
  title = {The Theory of Open Quantum Systems},
  ISBN = {9780191706349},
  url = {http://dx.doi.org/10.1093/acprof:oso/9780199213900.001.0001},
  DOI = {10.1093/acprof:oso/9780199213900.001.0001},
  publisher = {Oxford University PressOxford},
  author = {Breuer,  Heinz-Peter and Petruccione,  Francesco},
  year = {2007},
  month = jan 
}

@article{RevModPhys.58.699,
  title = {The semiclassical theory of laser cooling},
  author = {Stenholm, Stig},
  journal = {Rev. Mod. Phys.},
  volume = {58},
  issue = {3},
  pages = {699--739},
  numpages = {0},
  year = {1986},
  month = {Jul},
  publisher = {American Physical Society},
  doi = {10.1103/RevModPhys.58.699},
  url = {https://link.aps.org/doi/10.1103/RevModPhys.58.699}
}

@article{RevModPhys.82.1155,
  title = {Introduction to quantum noise, measurement, and amplification},
  author = {Clerk, A. A. and Devoret, M. H. and Girvin, S. M. and Marquardt, Florian and Schoelkopf, R. J.},
  journal = {Rev. Mod. Phys.},
  volume = {82},
  issue = {2},
  pages = {1155--1208},
  numpages = {0},
  year = {2010},
  month = {Apr},
  publisher = {American Physical Society},
  doi = {10.1103/RevModPhys.82.1155},
  url = {https://link.aps.org/doi/10.1103/RevModPhys.82.1155}
}

@article{PRXQuantum.3.040324,
  title = {Simulating Dynamical Phases of Chiral $p+ip$ Superconductors with a Trapped ion Magnet},
  author = {Shankar, Athreya and Yuzbashyan, Emil A. and Gurarie, Victor and Zoller, Peter and Bollinger, John J. and Rey, Ana Maria},
  journal = {PRX Quantum},
  volume = {3},
  issue = {4},
  pages = {040324},
  numpages = {23},
  year = {2022},
  month = {Nov},
  publisher = {American Physical Society},
  doi = {10.1103/PRXQuantum.3.040324},
  url = {https://link.aps.org/doi/10.1103/PRXQuantum.3.040324}
}

@article{Zhang2017,
  doi = {10.1038/nature24654},
  url = {https://doi.org/10.1038/nature24654},
  year = {2017},
  month = nov,
  publisher = {Springer Science and Business Media {LLC}},
  volume = {551},
  number = {7682},
  pages = {601--604},
  author = {J. Zhang and G. Pagano and P. W. Hess and A. Kyprianidis and P. Becker and H. Kaplan and A. V. Gorshkov and Z.-X. Gong and C. Monroe},
  title = {Observation of a many-body dynamical phase transition with a 53-qubit quantum simulator},
  journal = {Nature}
}

@article{PhysRevX.2.041014,
  title = {Cooling a Single Atom in an Optical Tweezer to Its Quantum Ground State},
  author = {Kaufman, A. M. and Lester, B. J. and Regal, C. A.},
  journal = {Phys. Rev. X},
  volume = {2},
  issue = {4},
  pages = {041014},
  numpages = {7},
  year = {2012},
  month = {Nov},
  publisher = {American Physical Society},
  doi = {10.1103/PhysRevX.2.041014},
  url = {https://link.aps.org/doi/10.1103/PhysRevX.2.041014}
}

@article{PhysRevB.82.165320,
  title = {Cooling of mechanical motion with a two-level system: The high-temperature regime},
  author = {Rabl, P.},
  journal = {Phys. Rev. B},
  volume = {82},
  issue = {16},
  pages = {165320},
  numpages = {11},
  year = {2010},
  month = {Oct},
  publisher = {American Physical Society},
  doi = {10.1103/PhysRevB.82.165320},
  url = {https://link.aps.org/doi/10.1103/PhysRevB.82.165320}
}

@article{Albrecht_2011,
doi = {10.1088/1367-2630/13/3/033009},
url = {https://dx.doi.org/10.1088/1367-2630/13/3/033009},
year = {2011},
month = {mar},
publisher = {},
volume = {13},
number = {3},
pages = {033009},
author = {A Albrecht and A Retzker and C Wunderlich and M B Plenio},
title = {Enhancement of laser cooling by the use of magnetic gradients},
journal = {New J. Phys.}
}

@article{Nakajima1958,
  doi = {10.1143/ptp.20.948},
  url = {https://doi.org/10.1143/ptp.20.948},
  year = {1958},
  month = dec,
  publisher = {Oxford University Press ({OUP})},
  volume = {20},
  number = {6},
  pages = {948--959},
  author = {Sadao Nakajima},
  title = {On Quantum Theory of Transport Phenomena},
  journal = {Prog. Theor. Phys.}
}

@article{PhysRev.124.983,
  title = {Memory Effects in Irreversible Thermodynamics},
  author = {Zwanzig, Robert},
  journal = {Phys. Rev.},
  volume = {124},
  issue = {4},
  pages = {983--992},
  numpages = {0},
  year = {1961},
  month = {Nov},
  publisher = {American Physical Society},
  doi = {10.1103/PhysRev.124.983},
  url = {https://link.aps.org/doi/10.1103/PhysRev.124.983}
}

@article{PhysRevA.86.012126,
  title = {Generalized Schrieffer-Wolff formalism for dissipative systems},
  author = {Kessler, E. M.},
  journal = {Phys. Rev. A},
  volume = {86},
  issue = {1},
  pages = {012126},
  numpages = {10},
  year = {2012},
  month = {Jul},
  publisher = {American Physical Society},
  doi = {10.1103/PhysRevA.86.012126},
  url = {https://link.aps.org/doi/10.1103/PhysRevA.86.012126}
}

@article{Jaehne_2008,
doi = {10.1088/1367-2630/10/9/095019},
url = {https://dx.doi.org/10.1088/1367-2630/10/9/095019},
year = {2008},
month = {sep},
publisher = {IOP Publishing},
volume = {10},
number = {9},
pages = {095019},
author = {Konstanze Jaehne and Klemens Hammerer and Margareta Wallquist},
title = {Ground-state cooling of a nanomechanical resonator via a Cooper-pair box qubit},
journal = {New J. Phys.}
}

@article{PhysRevA.46.2668,
  title = {Laser cooling of trapped ions in a standing wave},
  author = {Cirac, J. I. and Blatt, R. and Zoller, P. and Phillips, W. D.},
  journal = {Phys. Rev. A},
  volume = {46},
  issue = {5},
  pages = {2668--2681},
  numpages = {0},
  year = {1992},
  month = {Sep},
  publisher = {American Physical Society},
  doi = {10.1103/PhysRevA.46.2668},
  url = {https://link.aps.org/doi/10.1103/PhysRevA.46.2668}
}

@article{Zhang_2022,
doi = {10.1088/2058-9565/ac676c},
url = {https://dx.doi.org/10.1088/2058-9565/ac676c},
year = {2022},
month = {may},
publisher = {IOP Publishing},
volume = {7},
number = {3},
pages = {035006},
author = {Jessie T Zhang and Lewis R B Picard and William B Cairncross and Kenneth Wang and Yichao Yu and Fang Fang and Kang-Kuen Ni},
title = {An optical tweezer array of ground-state polar molecules},
journal = {Quantum Science and Technology}
}

@article{Gross2021,
  title = {Quantum gas microscopy for single atom and spin detection},
  volume = {17},
  ISSN = {1745-2481},
  url = {http://dx.doi.org/10.1038/s41567-021-01370-5},
  DOI = {10.1038/s41567-021-01370-5},
  number = {12},
  journal = {Nature Physics},
  publisher = {Springer Science and Business Media LLC},
  author = {Gross,  Christian and Bakr,  Waseem S.},
  year = {2021},
  month = nov,
  pages = {1316–1323}
}

@article{PhysRevA.49.2771,
  title = {Laser cooling of trapped three-level ions: Designing two-level systems for sideband cooling},
  author = {Marzoli, I. and Cirac, J. I. and Blatt, R. and Zoller, P.},
  journal = {Phys. Rev. A},
  volume = {49},
  issue = {4},
  pages = {2771--2779},
  numpages = {0},
  year = {1994},
  month = {Apr},
  publisher = {American Physical Society},
  doi = {10.1103/PhysRevA.49.2771},
  url = {https://link.aps.org/doi/10.1103/PhysRevA.49.2771}
}

@article{PhysRevLett.122.053603,
  title = {Near Ground-State Cooling of Two-Dimensional Trapped-Ion Crystals with More than 100 Ions},
  author = {Jordan, Elena and Gilmore, Kevin A. and Shankar, Athreya and Safavi-Naini, Arghavan and Bohnet, Justin G. and Holland, Murray J. and Bollinger, John J.},
  journal = {Phys. Rev. Lett.},
  volume = {122},
  issue = {5},
  pages = {053603},
  numpages = {5},
  year = {2019},
  month = {Feb},
  publisher = {American Physical Society},
  doi = {10.1103/PhysRevLett.122.053603},
  url = {https://link.aps.org/doi/10.1103/PhysRevLett.122.053603}
}

@article{phatak2024generalized,
	author = {Phatak, Saumitra S. and Blodgett, Karl N. and Peana, David and Chen, Meng Raymond and Hood, Jonathan D.},
	title = {{Generalized theory for optical cooling of a trapped atom with spin}},
	journal = {Phys. Rev. A},
	volume = {110},
	number = {4},
	pages = {043116},
	year = {2024},
	month = oct,
	publisher = {American Physical Society},
	doi = {10.1103/PhysRevA.110.043116}
}

@article{foss2025progress,
	author = {Foss-Feig, Michael and Pagano, Guido and Potter, Andrew C. and Yao, Norman Y.},
	title = {{Progress in Trapped-Ion Quantum Simulation}},
	journal = {Annu. Rev. Condens. Matter Phys.},
	volume = {16},
	pages = {145--172},
	year = {2025},
	month = mar,
	publisher = {Annual Reviews},
	doi = {10.1146/annurev-conmatphys-032822-045619}
}

@article{bharti2024strong,
	author = {Bharti, V. and Sugawa, S. and Kunimi, M. and Chauhan, V. S. and Mahesh, T. P. and Mizoguchi, M. and Matsubara, T. and Tomita, T. and de L{\ifmmode\acute{e}\else\'{e}\fi}s{\ifmmode\acute{e}\else\'{e}\fi}leuc, S. and Ohmori, K.},
	title = {{Strong Spin-Motion Coupling in the Ultrafast Dynamics of Rydberg Atoms}},
	journal = {Phys. Rev. Lett.},
	volume = {133},
	number = {9},
	pages = {093405},
	year = {2024},
	month = aug,
	publisher = {American Physical Society},
	doi = {10.1103/PhysRevLett.133.093405}
}

@article{picard2025entanglement,
	author = {Picard, Lewis R. B. and Park, Annie J. and Patenotte, Gabriel E. and Gebretsadkan, Samuel and Wellnitz, David and Rey, Ana Maria and Ni, Kang-Kuen},
	title = {{Entanglement and iSWAP gate between molecular qubits}},
	journal = {Nature},
	volume = {637},
	pages = {821--826},
	year = {2025},
	month = jan,
	issn = {1476-4687},
	publisher = {Nature Publishing Group},
	doi = {10.1038/s41586-024-08177-3}
}

@article{PhysRevLett.85.4458,
  title = {Ground State Laser Cooling Using Electromagnetically Induced Transparency},
  author = {Morigi, Giovanna and Eschner, J\"urgen and Keitel, Christoph H.},
  journal = {Phys. Rev. Lett.},
  volume = {85},
  issue = {21},
  pages = {4458--4461},
  numpages = {0},
  year = {2000},
  month = {Nov},
  publisher = {American Physical Society},
  doi = {10.1103/PhysRevLett.85.4458},
  url = {https://link.aps.org/doi/10.1103/PhysRevLett.85.4458}
}

@article{PhysRevA.91.012319,
  title = {Generation of spin-motion entanglement in a trapped ion using long-wavelength radiation},
  author = {Lake, K. and Weidt, S. and Randall, J. and Standing, E. D. and Webster, S. C. and Hensinger, W. K.},
  journal = {Phys. Rev. A},
  volume = {91},
  issue = {1},
  pages = {012319},
  numpages = {5},
  year = {2015},
  month = {Jan},
  publisher = {American Physical Society},
  doi = {10.1103/PhysRevA.91.012319},
  url = {https://link.aps.org/doi/10.1103/PhysRevA.91.012319}
}

@article{PhysRevLett.104.043003,
  title = {Fast and Robust Laser Cooling of Trapped Systems},
  author = {Cerrillo, J. and Retzker, A. and Plenio, M. B.},
  journal = {Phys. Rev. Lett.},
  volume = {104},
  issue = {4},
  pages = {043003},
  numpages = {4},
  year = {2010},
  month = {Jan},
  publisher = {American Physical Society},
  doi = {10.1103/PhysRevLett.104.043003},
  url = {https://link.aps.org/doi/10.1103/PhysRevLett.104.043003}
}

@article{PhysRevLett.125.053001,
  title = {Efficient Ground-State Cooling of Large Trapped-Ion Chains with an Electromagnetically-Induced-Transparency Tripod Scheme},
  author = {Feng, L. and Tan, W. L. and De, A. and Menon, A. and Chu, A. and Pagano, G. and Monroe, C.},
  journal = {Phys. Rev. Lett.},
  volume = {125},
  issue = {5},
  pages = {053001},
  numpages = {5},
  year = {2020},
  month = {Jul},
  publisher = {American Physical Society},
  doi = {10.1103/PhysRevLett.125.053001},
  url = {https://link.aps.org/doi/10.1103/PhysRevLett.125.053001}
}

@article{PhysRevLett.110.153002,
  title = {Sympathetic Electromagnetically-Induced-Transparency Laser Cooling of Motional Modes in an Ion Chain},
  author = {Lin, Y. and Gaebler, J. P. and Tan, T. R. and Bowler, R. and Jost, J. D. and Leibfried, D. and Wineland, D. J.},
  journal = {Phys. Rev. Lett.},
  volume = {110},
  issue = {15},
  pages = {153002},
  numpages = {5},
  year = {2013},
  month = {Apr},
  publisher = {American Physical Society},
  doi = {10.1103/PhysRevLett.110.153002},
  url = {https://link.aps.org/doi/10.1103/PhysRevLett.110.153002}
}

@article{PhysRevLett.85.5547,
  title = {Experimental Demonstration of Ground State Laser Cooling with Electromagnetically Induced Transparency},
  author = {Roos, C. F. and Leibfried, D. and Mundt, A. and Schmidt-Kaler, F. and Eschner, J. and Blatt, R.},
  journal = {Phys. Rev. Lett.},
  volume = {85},
  issue = {26},
  pages = {5547--5550},
  numpages = {0},
  year = {2000},
  month = {Dec},
  publisher = {American Physical Society},
  doi = {10.1103/PhysRevLett.85.5547},
  url = {https://link.aps.org/doi/10.1103/PhysRevLett.85.5547}
}

@article{Zhang2021,
  doi = {10.1088/1367-2630/abe273},
  url = {https://doi.org/10.1088/1367-2630/abe273},
  year = {2021},
  month = feb,
  publisher = {{IOP} Publishing},
  volume = {23},
  number = {2},
  pages = {023018},
  author = {Shuo Zhang and Jian-Qi Zhang and Wei Wu and Wan-Su Bao and Chu Guo},
  title = {Fast cooling of trapped ion in strong sideband coupling regime},
  journal = {New J. Phys.}
}

@article{PhysRevA.104.013117,
  title = {Steady-state phonon occupation of electromagnetically-induced-transparency cooling: Higher-order calculations},
  author = {Zhang, Shuo and Tian, Tian-Ci and Wu, Zheng-Yang and Zhang, Zong-Sheng and Wang, Xin-Hai and Wu, Wei and Bao, Wan-Su and Guo, Chu},
  journal = {Phys. Rev. A},
  volume = {104},
  issue = {1},
  pages = {013117},
  numpages = {7},
  year = {2021},
  month = {Jul},
  publisher = {American Physical Society},
  doi = {10.1103/PhysRevA.104.013117},
  url = {https://link.aps.org/doi/10.1103/PhysRevA.104.013117}
}

@article{PhysRevA.93.053401,
  title = {Electromagnetically-induced-transparency ground-state cooling of long ion strings},
  author = {Lechner, Regina and Maier, Christine and Hempel, Cornelius and Jurcevic, Petar and Lanyon, Ben P. and Monz, Thomas and Brownnutt, Michael and Blatt, Rainer and Roos, Christian F.},
  journal = {Phys. Rev. A},
  volume = {93},
  issue = {5},
  pages = {053401},
  numpages = {10},
  year = {2016},
  month = {May},
  publisher = {American Physical Society},
  doi = {10.1103/PhysRevA.93.053401},
  url = {https://link.aps.org/doi/10.1103/PhysRevA.93.053401}
}

@article{Lounis1992,
  doi = {10.1051/jp2:1992153},
  url = {https://doi.org/10.1051/jp2:1992153},
  year = {1992},
  month = apr,
  publisher = {{EDP} Sciences},
  volume = {2},
  number = {4},
  pages = {579--592},
  author = {B. Lounis and C. Cohen-Tannoudji},
  title = {Coherent population trapping and Fano profiles},
  journal = {Journal de Physique {II}}
}

@article{Gray:78,
author = {H. R. Gray and R. M. Whitley and C. R. Stroud},
journal = {Opt. Lett.},
number = {6},
pages = {218--220},
publisher = {Optica Publishing Group},
title = {Coherent trapping of atomic populations},
volume = {3},
month = {Dec},
year = {1978},
url = {https://opg.optica.org/ol/abstract.cfm?URI=ol-3-6-218},
doi = {10.1364/OL.3.000218},
}

@article{Aspect:89,
author = {A. Aspect and E. Arimondo and R. Kaiser and N. Vansteenkiste and C. Cohen-Tannoudji},
journal = {J. Opt. Soc. Am. B},
number = {11},
pages = {2112--2124},
publisher = {Optica Publishing Group},
title = {Laser cooling below the one-photon recoil energy by velocity-selective coherent population trapping: theoretical analysis},
volume = {6},
month = {Nov},
year = {1989},
url = {https://opg.optica.org/josab/abstract.cfm?URI=josab-6-11-2112},
doi = {10.1364/JOSAB.6.002112},
}

@article{PhysRevLett.75.4011,
  title = {Resolved-Sideband Raman Cooling of a Bound Atom to the 3D Zero-Point Energy},
  author = {Monroe, C. and Meekhof, D. M. and King, B. E. and Jefferts, S. R. and Itano, W. M. and Wineland, D. J. and Gould, P.},
  journal = {Phys. Rev. Lett.},
  volume = {75},
  issue = {22},
  pages = {4011--4014},
  numpages = {0},
  year = {1995},
  month = {Nov},
  publisher = {American Physical Society},
  doi = {10.1103/PhysRevLett.75.4011},
  url = {https://link.aps.org/doi/10.1103/PhysRevLett.75.4011}
}

@article{PhysRevA.67.033402,
  title = {Cooling atomic motion with quantum interference},
  author = {Morigi, Giovanna},
  journal = {Phys. Rev. A},
  volume = {67},
  issue = {3},
  pages = {033402},
  numpages = {9},
  year = {2003},
  month = {Mar},
  publisher = {American Physical Society},
  doi = {10.1103/PhysRevA.67.033402},
  url = {https://link.aps.org/doi/10.1103/PhysRevA.67.033402}
}

@article{PhysRevA.98.013423,
  title = {Double-path dark-state laser cooling in a three-level system},
  author = {Cerrillo, J. and Retzker, A. and Plenio, M. B.},
  journal = {Phys. Rev. A},
  volume = {98},
  issue = {1},
  pages = {013423},
  numpages = {10},
  year = {2018},
  month = {Jul},
  publisher = {American Physical Society},
  doi = {10.1103/PhysRevA.98.013423},
  url = {https://link.aps.org/doi/10.1103/PhysRevA.98.013423}
}

@article{retzker2007fast,
  title={Fast cooling of trapped ions using the dynamical Stark shift},
  author={Retzker, A and Plenio, MB},
  journal={New J. Phys.},
  volume={9},
  number={8},
  pages={279},
  year={2007},
  publisher={IOP Publishing}
}

@article{PhysRevA.85.032111,
  title = {Effective operator formalism for open quantum systems},
  author = {Reiter, Florentin and S\o{}rensen, Anders S.},
  journal = {Phys. Rev. A},
  volume = {85},
  issue = {3},
  pages = {032111},
  numpages = {11},
  year = {2012},
  month = {Mar},
  publisher = {American Physical Society},
  doi = {10.1103/PhysRevA.85.032111},
  url = {https://link.aps.org/doi/10.1103/PhysRevA.85.032111}
}

@article{RevModPhys.93.025001,
  title = {Programmable quantum simulations of spin systems with trapped ions},
  author = {Monroe, C. and Campbell, W. C. and Duan, L.-M. and Gong, Z.-X. and Gorshkov, A. V. and Hess, P. W. and Islam, R. and Kim, K. and Linke, N. M. and Pagano, G. and Richerme, P. and Senko, C. and Yao, N. Y.},
  journal = {Rev. Mod. Phys.},
  volume = {93},
  issue = {2},
  pages = {025001},
  numpages = {57},
  year = {2021},
  month = {Apr},
  publisher = {American Physical Society},
  doi = {10.1103/RevModPhys.93.025001},
  url = {https://link.aps.org/doi/10.1103/RevModPhys.93.025001}
}

@article{PhysRevLett.128.120404,
  title = {Spin-Holstein Models in Trapped-Ion Systems},
  author = {Kn\"orzer, J. and Shi, T. and Demler, E. and Cirac, J. I.},
  journal = {Phys. Rev. Lett.},
  volume = {128},
  issue = {12},
  pages = {120404},
  numpages = {5},
  year = {2022},
  month = {Mar},
  publisher = {American Physical Society},
  doi = {10.1103/PhysRevLett.128.120404},
  url = {https://link.aps.org/doi/10.1103/PhysRevLett.128.120404}
}

@article{Britton2012,
  doi = {10.1038/nature10981},
  url = {https://doi.org/10.1038/nature10981},
  year = {2012},
  month = apr,
  publisher = {Springer Science and Business Media {LLC}},
  volume = {484},
  number = {7395},
  pages = {489--492},
  author = {Joseph W. Britton and Brian C. Sawyer and Adam C. Keith and C.-C. Joseph Wang and James K. Freericks and Hermann Uys and Michael J. Biercuk and John J. Bollinger},
  title = {Engineered two-dimensional Ising interactions in a trapped-ion quantum simulator with hundreds of spins},
  journal = {Nature}
}

@article{PhysRevLett.121.040503,
  title = {Verification of a Many-Ion Simulator of the Dicke Model Through Slow Quenches across a Phase Transition},
  author = {Safavi-Naini, A. and Lewis-Swan, R. J. and Bohnet, J. G. and G\"arttner, M. and Gilmore, K. A. and Jordan, J. E. and Cohn, J. and Freericks, J. K. and Rey, A. M. and Bollinger, J. J.},
  journal = {Phys. Rev. Lett.},
  volume = {121},
  issue = {4},
  pages = {040503},
  numpages = {6},
  year = {2018},
  month = {Jul},
  publisher = {American Physical Society},
  doi = {10.1103/PhysRevLett.121.040503},
  url = {https://link.aps.org/doi/10.1103/PhysRevLett.121.040503}
}

@article{Gilmore2021,
  doi = {10.1126/science.abi5226},
  url = {https://doi.org/10.1126/science.abi5226},
  year = {2021},
  month = aug,
  publisher = {American Association for the Advancement of Science ({AAAS})},
  volume = {373},
  number = {6555},
  pages = {673--678},
  author = {Kevin A. Gilmore and Matthew Affolter and Robert J. Lewis-Swan and Diego Barberena and Elena Jordan and Ana Maria Rey and John J. Bollinger},
  title = {Quantum-enhanced sensing of displacements and electric fields with two-dimensional trapped-ion crystals},
  journal = {Science}
}

@article{PhysRevA.99.023409,
  title = {Modeling near ground-state cooling of two-dimensional ion crystals in a Penning trap using electromagnetically induced transparency},
  author = {Shankar, Athreya and Jordan, Elena and Gilmore, Kevin A. and Safavi-Naini, Arghavan and Bollinger, John J. and Holland, Murray J.},
  journal = {Phys. Rev. A},
  volume = {99},
  issue = {2},
  pages = {023409},
  numpages = {13},
  year = {2019},
  month = {Feb},
  publisher = {American Physical Society},
  doi = {10.1103/PhysRevA.99.023409},
  url = {https://link.aps.org/doi/10.1103/PhysRevA.99.023409}
}

\end{document}